%% file: arXiv.tex
\documentclass{jfrarticle}
\usepackage{color}
\usepackage{ProofPower}
\vertbarfalse 
\usepackage{graphicx}
\usepackage{mathpartir}
\usepackage{float}
\usepackage{url}
\usepackage{wrapfig}
\usepackage{amssymb,amsmath}
\usepackage{stmaryrd}

\usepackage[raiselinks=false,plainpages=false,pdfpagelabels,naturalnames=true,colorlinks=true,citecolor=blue,urlcolor=blue,linkcolor=blue,bookmarksopen=true,pagebackref,hyperindex]{hyperref}

\newcommand{\Hide}[1]{\relax}

\newtheorem{definition}{Definition}

\newcommand{\rewritesto}{\hookrightarrow}

\newcommand{\elit}{\Downarrow_c}
\newcommand{\ebod}{\Downarrow_b}

\def\owr{\dagger}
\newcommand{\ggoverride}[2]{#1\;\owr #2}
\newcommand{\emptymap}{\{\}}
\newcommand{\map}[1]{\{ #1 \}}
\newcommand{\dom}[1]{\textit{dom}( #1 )}
\newcommand{\len}[1]{\textit{length}( #1 )}

\newcommand{\tac}[1]{\textbf{\textit{#1}}}
\newcommand{\stac}[1]{{\textit{#1}}}

\renewcommand{\tac}[1]{\PrNL{}#1\PrNN{}}
\renewcommand{\tac}[1]{$#1$}
\newcommand{\gt}[1]{\textsf{#1}}
\newcommand{\node}[1]{\textsf{#1}}
\newcommand{\code}[1]{\textit{#1}}

\newcommand{\DRisQ}{D\mbox{-}RisQ}

\newcommand{\rev}[1]{\textcolor{blue}{#1}}
\renewcommand{\rev}[1]{#1}

\title[Understanding and maintaining tactics graphically]{Understanding and maintaining tactics graphically
OR how we are learning that a diagram can be worth more than 10K LoC}

\author{Lin, Grov and Arthan}
{Yuhui Lin, Gudmund Grov \\ Heriot-Watt University, UK \and
Rob Arthan\\ Lemma1, UK}

\begin{abstract}
The use of a functional language to implement proof strategies as \emph{proof tactics} in interactive theorem provers,
often provides short, concise and elegant implementations. Whilst being elegant, the use of higher order features and 
combinator languages often results in a very procedural view of a strategy,
which may deviate significantly from the high-level ideas behind it. This can make a tactic hard to
understand and hence difficult to  
to debug and maintain for experts and non-experts alike: one often has to tear apart complex combinations of lower level tactics manually in order to analyse a failure in the overall strategy.

In an industrial technology transfer project, 
 we have been working on porting a very
large and complex proof tactic into PSGraph, 
a graphical language for representing proof strategies, supported by the Tinker tool.
The goal of this work is to improve understandability and maintainability of tactics. 
Motivated by some initial successes with this, we here extend PSGraph with additional features for development
and debugging. Through the re-implementation and refactoring of several existing tactics, we
demonstrates the advantages of PSGraph compared with a typical linear (term-based) tactic language with respect to debugging,
readability and maintenance. In order to act as guidance for others, we give a fairly detailed  comparison
of the user experience with the two approaches. The paper is supported by  a web page 
providing further details about the implementation as well as  interactive illustrations of the examples.
\end{abstract}

\begin{document}


\begin{bottomstuff}
This work has been predominantly supported by EPSRC platform grants EP/J001058 and \linebreak EP/N014758 and IAA grant EP/K503915.
The second author is supported by a SICSA industrial fellowship and the first author
by EPSRC grant EP/M018407. The development of PSGraph was started in the AI4FM EPSRC grant (EP/H023852 and EP/H024204). We would
also like to thank {\DRisQ}, in particular Colin O'Halloran and Priiya G, for excellent discussions. \rev{We would also like to thank
the anonymous reviewers for the suggested improvements to the paper.}


\end{bottomstuff}

\maketitle

\section{Introduction}\label{sec:intro}

\emph{Proof tactics} have played an important role in reducing user interaction and proof development time for interactive theorem provers. However, tactics tend to be difficult to debug and maintain: (1) they may not fail outright and  instead generate undesirable subgoals; (2) each layer of a (reasonably) large and powerful (hierarchically composed) tactic may involve search which makes it hard to identify \emph{why} it failed and \emph{where} the culprit for the failure resides.

We will illustrate the maintenance issue with an industrial example:
{\DRisQ} Software Systems\footnote{\href{www.drisq.com}{http://www.drisq.com}} deploys a very powerful tactic 
to automate formal proofs of correctness of code auto-generated from Simulink models \cite{OHalloran13}. This tactic has been developed over a number of years and now constitutes around $10K$ lines of dense ML code ($50K$ LoC if scripts to prove supporting lemmas are included).
Both a high degree of automation and \emph{ease of maintenance} is crucial for {\DRisQ}'s business model: when a conjecture fails a developer must have an efficient way of finding and fixing the problem.  
The tactic must be intuitive to use and understand so that, as personnel move on, new developers can take over maintenance and further development. Proofs of low-level properties of automatically generated code are not interesting in themselves: what is important is the ability to produce proofs automatically as evidence that the code generator has not introduced bugs. To support this, the tactic developers will want to exploit insights from one failed and patched proof to increase the level of automation on other conjectures.

Crucial to such debugging and refactoring of tactics is a suitable \emph{tactic representation}. If one think of tactics as \emph{flow networks}, where subgoals flow between tactics, then there
is evidence that the human brain finds it more natural to understand such networks diagrammatically compared with linear (term-based) representations \cite{larkin1987diagram}. Our \emph{PSGraph} \cite{grov13} language was designed to make proof strategies more intuitive to understand and easier to debug and change than is the case with existing tactic languages. In PSGraph, the `flow graph' view is followed literally and tactics are represented as directed, typed and hierarchical graphs. Boxes are labelled by (smaller) tactics and wire labels are used to direct subgoals as they `flow' through the graph. This flow can be inspected step-by-step when debugging a graph. \rev{The PSGraph language is implemented in the \emph{Tinker} tool \cite{grov14,Lin2016}, which includes a graphical user interface to support the development and analysis of PSGraphs. The tool can support a range of theorem provers and has
currently been instantiated for Isabelle~\cite{grov14}, Rodin~\cite{Liang2016}
and  ProofPower~\cite{Arthan-Jones05}.  In this paper, we concentrate on
ProofPower, a system which is comparable with other provers in the HOL family
such as HOL4 \cite{hol4}, HOL Light \cite{hol-light} and Isabelle/HOL
\cite{isabelle-isar}.
}

Motivated by work with {\DRisQ} on their tactic in PSGraph \cite{ABZ16}, our main hypothesis \rev{of this research} is that 
\begin{quote}\it
understanding, debugging and maintaining\footnote{
\rev{
Software \emph{maintainability} refers to the ease in which a software system or component can be modified or adapted to a changed environment \cite{IEEStandardSE}. This will naturally include
both \emph{understanding} and \emph{debugging} software. However, as the main loci is on these two aspects of maintainability, the hypothesis is explicit about them. 
}}
proof strategies is easier with PSGraph than with traditional linear tactic languages.
\end{quote}
\rev{
The hypothesis relates to our overall research vision. We have already reported some evidence for it in an industrial setting \cite{ABZ16}. 
This paper has an \emph{exploratory} objective, where the goal is to study PSGraph's relative strengths and weaknesses
with respect to our given hypothesis. The contributions of this paper  are three-fold.} 

\rev{The first contribution, discussed in  \S \ref{sec:framework}, comprises recent extensions to  PSGraph and the Tinker tool, with new features to improve development and debugging. The most notable extension is the introduction of a new language for specifying wire labels.}



\rev{The second and, in our view, main contribution is found in \S \ref{sec:case_studies}, 
where we address our hypothesis by means of three case studies, each with a distinctive flavour 
and level of complexity. 
Evaluation based upon case-studies is motivated by the work being \emph{exploratory} and \emph{improvement-driven} \cite{runeson2009guidelines}; the aim is to identify 
actionable limitations of PSGraph with respect to the hypothesis -- and to provide the necessary armoury to address D-RisQ's tactic in full. 
We reflect on alternative evaluation approaches in \S \ref{sec:related}.
}

To expose the differences between the user experience with the traditional
linear representation of proofs and proof procedures and the user
experience with our graphical representation, we work through several simple
but instructive aspects of the case studies in some detail. Our aim is to give
a good understanding of what goes on in the two approaches and to guide future
work on and with PSGraph by ourselves and others. This leads us to the third 
contribution of this paper: to provide a tutorial-like introduction to 
PSGraph and how to go about connecting Tinker to a theorem prover. To support this, we
therefore provide a fairly detailed background on ProofPower and PSGraph/Tinker
in \S \ref{sec:background}. Furthermore, \rev{\S \ref{sec:tinker} and \S \ref{sec:tacticlift} are fully devoted to prover integration, 
while parts of \S \ref{sec:proofpower} and \S \ref{sec:envtactic} discuss such integration. These parts can be skipped if desired.
Some aspects of the case studies are very detailed for the same reasons.}
However, space does not permit a discussion of
every detail and so particularly in the second case study, we have tried to
give the flavour of the bigger picture supported by enough information to help
interested readers find their way around the original source material.

After the description of each case study, we reflect and analyse our approach and provide recommendations which we hope can be used as a template
for other developments. Crucially, while two of the authors (Lin and Grov) are developers of PSGraph, the third author (Arthan) had never used
PSGraph before we started this work. Arthan was the developer of the original version of the case studies. 	

Our three case studies consider already extant proofs and proof procedures
implemented in ProofPower: they comprise {\em(i)} a proof procedure
for tautologies supplied as part of the standard proof infrastructure,
{\em(ii)} some application-specific tactics used to finesse a tricky
lemma forming part of the proof of security properties of a database
system and {\em(iii)} a decision procedure 
taken from a collection of case
studies on pure mathematics in ProofPower that automates problems
such as proving the continuity of real-valued functions.

In \S \ref{sec:related} we discuss related
work, and we conclude and discuss further work in \S \ref{sec:concl}. Additional supporting materials, including animations of the 
examples and detailed instructions are available from a dedicated webpage \cite{webpage}.


\section{Background}\label{sec:background}

\subsection{ProofPower}\label{sec:proofpower}




\input{proofpower}

\subsection{PSGraph}

In PSGraph, tactics are represented as directed, typed, hierarchical and open graphs. A graph consists of \textit{boxes}, representing ``processes'' and typed (labelled) \textit{wires} that 
connect them together. A process box is labelled by a tactic, which can either be an existing ProofPower tactic or another graph, where the latter introduces hierarchies. The graphs are \emph{open}
in that wires need not be connected to a box at both ends, but can be left open to represent graph inputs and graph outputs. 
Evaluation is achieved by adding input goals to a graph input wire. 
\rev{The goals will then flow through the graph; each step will apply a tactic to an incoming goal, consume the goal, and add the resulting subgoals
to its output wires. This process will continue until all subgoals appear on the graph output wires. These subgoals will then be returned.}
All wires are labelled by \emph{goal types}, which are predicates on a goal that are used to direct goals to the correct tactic.


\begin{figure}
\begin{center}
\includegraphics[width=0.5\textwidth]{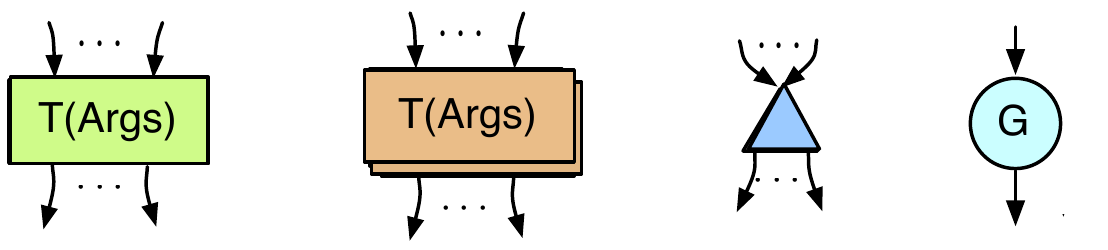} \\
 \hspace{-2em} atomic tactic \hspace{1em} graph tactic \hspace{0.1em} identity tactic \hspace{0.1em} goal
\end{center}
\vspace{-10pt}
\caption{Types of graph nodes for PSGraph}
\label{fig:psgraph:nodes}
\end{figure}


Fig.~\ref{fig:psgraph:nodes} shows the types of boxes that can appear in a PSGraph. An \emph{atomic} tactic is an existing ProofPower tactic, possibly parametrised. A parameter could for example be the name of a rewrite rule to apply or a term used to instantiate a variable. Note that if the list of parameters ($Args$) is empty, then we can write $T$ instead of $T()$. A \emph{graph} tactic is labelled by a named graph, which we can look up. For graph tactics, the arguments ($Args$) relate to the scope of the variables of the goal node environment, which we return to in \S \ref{sec:envtactic}. An \emph{identity} tactic is used to split and merge multiple wires and, as discussed below, will not have any side-effects on the proof state or goal nodes. The final type of box is a \emph{goal}. This is only used for evaluation and cannot be added to a PSGraph by the user. It contains sufficient information to evaluate and link to the ProofPower proof state, including:
\begin{itemize}
\item the name given by ProofPower for the goal;
\item the internal representation of the goal in ProofPower; and 
\item an environment that is used to support variables in the graph, which is discussed in detail in \S \ref{sec:envtactic}.
\end{itemize}
 When displaying the goals, we will only show the name (see e.g. Fig.~\ref{fig:demo eval simple ps}). We return to goals when discussing evaluation below. 



\begin{figure}[h]
\vspace{-5pt}
\centering
\includegraphics[width=0.3\textwidth]{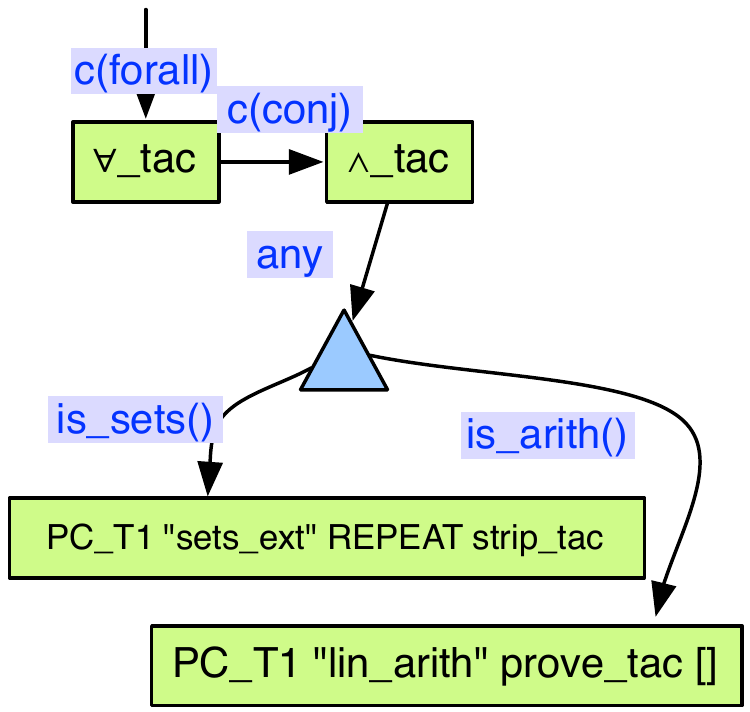}
\vspace{-10pt}
\caption{A PSGraph encoding of the proof discussed in \S \ref{sec:proofpower}}\label{fig:demo simple ps}
\end{figure}

Finally, the wires are labelled by \emph{goal types}, which are predicates defined on goals. Intuitively, these provide information about some characteristics, such as ``shape'', of a goal, which are used to influence the path a goal takes as it passes through the graph. We develop a language for expressing these in \S \ref{sec:goaltype}, and defer the details to that section.

A simple but complete example of a PSGraph is given in Fig.~\ref{fig:demo simple ps}. This is a PSGraph encoding of the proof discussed in \S \ref{sec:proofpower}. The input wire is labelled by 
$c(forall)$ which means that the conclusion ($c$) must be a universal quantifier. The $\forall\_tac$ tactic is applied to it followed by the $\wedge\_tac$ tactic, as long as the conclusion of the
goal is a conjunction. The $any$ goal type always succeeds. The \emph{identity} tactic is then used to separate the goals that are arithmetic ($is\_arith$) from those that are set theoretic ($is\_sets$),
where the suitable tactic is applied in both instances. 

\subsubsection{Evaluation}\label{sec:psgraph:eval}

\begin{figure}
$$
R_{eval} =  
\left\{
\raisebox{1.2cm}{
\inferrule
 {
 \langle \textit{defs,atoms} \rangle \vdash h_{10} : H_1  \cdots  \langle \textit{defs,atoms} \rangle \vdash h_{1m} : H_1 \\\\
 \vdots \\\\
  \langle \textit{defs,atoms} \rangle \vdash h_{n0} : H_n  \cdots \langle \textit{defs,atoms} \rangle \vdash h_{nk} : H_n \\\\
([h_{10},\cdots,h_{n0},\cdots,h_{n0},\cdots,h_{nk}],\_) = T(Args) ~g
 }{
\vcenter{\hbox{\includegraphics[width=1.7cm]{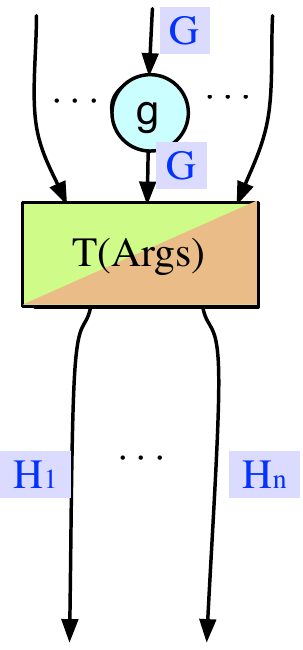}}}  
\quad
\rewritesto
\quad
\vcenter{\hbox{\includegraphics[width=1.7cm]{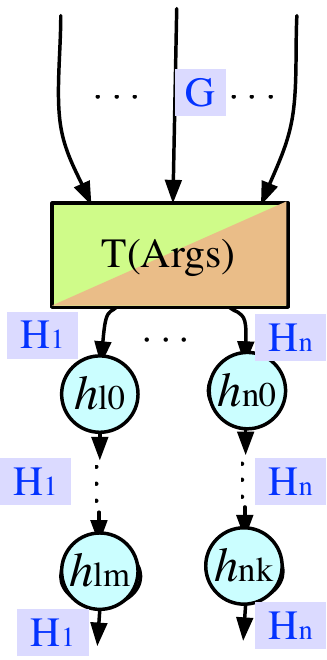}}} } }
 \quad , \quad
\raisebox{1.5cm}{
\inferrule
 {
 \langle \textit{defs,atoms} \rangle \vdash g : G
 }{
    \vcenter{\hbox{\includegraphics[width=1.2cm]{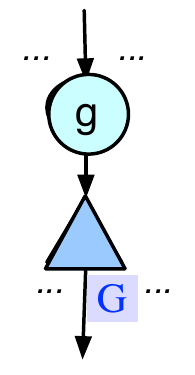}}} 
    \rewritesto 
     \vcenter{\hbox{\includegraphics[width=1.35cm]{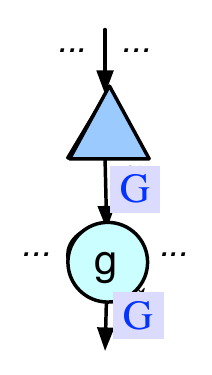}}}
 }} \right\}
$$
\caption{Evaluation of PSGraph.}
\label{fig:eval}
\end{figure}

In order to initialise the evaluation (i.e. proof of) of a subgoal by a PSGraph, a ProofPower goal must be provided. This is achieved by calling  \emph{top\_goal()}, which
will provide the first subgoal from the ProofPower subgoal package. From this subgoal, a goal node with an empty environment is created,
and added to a graph input wire where the goal type is satisfied. 
A step of the proof will apply a tactic box to a goal that is on an input wire. Each generated subgoal (if any) will be added to an output wire where the goal type is satisfied.
This process will continue until \emph{termination}:
\begin{definition}[(Termination)]\label{def:terminated}
A graph has \textit{terminated}, if for all goals $g$ of the graph, $g$ is either on a graph output wire or it is wired to another goal.
\end{definition}


It follows by induction over the number of goals present that all goals are (directly or indirectly) on graph output wires. 

It is worth noting that (as with other tactic languages) termination is not guaranteed. \rev{One example of non-termination is that an atomic tactic from the theorem prover may not terminate; another example is a PSGraph that contains a non-terminating loop, e.g. as a result of rewriting in presence of commutative operators.}

As the proof state is handled by ProofPower's subgoal package, evaluation is only concerned with how the subgoals ``flow'' from the 
graph input to the graph output wires. Two properties are crucial for \rev{a \emph{successful} evaluation step}:
\begin{itemize}
\item No subgoals are lost, that is if a tactic produces a subgoal then it will appear on the graph.
\item No subgoals are duplicated in the graph.
\end{itemize}

At the graph level, evaluation is achieved by graphical rewriting, where $l \rewritesto r$ is a rule that rewrites $l$ to $r$\footnote{When applied to a graph, this rule will match $l$ with a subgraph and replace this with $r$. For more details see \cite{paper:Dixon:10,EhrigEPT06}.}.
Fig.~\ref{fig:eval} gives the set $R_{eval}$ of rewrite rules used to evaluate a PSGraph. We use a notation where side-conditions are above the line and the rewrite is below the line. Each rule is non-deterministic in the sense that there may be several ways to apply it in a given situation.  Evaluation of a graph is achieved by applying rules from $R_{eval}$ repeatedly until no rules 
are applicable. At this point evaluation have either failed or successfully terminated.

The simplest case is the \emph{identity} box, shown rightmost in Fig.~\ref{fig:eval}. Here, the input goal and the output goal is the same as the node is essentially used to fork the goal to the correct target box.  $\langle \textit{defs,atoms} \rangle \vdash g : G$ is a predicate that holds if goal $g$ satisfies goal type $G$. The ellipses illustrate that there could be other input and output wires. Note that if there are more than one output wire where $g$ satisfies the goal type then there will be multiple rewrites. Each of these rewrites will be a separate branch of the search space. We 
will return to the goal type predicate in \S \ref{sec:goaltype}.

The leftmost rule of Fig.~\ref{fig:eval} shows the evaluation of an atomic or graph tactic. \rev{
For an atomic tactic, $T(Args) ~g$ is the result of applying the tactic (in ProofPower). This is described in \S \ref{sec:tacticlift}. For a graph tactic, this is the result of evaluating the nested graph as discussed below.}
For these tactics, there will be a side-effect on the proof state, which we return to in  \S \ref{sec:tacticlift}. The case when $T(Args)$ is an \emph{atomic} tactic can be summarised as follows:
\begin{enumerate}
  \item Apply tactic $T(Args)$ to obtain a list of subgoal nodes.
  \item Consume $g$ from the graph.
  \item Add all valid combination of the resulting subgoals to output wires. 
\end{enumerate}


\begin{figure}
\vspace{-10pt}
  $\vcenter{\hbox{\includegraphics[width=0.29\textwidth]{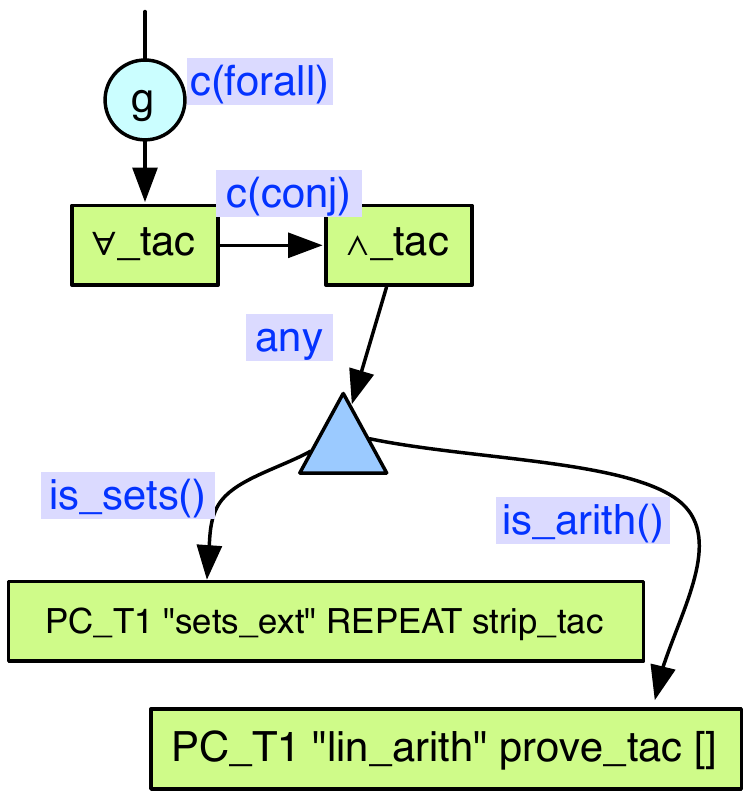}}}$
  \large{$\rightsquigarrow_2$}
  $\vcenter{\hbox{\includegraphics[width=0.29\textwidth]{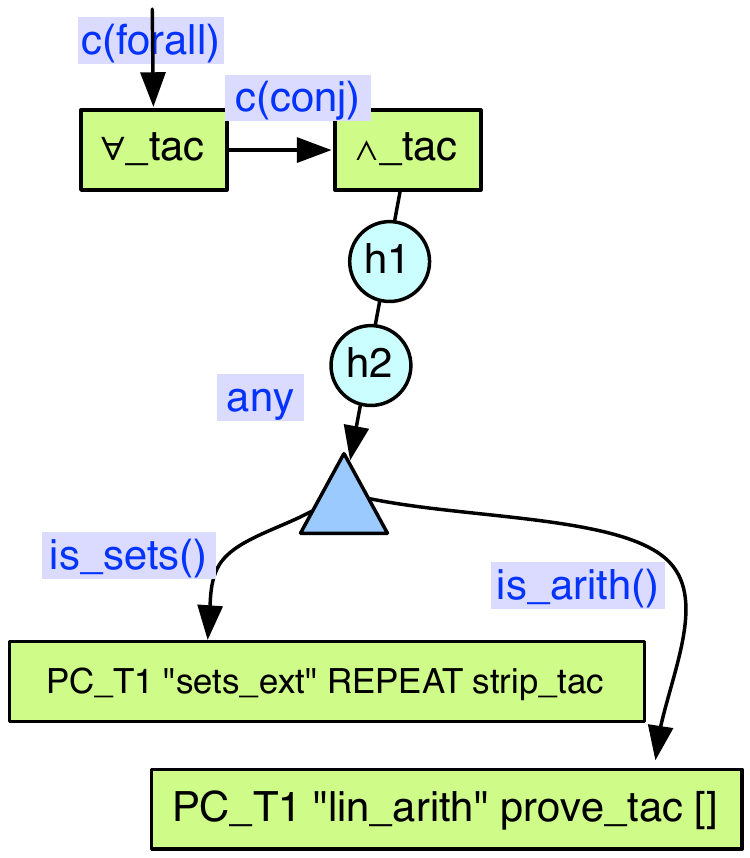}}}$
  \large{$\rightsquigarrow_2$}
  $\vcenter{\hbox{\includegraphics[width=0.29\textwidth]{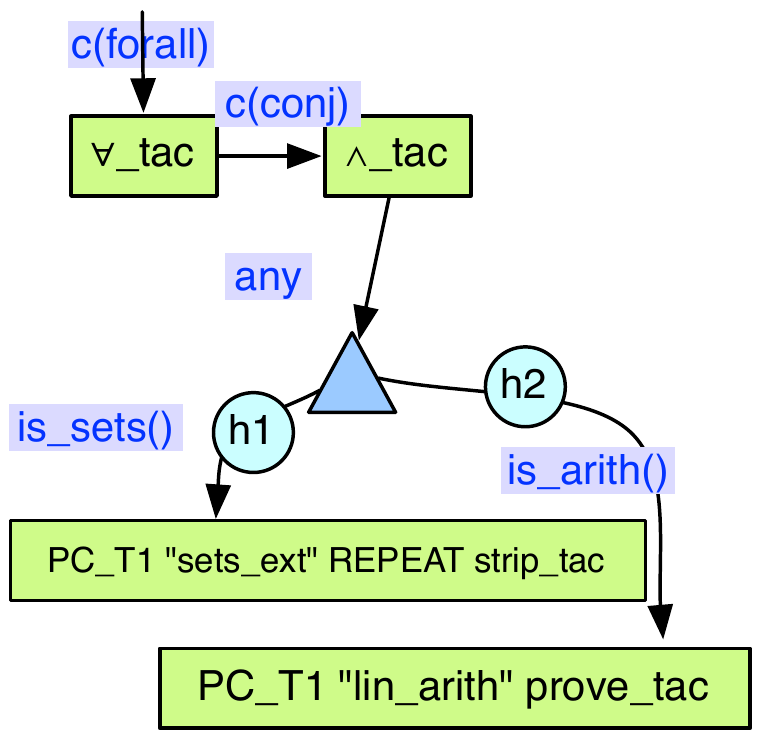}}}$
\caption{Example evaluation of PSGraph.}\label{fig:demo eval simple ps}
\end{figure}
Fig.~\ref{fig:demo eval simple ps} illustrates some of the steps of the flow through the proof strategy of Fig.~\ref{fig:demo simple ps} applied to the example of \S \ref{sec:proofpower}. In the left-most graph, $g$ holds the initial goal:
\begin{GFT}{}
\+(* ?\PrPE{} *)\PrKM{}\PrLF{} x\PrLH{} (1, x) \PrIN{} \{(a, b)|a = 1 \PrLC{} b \PrLO{} 2\} \PrLB{} (x \PrLO{} 3 \PrLC{} x \PrLM{} 4)\PrKO{}\\
\end{GFT}
It then applies $R_{eval}$ twice, which will first apply universal introduction followed by conjunction introduction, introducing two new subgoals, $h1$: 
\begin{GFT}{}
\+(* ?\PrPE{} *)\PrKM{}(1, x) \PrIN{} \{(a, b)|a = 1 \PrLC{} b \PrLO{} 2\}\PrKO{}\\
\end{GFT}
\noindent and $h2$:
\begin{GFT}{}
\+(* ?\PrPE{} *)\PrKM{}x \PrLO{} 3 \PrLC{} x \PrLM{} 4\PrKO{}\\
\end{GFT} 
\noindent  Next it applies the identity tactic to $h2$  and then $h1$, with the result shown in the right-most graph. This is used to route the goals to the correct tactic to complete the proof. $h1$ is set-theoretic and thus goes down the left branch while $h2$ is arithmetic and follows the right branch. Note that when there are multiple goals, as in the right most graph, the order in which goals are evaluated will have no impact on the end result\footnote{This would not have been the case in presence of shared meta-variables between goals, a feature that is not currently supported in either ProofPower's subgoal package or PSGraph.}.

When $T$ in the leftmost rule of Fig.~\ref{fig:eval} is a \emph{graph} tactic,  the arguments $Args$ of $T$ are used to introduce 
\emph{local scoping}: any variable not in $Args$ is ``fresh'' in the nested scope and will not have global effect. The evaluation
can be summarised as:
\begin{enumerate}
\item Consume $g$ from the graph. 
\item Lookup the graph $G$ which $T$ points to.
\item Constrain the environment of $g$ to variables in $Args$ and add this to an input wire of $G$ (such that the goal type is satisfied). If there are multiple satisfying input wires, then one branch will be generated for each. 
\item Evaluate $G$ until termination.
\item Add all valid combination of the goals on the output wires of $G$ to the output wires of the graph tactic $T$. 
\end{enumerate}
If any steps fail then evaluation of this node fails. Note that when adding a resulting subgoal to the output of $T$ in the last step, the subgoal will be given the environment of $g$, with values of $Args$ replaced by those in the resulting subgoal. We will 
return to how this works in \S \ref{sec:envtactic}, after introducing environments more formally.

\begin{figure}
\vspace{-10pt}
\centering
\includegraphics[width=0.4\textwidth]{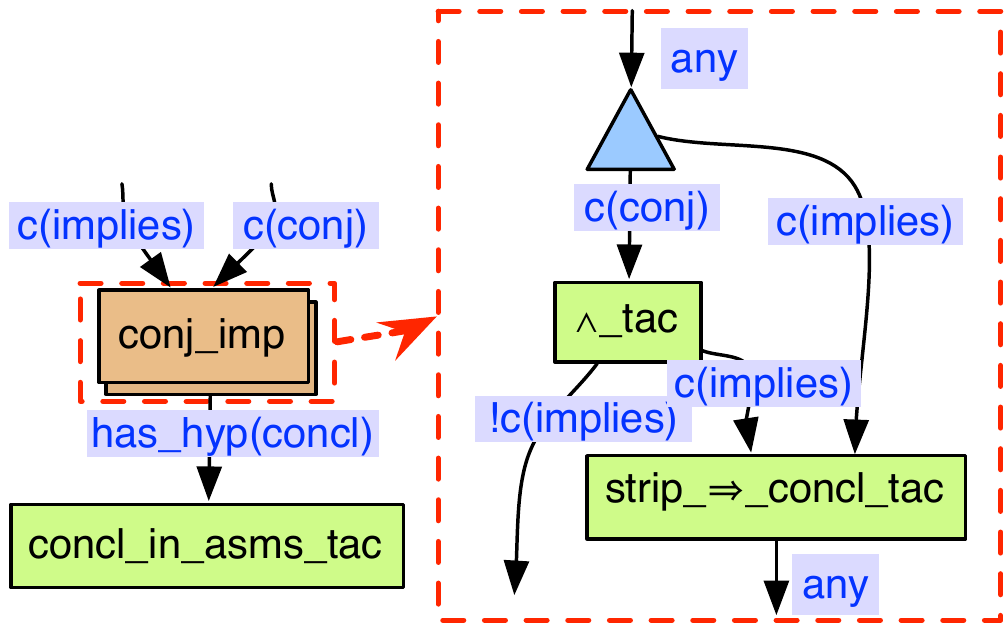}
\caption{Example PSGraph with graph tactic.}\label{fig:psgraph:nested:ex1}
\end{figure}

\begin{figure}
\begin{center}
  $\vcenter{\hbox{\includegraphics[width=0.19\textwidth]{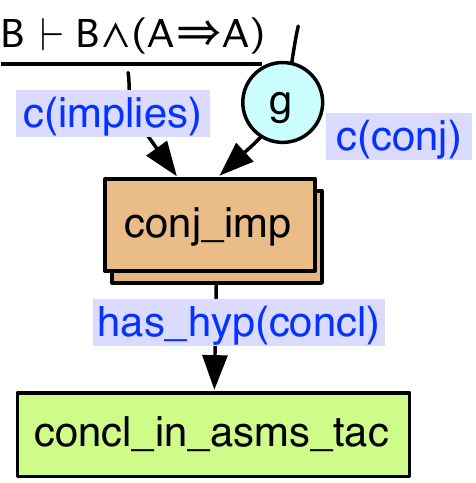}}}$
  \large{$\rightsquigarrow$}
  $\vcenter{\hbox{\includegraphics[width=0.2\textwidth]{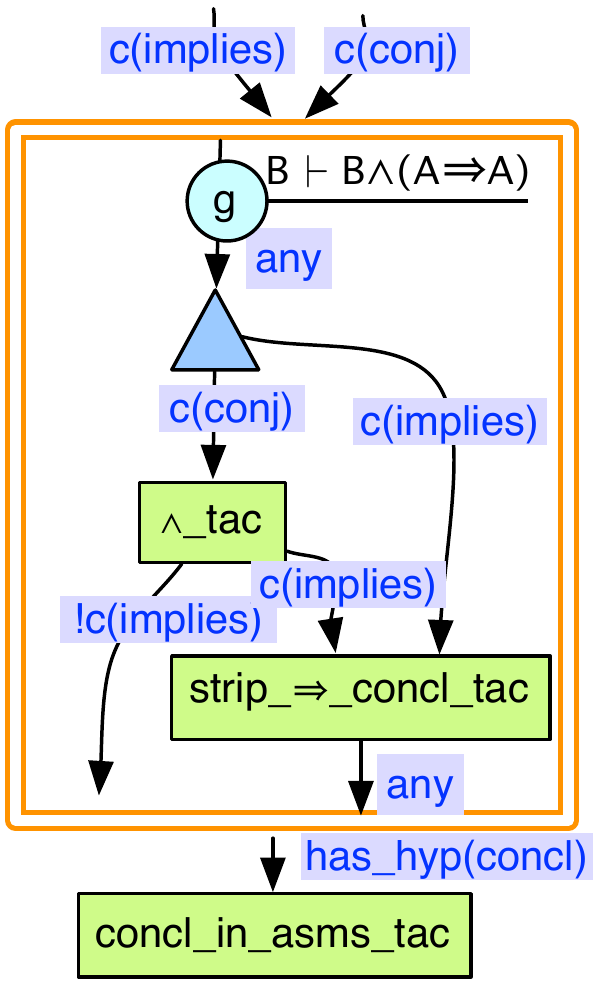}}}$
  \large{$\rightsquigarrow_3$}
  $\vcenter{\hbox{\includegraphics[width=0.2\textwidth]{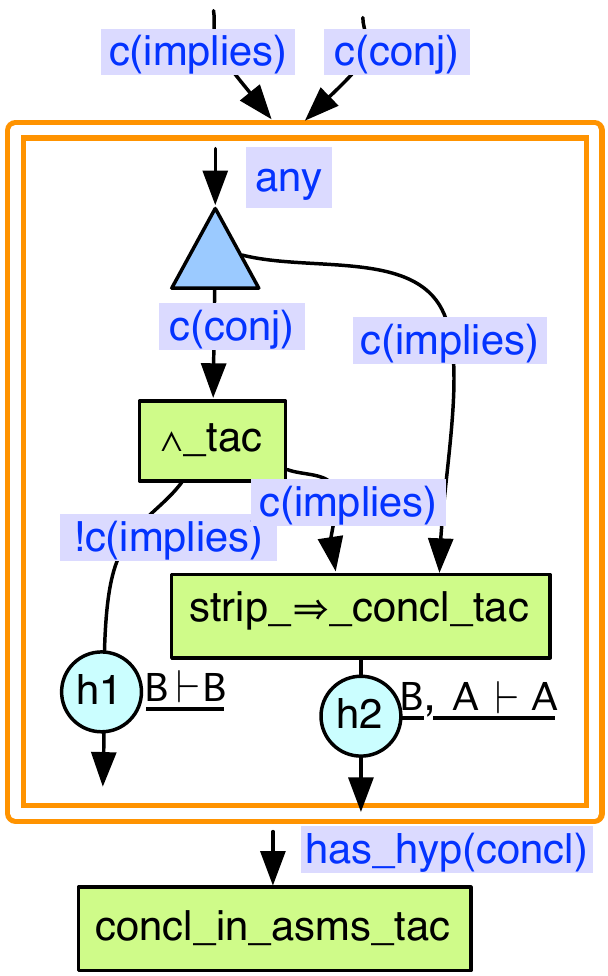}}}$
    \large{$\rightsquigarrow$}
  $\vcenter{\hbox{\includegraphics[width=0.17\textwidth]{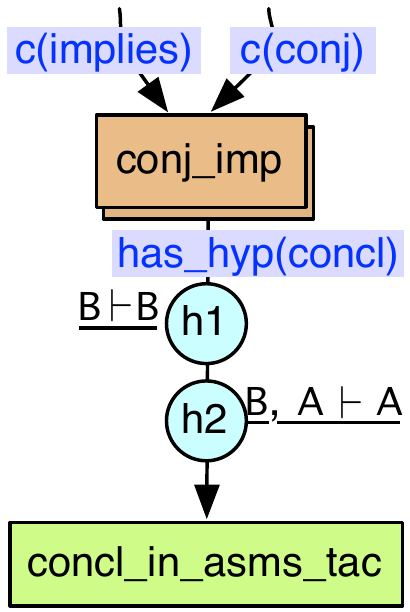}}}$
\end{center}
\caption{Example evaluation of graph tactic in PSGraph.}\label{fig:psgraph:nested:eval:ex1}
\end{figure}

\rev{
To illustrate how a graph tactic is evaluated, consider the PSGraph in Fig. \ref{fig:psgraph:nested:ex1}.
It contains a graph tactic \textit{conj\_imp}, with two input wires and one output wire. The input
wires require that the conclusion of the a goal is either an implication ($c(implies)$) or a conjunction ($c(conj)$). 
The output must have an hypothesis that is the same as the conclusion. It is then proven by assumption by \emph{concl\_in\_asms\_tac}.}

\rev{The graph
nested by \emph{conj\_imp} is shown in the red stippled box of Fig. \ref{fig:psgraph:nested:ex1} (right). 
Depending on the input goal, it will either break up the conjunction or the disjunction, and in the former case, it 
may also be followed by breaking up a disjunction. In order to illustrate several aspects of evaluation, the number of input/output wires of the nested graph,
and their goal types, deviates from the parent \textit{conj\_imp} graph tactic.}

\rev{
Consider Fig. \ref{fig:psgraph:nested:eval:ex1},
which shows the key evaluation steps for a goal $g$:}
\begin{GFT}{}
\+(*  1 *)\PrKM{}B\PrKO{}\\
\+(* ?\PrPE{} *)\PrKM{}B \PrLB{} (A \Rightarrow A)\PrKO{}\\
\end{GFT}
\noindent \rev{applied to one of the input wires of the graph of Fig. \ref{fig:psgraph:nested:ex1}.
}

\rev{
In the first step, this goal is consumed from the parent graph and added to one of the input wires of the nested graph. Note that for evaluation there is no correspondence between the input wires of the parent box (in this case labelled by $c(conj)$), and the input wire of the nested graph (here $any$). The goal $g$ is simply added to any input wire of the nested graph where the goal type holds (with a separate branch in the search space for each such wire). In this case there is only one possibility. It will then go through three steps of evaluation of the nested graph, and at the end there are two goals, $h1$ and $h2$, on the output wires of the nested graph. According to the definition of termination, the graph tactic has now terminated. The (nested) graph will then ``return'' the list of goals $[h1,h2]$\footnote{The order of the goals returned is irrelevant.}. These are then added to the output wires where the goal type is satisfied, as the case is for an atomic tactic. In this case, both matches the goal type of  \textit{conj\_imp}'s output wire ($has\_hyp(concl)$) since the conclusion is found in the list of hypothesis. Again, note that there is no relationship between the goal types of the output wires of the nested graph ($!c(implies$ and $any$), and those of the nesting graph tactic ($has\_hyp(concl)$). The \textit{concl\_in\_assms\_tac} tactic will then discharge both $h1$ and $h2$ by assumption.
}




\subsubsection{Architecture \& GUI of the Tinker tool}\label{sec:tinker}

\begin{figure}
\begin{center}
\includegraphics[width=0.5\textwidth]{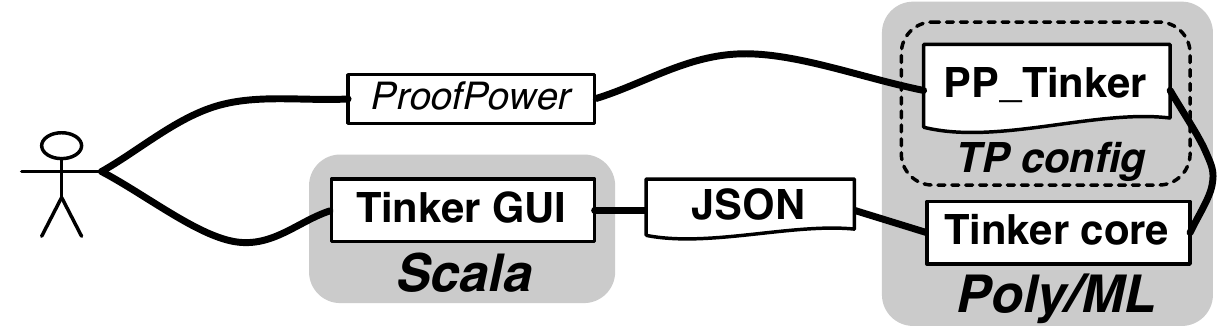}
\end{center}
\vspace{-10pt}
\caption{Tinker architecture.}\label{fig:arch}
\end{figure}

The Tinker tool \cite{grov14,Lin2016} implements PSGraph with support for the \emph{Isabelle}, \emph{Rodin} and \emph{ProofPower} theorem provers. Here, we will focus on the ProofPower version only. Tinker consists of two parts: the \textit{CORE} and the \textit{GUI}. These are shaded in separate boxes in Fig.~\ref{fig:arch}. The core implements the main functions of Tinker. Most of the functions are implemented using ML functors to achieve theorem prover independence. In order to connect a theorem prover to Tinker, and use its GUI and basic functionality, a ML structure that implements a provided ML signature called \textit{PROVER} has to be provided. This will enable basic usage of Tinker and the GUI. In Fig.~\ref{fig:arch} the structure implementing this signature is called 
\textit{PP\_Tinker}.

\rev{Note that our ambition is not to replace the existing tactic language, but to offer a different view of tactics. Tinker is designed
to support a dynamic interplay between a PSGraph and the existing tactic language, where the level of atomicity of the atomic tactics used
in PSGraph is flexible. This enables developers to decide themselves which parts are best to express in PSGraph and which are not.}

\rev{The remaining of this subsection, as well as the next subsection, are intended for readers interested in the details of how Tinker connects to theorem provers. Other re.}

The \emph{PROVER} signature includes both the types and functions required. It has to know how types, terms, theorems and contexts are represented:
\begin{GFT}{}
\+type typ\\
\+type term\\
\+type thm\\
\+type context\\
\end{GFT}
\noindent To illustrate, in ProofPower these are instantiated to:
\begin{GFT}{}
\+type typ = TYPE\\
\+type term = TERM\\
\+type thm = THM\\
\+type context = string list * string\\
\end{GFT}
The CORE communicates with the GUI (written in Scala) over a JSON socket protocol, which requires serialisation functions for some of this types (via strings), e.g.:
\begin{GFT}{}
\+val trm\_of\_string : context -> string -> term\\
\+val string\_of\_trm : context -> term -> string\\
\end{GFT}  
The GUI allows users to develop proof strategies in a mostly graphical approach, and to debug proof strategies with controlled interactive inspections. 

\begin{figure}[ht]
\begin{center}
\includegraphics[width=0.9\textwidth]{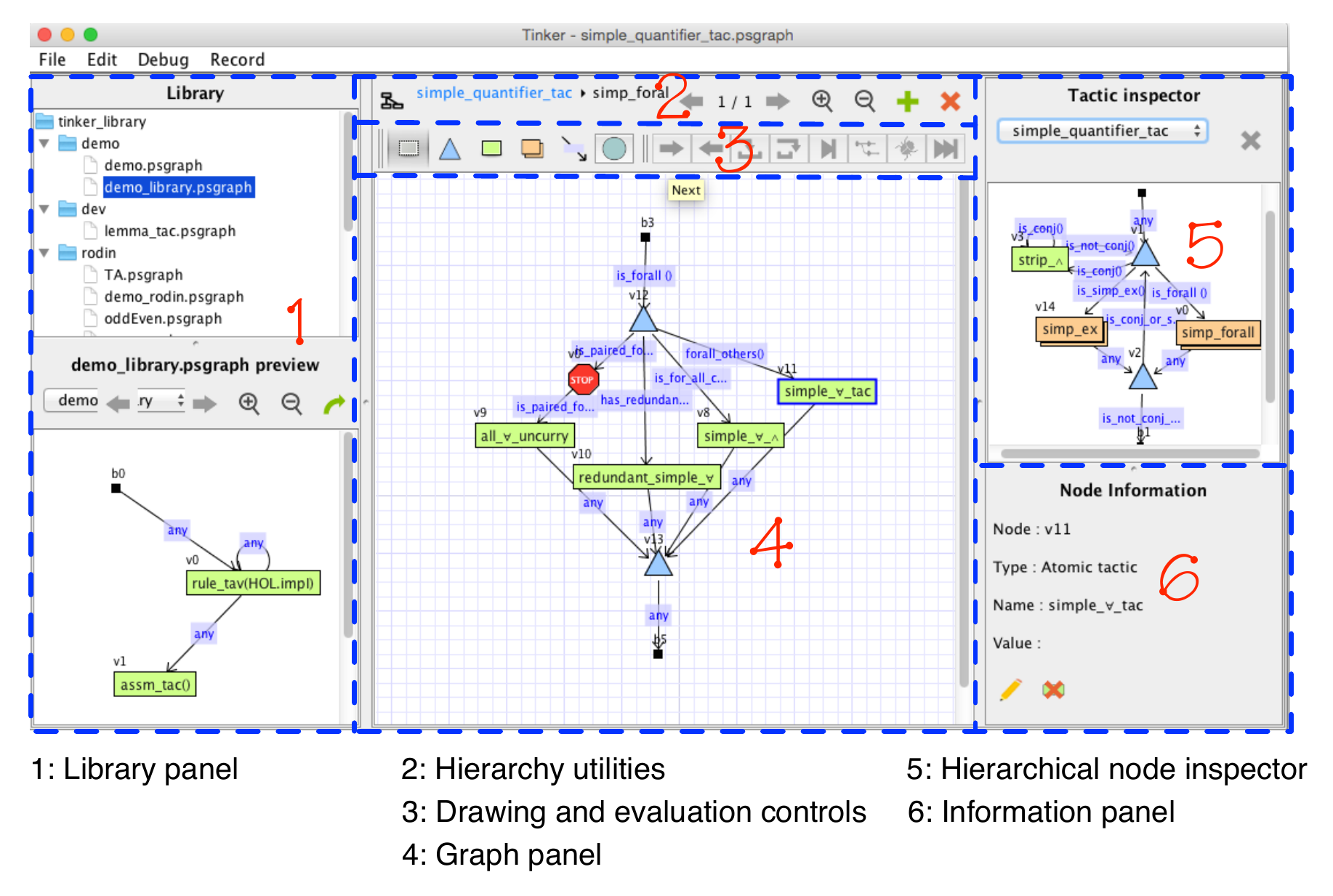}
\end{center}
\vspace{-20pt}
\caption{The Tinker GUI \cite{Lin2016}.} 
\label{fig:tinkeroverview}
\end{figure}

Fig.~\ref{fig:tinkeroverview} shows the components of the GUI. The \emph{graph panel} is the main area where users view and edit the graph of current proof strategies. With the interactive options in the \emph{drawing and evaluation control panel}, users can develop graphs in a click and drag style, and step through evaluation with controls such as step over a graph tactic. When a user selects a node or edge in the graph panel, the detail information of the node or edge will be showed in the \emph{information panel}. To facilitate developing hierarchical graphs, the \emph{hierarchical node inspector} panel allows users to preview the sub-graphs of a graph tactic; and the \emph{hierarchy utility} panel shows the depth and path of the graph in the graph panel. For reusing existing proof strategies, there is also a \emph{library panel} to preview existing PSGraphs, and import them to the graph panel. The core is implemented on top of the Quantomatic graphical rewrite system \cite{kissinger2015quantomatic}. Videos of interaction with the different features of the GUI are available from \cite{webpage}.

Tinker also needs to know about tactics and their execution, also provided via implementation of the \textit{PROVER} signature:
\begin{GFT}{}
\+type tactic\\
\end{GFT}
\noindent Tinker uses the underlying prover's proof state and goal representation augmented with some additional book-keeping information represented
in the following ML types:
\def\VCDate{2016/08/13}\def\VCVersion{(Current)}
\begin{GFT}{}
\+type pplan \\
\+type pnode\\
\end{GFT}
\noindent The proof state (type \emph{pplan}) is mainly used to link with the proof state of the theorem prover and keep track of the goals that are ``active'' in the PSGraph.
To illustrate, in ProofPower this is a record where the main fields are the underlying goal state of ProofPower and the goals that the PSGraph are allowed to work on:
\begin{GFT}{}
\+type pplan = \{gstate: GOAL\_STATE, opengs : pnode table, ...\} \\
\end{GFT}
\noindent The type \emph{pnode table} is a map from a \emph{string} to a \emph{pnode}. The key fields of the goal representation (\emph{pnode}) are the name of the goal, its internal representation
and an environment:
\begin{GFT}{}
\+type pnode = \{pname : string, g: GOAL, env : env, ...\} \\ 
\end{GFT}

\noindent The Tinker representation of a tactic uses these types, and therefore has type:
\begin{GFT}{}
\+type appf = pnode * pplan -> (pnode list * pplan) Seq.seq\\
\end{GFT}
\noindent To be used in Tinker, the underlying prover's tactics (i.e. functions of type \emph{tactic}) have to be ``lifted" to \emph{application functions} with the above type \emph{appf}.


\subsubsection{Tactic ``lifting''}\label{sec:tacticlift}

An atomic tactic $T(Args)$ encapsulates a tactic of the underlying theorem prover, i.e., ProofPower for the purposes of the present paper. Recall from \S \ref{sec:proofpower} that a ProofPower tactic maps a goal to a pair of new subgoals and a validation function. PSGraph uses ProofPower's existing package for handling the subgoal state, meaning we can ignore the validation function at this level. We write 
$$(gs,\_) = T (Args) ~g$$ 
to denote that $T(Args)$ returns the list of subgoals $gs$ when applied to $g$. 

We need to be able to connect ProofPower tactics, which works on goals, to our atomic tactic boxes,  which works on goal nodes
(i.e. type \emph{pnode}). The simplest case is when there are no arguments, which we can just write $T$.
All the atomic tactics of Fig.~\ref{fig:demo simple ps} are examples of this case.
Here, the label of the atomic box, e.g.  \textit{PC\_T1 "lin\_arith" prove\_tac[]}, is wrapped into a function that will
take a goal node and produce a list of new goal nodes as follows. It will extract the goal name from the input goal node
and set this to be the top goal in ProofPower's goal stack, using a ProofPower function called \emph{set\_labelled\_goal}.
It will then apply the wrapped tactic (to the goal at the top of the stack), and finally it will put each goal name and goal into a goal node, together with the environment of the input node. We call this process ``lifting'' of the ProofPower tactic to PSGraph. 

As there are no arguments, the string of the atomic tactic box is simply parsed as ML code and applied. 
Such parsing has to be provided in the \emph{PROVER} signature:
\begin{GFT}{}
\+val exec\_str : string -> unit\\
\end{GFT}
\noindent \rev{This is a generic parser interface, which is also used for other parsing tasks.
Internally, Tinker provides functionality to cast it to the correct type after some minor user configurations.
For tactics, it will cast it to the type \textit{tactic}.}
The ``lifting'' into the \emph{appf} type is trivial.

When there are arguments, then we need to manually provide some ML code for this\footnote{We hope to introduce some level of automation for this process in the future.}. To illustrate, consider the  ProofPower tactic \emph{prove\_tac} discussed above.  We would like to parameterise over the proof context, which we can do by providing the name of a proof context as a parameter:
\begin{GFT}{}
\+fun prove\_with\_ctxt0 ctxt = PC\_T1 ctxt prove\_tac []; \\
\end{GFT}
However, this will not work in PSGraph. In order to use such parametrised tactics, the function needs to have a different type.
Internally, the arguments of an atomic tactic are represented using a deep embedding, i.e. as a list of an inductive datatype with a constructor for each type:
\begin{GFT}{}
\+datatype arg\_data =  A\_Trms of term list | A\_Var of string | A\_Str of string | \cdots \\
\end{GFT}
Arguments must be passed as a list of \textit{arg\_data}, and such arguments have to be reflected in the tactic type of \emph{PROVER}; for ProofPower it is\footnote{For example, Isabelle in addition needs the context and the index of the subgoal as arguments.}:
\begin{GFT}{}
\+type tactic = arg\_data list -> TACTIC\\
\end{GFT}
With this type, the signature has to be provided an interpretation of tactics in terms of the defined application function (type \textit{apps}):
\begin{GFT}{}
\+val apply\_tactic : arg\_data list -> tactic -> appf\\
\end{GFT}
Returning to our example, we represent the context as a string, so we provide the following ``lifting'' function: 
\begin{GFT}{}
\+fun prove\_with\_ctxt [A\_Str ctxt] = PC\_T1 ctxt prove\_tac [] \\
\+  |  prove\_with\_ctxt \_ = fail\_tac; \\
\end{GFT}
As a result, \emph{prove\_with\_ctxt(A\_Str lin\_arith)} will apply this function, with \emph{lin\_arith} parsed as a string. 

We will introduce a new type of tactic in \S \ref{sec:envtactic}, while \S \ref{sec:case_studies} contains many examples of tactics with and without arguments.

\section{A (mostly) graphical development \& debugging framework}\label{sec:framework}

In \S \ref{sec:case_studies} we will showcase PSGraph and the Tinker tool by developing, debugging and refactoring several 
case studies adapted from the existing ProofPower developments. To support this we first extend PSGraph and Tinker with new features.

In \S\ref{sec:envtactic} and \S \ref{sec:goaltype} we add features that are mainly beneficial for development: in 
\S\ref{sec:envtactic} we introduce a new family of tactics used to exchange information and constraints between tactics and 
goal types; while in \S \ref{sec:goaltype} we develop a goal type that allows us to hide low-level details in the graphs to improve readability. 

In \S \ref{sec:breakpoint} and \S \ref{sec:logging} we develop support for debugging: \S \ref{sec:breakpoint} introduces breakpoints to PSGraph, while \S \ref{sec:logging} describes a simple, yet useful,
logging mechanism for Tinker.

\subsection{``Environment'' tactics}\label{sec:envtactic}



\def\Env{\mathsf{Env}}
\def\Var{\mathsf{Var}}
\def\GVar{\mathsf{GVar}}
\def\EnvVal{\mathsf{EnvVal}}

Recall that the type of a tactic is from a goal to a pair consisting of a list of new subgoals and a validation function. In an atomic tactic box, such tactic is then ``lifted" to work on the goal nodes that are in the graph.  In addition to the actual goal, such a goal node also contains an \emph{environment}, which we introduce here:
\begin{definition}[(Environment)]\label{def:psgraph:env}
An environment $\Env$ is a function
$$
 \Var \rightarrow \EnvVal
$$
where $\Var$ is a finite set of variables named by strings prefixed with a `?' and $\EnvVal$ is the disjoint union
$$
T \uplus N \uplus T^* \uplus N^*
$$
where $T$ and $N$ denote the set of all terms and names respectively,
where a name is an arbitrary uncapitalised string and $X^*$ denotes the set of
lists of elements of $X$\footnote{\rev{Note that $T$ and $N$ are redundant as they can be represented as singleton sequences
of $T^*$ and $N^*$, respectively. We find it more natural to separate them, as they are often treated differently
(e.g. some tactics only work on a single term). This will also simplify static checking, which we plan to add in the future.}}.
\end{definition}


The validity of a  name depends on the context. For example, it could be a named lemma, which will only be valid if that lemma exists. Two special names are: \textit{concl}, which refers to the conclusion of a given goal; and \textit{hyps}, which refers to the (list of) hypothesis of the goal.

\begin{figure}
\begin{center}
\includegraphics[width=0.28\textwidth]{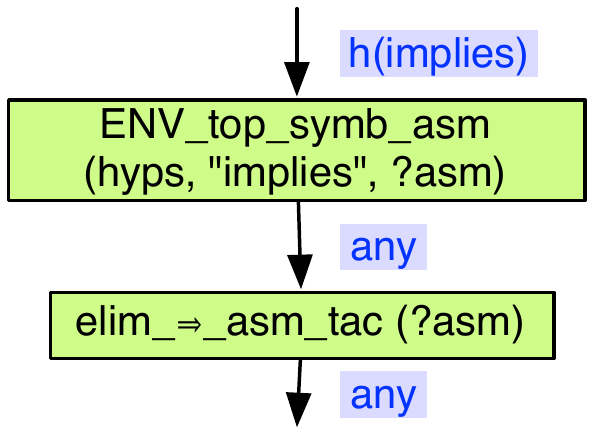} 
\end{center}
\vspace{-15pt}
\caption{An example of ``environment'' tactic}
\label{fig:env tac exmaple}
\end{figure}

An environment is used to pass information between tactics and between tactics and goal types. This could be to 
extract some information at one point in the proof and use it later. For example, consider Fig.~\ref{fig:env tac exmaple}.
Here, \emph{ENV\_top\_symb\_asm} looks for a hypothesis that starts with an implication, and binds it to a variable $?asm$.
In the next atomic tactic, $elim\_\Rightarrow\_asm\_tac$ will apply implication elimination to the hypothesis that $?asm$ is bound to.

Within graph tactics, local scoping of the environment is achieved by using the arguments $Args$ of the box. Recall the
evaluation steps of graph tactics, as given in \S \ref{sec:psgraph:eval}. We are given a goal with an environment $\{?x \mapsto v_1,?y \mapsto v_2\}$ and a graph tactic $t(?x)$. When evaluating $g$ for the graph $t$ references, the environment  $\{?x \mapsto v_1\}$ is provided. Assume that on termination of $t$ that a new goal $g'$ has the environment  $\{?x \mapsto v_3,?z \mapsto v_4 \}$. This will return a goal $g'$ with environment  $\{?x \mapsto v_3,?y \mapsto v_2\}$. This illustrates how environments are constrains for ``local computation'' within graph tactics.

Now, the problem with \emph{ENV\_top\_symb\_asm}, discussed above, is that it binds $?asm$, meaning the result of applying it is a change to the environment, while a ``lifted'' tactic will change the goal (and proof state) and cannot change the environment. In this case, this issue could be overcome by combining these two atomic tactics into a single ProofPower tactic. However, part of the reason to have them as a separate boxes is to enable users to inspect the flow, and use Tinker's debugging features if a tactic application fails. By combining them into a single tactic the granularity becomes too terse for such analysis.
A second problem is that there are more complex examples where there are tactic in between binding and using a variable, which we will see examples of in \S \ref{sec:case_studies}. For these cases the solution of merging boxes will not work. 

Instead we introduce a type of atomic tactic that works on the environment, which we call \emph{environment tactics}\footnote{\rev{An alternative to environment tactics,
is to bind variables during matching in goal types, which is explained in \S \ref{sec:goaltype}. We have made a design decision to treat goal types as predicates,
and therefore to not support this. This is a deliberately compromise made in order to cleanly separate concerns and to simplify the semantics of PSGraph composition (see \cite{grov13}).}}:
\begin{definition}[(Environment tactic)] An \emph{environment tactic}, is an atomic tactic, with a name prefixed by `ENV\_', whose underlying function is a function 
$$
 \Env \rightarrow \Env^*
$$
\end{definition}
\rev{The rest of this section focuses on implementation issues of environments and environment tactics. This material
is intended for readers interested in the technical details of the connection between Tinker and a theorem prover. It can can be skipped if desired.}
To support environment tactics, the \textit{PROVER} signature is augmented with new types for an environment and an environment tactic:
\begin{GFT}{}
\+type env = env\_data table \\
\+type env\_tac\\
\end{GFT}
\noindent The type \textit{env\_data table} is a map from a string to the type \textit{env\_data}, which holds the types an environment may contain
\begin{GFT}{}
\+datatype env\_data = E\_Str of string | E\_Trm of term  | \cdots \\
\end{GFT}

As environment tactics may  also have arguments ($Args$), they have a dual type and application function to ProofPower tactics:
\begin{GFT}{}
\+type env\_tac = arg\_data list -> env -> env list\\
\+val apply\_env\_tactic :  arg\_data list  -> env\_tac -> appf\\
\end{GFT}
\noindent From the types one can see that an environment tactic will not change the underlying proof state; it will only change the environment of a goal node (type \textit{pnode}). However, they may still require features of the provers, such as matching of terms.

\emph{ENV\_top\_symb\_asm} is an example of an environment tactic. As we can see from Fig.~\ref{fig:env tac exmaple}, it takes three arguments: the first (\emph{hyps}) is a list of terms, the second (\emph{"implies"}) is a string, and the third (\emph{?asm}) is a variable. The underlying function that has to be provided by the user will look something like:
\begin{GFT}{}
\+fun ENV\_top\_symb\_asm [A\_Trms ts, A\_Str s, A\_Var var] env = \cdots \\
\end{GFT}
Note that Tinker will automatically lookup \emph{hyps}, which is the list of hypothesis, before calling this function. 
We will see many examples of environment tactics in \S \ref{sec:case_studies}.

\subsection{A constraint language to express goal types}\label{sec:goaltype}


Goal types are crucial in order to achieve maintainable proof strategies and to reduce the search space. These are
represented as \emph{constraints} on the goal, and to represent these, we develop goal types as a Prolog inspired \emph{constraint language}. 
\rev{Prolog is a natural starting point as constraints can be combined in an elegant and declarative manner\footnote{Compared with for example using the underlying implementation language of Tinker (ML).}, and enables support for machine learning goal types using a technique based upon logic-based learning \cite{farquhar2015typed}.}
\rev{Analogous to how graphs are used to compose tactics, this language acts as a way of combining and re-using low-level atomic constraints, which the underlying prover needs to provide. By supporting recursive definitions, expressive constraints can be 
encoded, and lower-level details can be hidden in the goal types appearing on the graphs.}

\rev{
A relation may have (goal type) variables that are instantiated. This is discussed below. 
However, a goal type appearing in a graph cannot have any such variables. We therefore distinguish \emph{goal type schemas}, which may have goal type variables, from \emph{goal types}, which does not allow the use of such variables:}
\begin{definition}[(Goal type \& goal type schema)]\label{def:psgraph:gt} A \emph{goal type schema} (GTS) is a predicate on a \textit{goal}, defined by the following BNF:
$$
\begin{array}{rclcrcl}
GTS & ::= &  C , GTS ~|~ C . & \qquad & Oa & ::= &  As ~|~ \epsilon \\ 
C & ::= &   L ~|~ !L && As & ::= &  A ~|~ A , As \\ 
L & ::= &   F( Oa ) && A & ::= &  T ~|~ N  ~|~ \GVar \\ 
\end{array}
$$
Following from Definition \ref{def:psgraph:env} (Environment): $T$ denotes a term; $F$ denotes a named fact and $N$ denotes a name, which
in either case is an arbitrary uncapitalised string; and $\GVar$ denotes a goal type variable, which is an arbitrary capitalised string.
A \emph{goal type} is a goal type schema without any goal type variables, i.e. with $A ::=  T ~|~ N$ 
\end{definition}
\noindent To illustrate, the goal type schema
$$
\textit{top\_symbol}(T,Y).
$$
expresses that the top level symbol of (goal type variable) $T$ has to be  (goal type variable) $Y$. 
An example goal type using this schema: 
$$
\textit{top\_symbol}(concl,\wedge), \textit{has\_top\_symbol}(hyps,\wedge).
$$
It states that the conclusion and one hypothesis of the goal has to be a conjunction; as we shall see later, \textit{has\_top\_symbol} can be defined in terms of \textit{top\_symbol}. 



In practice, we have found that most goal types, \rev{such as $\textit{top\_symbol}(T,Y)$,} are constraints over terms.
These may have to be provided by the underlying theorem prover, and we call them \emph{atomic goal types}.
One example atomic goal type is $\textit{top\_symbol}(T,Y)$, while $any()$, which we can also write $any$, is provided by default. The $any$ predicate will always succeed. 

As will be seen in case studies, many goal types combines such atomic goal types. To achieve readable and intuitive proof strategies, low-level implementation details needs to be
hidden to highlight the high-level concepts of the graph. To support that, a user can define a new goal type schemas, which can be used by a goal type:
\begin{definition}[(Goal type schema definition)]
A \emph{goal type schema definition} (GTSD) is a \emph{rule} defined by:
$$
\begin{array}{rclcrcl}
GTSD & ::= &  N( Ov ) \leftarrow GTS & \qquad & Ov & ::= &  Vs ~|~ \epsilon \\
&  | & N( Ov ) \leftarrow GTS ~~ GTSD  & & Vs & ::= &  \GVar ~|~ \GVar , Vs 
\end{array}
$$
\end{definition}
We often call the left hand side of `$\leftarrow$'  the \emph{head} and the right hand side the \emph{body}.

To illustrate, the goal type scheme definition:
$$
g\_h(X,Y) \leftarrow top\_symbol(concl,X),has\_top\_symbol(hyps,Y).
$$
requires that the first argument is the top symbols of the conclusion and that there exists a hypothesis that has the top symbol of the second argument. This is used in the following definition:
$$
\begin{array}{l}
g(X) \leftarrow g\_h(X,\wedge). \\
g(X) \leftarrow g\_h(X,\vee).
\end{array}
$$
Here, $g(X)$ is defined to be a goal type scheme where the conclusion has the top symbol given by the argument $X$, and there is a hypothesis with either $\wedge$ or $\vee$ as top symbol.  $g(\wedge)$ is an example of a a valid \emph{goal type} using this schema. 

One can also use variables in the body not present in the head. This is used to pass arguments between literals of the body. For example, 
$$
concl\_top\_in\_hyp() \leftarrow top\_symbol(concl,X),has\_top\_symbol(hyps,X).
$$
expresses that there is a hypothesis with the same top symbol as the conclusion. 

Recall from Fig.~\ref{fig:eval} (\S \ref{sec:psgraph:eval}) that in an evaluation step of a PSGraph, we need to check if a goal $g$ satisfies a goal type $G$. This depends on the provided atomic goal types \emph{atoms}  and goal type definitions \emph{defs}. We write 
$$
\langle \textit{defs,atoms} \rangle \vdash g : G
$$
to express that such relation holds. 
In order to determine this, information about the values of variables has to be passed between the clauses. For example, consider \textit{concl\_top\_in\_hyp()}. Here, the $X$ of both the clauses in the body has to be the same; in other words, $has\_top\_symbol(hyps,X)$ needs to know the value of $X$ from $top\_symbol(concl,X)$. 
To achieve this, an environment is passed between the literals. This environment is different from the
environment in the goal node in that goal type variables are bound, and we call it a \emph{goal type environment}:
\begin{definition}[(Goal type environment)]\label{def:psgraph:gtenv}
A goal type environment (GTEnv) is a function:
$$
 \GVar \rightarrow \EnvVal.
$$
\end{definition}
For $concl\_top\_in\_hyp()$, the result of applying the first $top\_symbol$ will be an environment with $X$ bound, which is then used in the second application of $top\_symbol$. 

In order to specify $\langle \textit{defs,atoms} \rangle \vdash g : G$,
we introduce a relation that generates a goal type environment:
$$
\langle \textit{defs,atoms} \rangle \vdash \langle g,G \rangle \Downarrow gtenv
$$
This should be read as: given a \emph{context} consisting of a pair of the atomic goal types \emph{atoms}  and goal type definitions \emph{defs}, and an \emph{input} consisting of a pair of a goal $g$ and a goal type $G$, a goal type environment \textit{gtenv} is produced. $\langle \textit{defs,atoms} \rangle \vdash g : G$ can then be defined in terms of the existence of such a goal type environment:
$$
(\langle \textit{defs,atoms} \rangle \vdash g : G) \Leftrightarrow (\exists~ gtenv.\; \langle \textit{defs,atoms} \rangle \vdash \langle g,G \rangle \Downarrow gtenv)
$$
\rev{
The semantics of the $\Downarrow$ relation is inspired by Prolog with some key differences. Firstly, due to the
``lifted'' nature over atomic goal types, most of the unification work is provided by the underlying prover.
Secondly, we have to work with two distinct environments. Thirdly, we have to communicate with the underlying
theorem prover. We can therefore not use Prolog directly in a natural way, and decided to develop a domain specific
version,} which we provide big-step operational semantics for next.

The semantics of $\Downarrow$ are non-deterministic, in that it can generate multiple valid goal type environments (\emph{gtenv}).
As $\langle \textit{defs,atoms} \rangle \vdash g : G$ is only concerned with the existence of a valid goal type environment, it does not matter which of the valid ones is found.
The definition of the semantics uses auxiliary relations $\Downarrow_b$ and $\Downarrow_c$.

The evaluation of a goal type is a special case of the evaluation of the body of a goal type schema, the difference being that the relation is over a goal type environment, which is updated. This is evaluated by the relation $\ebod$, where $b$ stands for `body'. This is evaluated over an environment and a clause body. As  there are no side-effects on the goal, the goal and its environment are moved to the context:
$$
 \inferrule
 {
   \langle env(g),defs,atoms , g \rangle \vdash \langle \emptymap,G \rangle \ebod{} gtenv
 }{
\langle \textit{defs,atoms} \rangle \vdash \langle g,G \rangle \Downarrow gtenv
 } 
 $$
The body is either a single clause `$C.$', or a clause followed by more clauses `$C, GTS$'. As a result there are four cases: 
`$C, GTS$', `$C.$', atomic goal types and negated literals. 

For the second case, the clause is evaluated. We write this as $f\;vs$ where $vs$ is a list of the arguments. To evaluate this we try to find a definition of $f$ in all the definitions. This is achieved by
the $\elit{}$ relation, where $c$ stands for `clause':
$$
 \inferrule
 {
   \langle env,defs,atoms , g \rangle \vdash \langle gtenv,f\;vs,\textit{defs} \rangle \elit{} gtenv'
 }{
  \langle env,defs,atoms , g \rangle \vdash \langle gtenv,f\;vs. \rangle \ebod{} gtenv'
 } 
 $$
 Each definition is terminated by `.'; we therefore ensure that all clauses are evaluated:
 $$
 \inferrule
 {
   \langle env,defs,atoms , g \rangle \vdash \langle gtenv,f\;vs,\textit{clauses} \rangle \elit{} gtenv'
 }{
  \langle env,defs,atoms , g \rangle \vdash \langle gtenv,f\;vs,\textit{clause}.\;\textit{clauses} \rangle \elit{} gtenv'
 } 
$$
The main work happens in the case where we find a definition with a head $f\;\textit{Vs}$ of the same name. The formal parameter list \textit{Vs} is a list of variable names of the same length as \emph{vs}. To illustrate, assume we are evaluating the underlined $f$ in
$$
\cdots \leftarrow h(X),\underline{f(concl,X,Y,?x)},\cdots.
$$
As a result of evaluating $h$, $X$ may be bound to some value $e_1$ in the goal type environment. Furthermore, assume that $concl$ is $e_2$ and $?x$ is bound to $e_3$ in the environment. When a definition of the same name, such as

$$
f(A,B,C,D) \leftarrow \textit{body}.
$$
is found, then the formal parameters must be instantiated. In this case it should generate an environment $\map{A \mapsto e_2,B\mapsto e_1,D \mapsto e_3}$. Note that these are \emph{constraints} of the variables; as $Y$ is not bound it is unconstrained, and is therefore not included.  To achieve this instantiation we first introduce the partial function \emph{lookup}:
$$
\textit{lookup}(\textit{goal},\textit{env},\textit{gtenv},v) =\left\{
\begin{array}{ll}
\textit{get\_name}(\textit{goal},v) & \textbf{if}\,v \in \textit{name} \\
\textit{env}(\textit{v}) & \textbf{if}\,v \in \textbf{dom}(\textit{env}) \\
\textit{gtenv}(\textit{v}) & \textbf{if}\,v \in \textbf{dom}(\textit{gtenv}) \\
\end{array}\right.
$$
This function looks up values of names or variables when present in one of the environments. Note that
$concl$ and $hyps$ are examples of names, and \emph{get\_name} will in those cases return the underlying 
conclusion (a term) or list of hypothesis (list of terms), respectively.
We then define a function that is used to apply the instantiations:
$$
\begin{array}{lll}
\textit{inst\_gtenv}(\textit{goal},\textit{env},\textit{gtenv},[V_1,\cdots,V_n],[v_i,\cdots,v_n])   := \\
\;\; \{ V_i \mapsto \textit{lookup}(\textit{goal},\textit{env},\textit{gtenv},v_i) |
(\textit{goal},\textit{env},\textit{gtenv},v_i) \in \dom{lookup} \}
\end{array}
$$

Returning to our example, the \textit{body} of $f$ is then evaluated, starting with the initial environment 
$\map{A \mapsto e_2,B\mapsto e_1,D \mapsto e_3}$ generated. A result of this evaluation is a new environment. Let's assume that this binds $C$ and
a new variable $F$: $\map{A \mapsto e_2,B\mapsto e_1,D \mapsto e_3,\underline{C \mapsto e_4, F \mapsto e_5}}$. The clause should return an environment
with only the variables in the actual parameters bound. In this case, the call made was $f(concl,X,Y,?x)$, meaning only $X$ and $Y$ should be in the domain -- which
corresponds to variables $B$ and $C$:  $\map{X\mapsto e_1,Y \mapsto e_4}$. This functionality is handled by the 
\emph{res\_gtenv} function:
$$
\begin{array}{lll}
\textit{res\_gtenv}(\textit{gtenv},[V_1,\cdots,V_n],[v_i,\cdots,v_n])   := \\
\qquad \{ v_i \mapsto \textit{gtenv}(V_i)  | v_i \in var \wedge V_i \in \dom{gtenv} \}.
\end{array}
$$
As a result, the derivation rule for evaluating a single goal type schema becomes:
$$
 \inferrule
 {
   \len{vs} = \len{\textit{Vs}} \\
   gtenv_0 = inst\_gtenv(g,\textit{env},gtenv,\textit{Vs},vs) \\
   \langle env,defs,atoms , g \rangle \vdash \langle gtenv_0,\textit{body} \rangle \ebod{} gtenv_1 \\
   gtenv' =  res\_gtenv(g,gtenv_1,\textit{Vs},vs) 
 }{
  \langle env,defs,atoms , g \rangle \vdash \langle gtenv,f\;vs,f\;\textit{Vs} \leftarrow \textit{body}.\;\textit{clauses} \rangle \elit{} gtenv'
 } 
$$
with a special case when $f$ is the last definition:
  $$
 \inferrule
 {
   \len{vs} = \len{\textit{Vs}} \\
   gtenv_0 = inst\_gtenv(g,\textit{env},gtenv,\textit{Vs},vs) \\
   \langle env,defs,atoms , g \rangle \vdash \langle gtenv_0,\textit{body} \rangle \ebod{} gtenv_1 \\
   gtenv' =  res\_gtenv(g,gtenv_1,\textit{Vs},vs) 
 }{
  \langle env,defs,atoms , g \rangle \vdash \langle gtenv,f\;vs,f\;\textit{Vs} \leftarrow \textit{body}. \rangle \elit{} gtenv'
 } 
 $$
 Another option is that $f$ is an atomic goal type, which is then applied to generate the new environment:
 $$
  \inferrule
 {
    f \in \dom{atoms} \\ gtenv' \in atoms(f)\;gtenv\;(v_1,\cdots, v_n)\; g
 }{
    \langle env,defs,atoms , g \rangle \vdash \langle gtenv,f\;(v_1,\cdots,v_n) \rangle \ebod{} gtenv' 
 } 
$$
We have already mentioned $any$ and $top\_symbol(X,Y)$ which are both atomic goal types. 
Other generic atomic goal types include:
\begin{itemize}
\item $trm\_var(X)$ holds if term $X$ is a variable.
\item $member(XS,X)$ holds if $X$ is a member of list $XS$.
\item $eq\_trm(X,Y)$ holds if the terms $X$ and $Y$ are syntactically equal ($\alpha$-equivalence).
\end{itemize}
These can for example be used in a schema to check if a given term is the same as a hypothesis or the conclusion:
$$
\begin{array}{lcl}
 is\_goal(X) & \leftarrow & eq\_trm(concl,X). \\
 has\_hyp(X) & \leftarrow & member(hyps,Y), eq\_trm(X,Y). 
\end{array}
$$
\rev{
We can also use $member$ to define the above $has\_top\_symbol(X,Y)$:
$$
\begin{array}{lcl}
 has\_top\_symbol(X,Y) & \leftarrow & member(X,Z), top\_symbol(Z,Y). 
\end{array}
$$
}
A third case is negation. This case will behave as an identity function on the environment, if the non-negated version fails:
 $$
  \inferrule
 {
    \lnot\big(\exists~gtenv'. {\langle env,defs,atoms , g \rangle \vdash \langle gtenv,f\;(v_1,\cdots,v_n) \rangle \ebod{} gtenv'}\big)
 }{
    \langle env,defs,atoms , g \rangle \vdash \langle gtenv,!f\;(v_1,\cdots,v_n) \rangle \ebod{} gtenv
 } 
$$

The final case is the evaluation of the body of a goal type of multiple clauses, i.e. the case: `$C, GTS$'. Here, evaluation
is sequential: first $C$ is evaluated 
by $\elit$, then the rest $GTS$ is evaluated recursively by $\ebod{}$. However, the goal type environments cannot just be passed sequentially as the following example illustrates.
Consider:
$$
p(X,Y,Z) \leftarrow a(X,Z), b(Y), c(Z).
$$
Assume we have the following application: $p(t_1,t_2,A)$. Here, the initial environment will be
$$
\{ X \mapsto t_1, Y \mapsto t_2 \}
$$
However, when applying $a$ this is restricted to $X$ and $Z$, so $a$ will only return an environment with $X$ and $Z$ in.
If we use this directly the $Y$ binding is lost. 

Instead, the environment is \emph{updated} with the new values. Note that the constraints
are checked in each element -- thus it is safe to override. Finally, the two environments are combined, where the latter overrides\footnote{
$\ggoverride{A}{B}$ denotes that $B$ overrides $A$} the former:
$$
 \inferrule
 {
   \langle env,defs,atoms , g \rangle \vdash \langle gtenv,c_1,\textit{defs} \rangle \elit{} gtenv'' \\
   \langle env,defs,atoms , g \rangle \vdash \langle \ggoverride{\textit{gtenv}}{\textit{gtenv}''},(c_1,c_2) \rangle \ebod{} gtenv''' \\
   \textit{gtenv}' = \ggoverride{(\ggoverride{\textit{gtenv}}{\textit{gtenv}'')}}{\textit{gtenv}'''}
 }{
  \langle env,defs,atoms , g \rangle \vdash \langle gtenv,(c_1,c_2) \rangle \ebod{} gtenv'
 } 
 $$
As the example above illustrates, the environment following $b$ only contains $Y$. Thus $X$ and $Z$ are added, whilst $X$ and $Z$ have to be added after evaluating $c$. 
This completes the evaluation semantics of goal types.

\rev{
To illustrate more complex usage of goal types, we provide an example from ongoing work by Farquhar and others on machine learning PSGraphs and goal types using 
a technique called \emph{meta interpretive learning} \cite{farquhar2015typed}.
 Here, we provide low-level operations on terms and, using a small set of examples, learn suitable goal type definitions
from them. 
}

\rev{To achieve this, we introduce an atomic goal type
$$dest\_trm(X, L, R),$$
which holds if term $X$ is ``destructed'' into its left $L$ and right $R$ sub-terms, when such exists\footnote{Applications may also instantiate variables. For example, $dest\_trm(\PrKM{}f~x~y\PrKO{},V_1,V_2)$ will
instantiate $V_1$ to $\PrKM{}f~x\PrKO{}$ and $V_2$ to $\PrKM{}y\PrKO{}$, if they are not bound in the goal type environment.}
For example $dest\_trm(\PrKM{}f~x~y\PrKO{}$,$\PrKM{}f~x\PrKO{}$,$\PrKM{}y\PrKO{})$ holds, meaning that
$\PrKM{}f~x\PrKO{}$ is the left sub-term and $\PrKM{}y\PrKO{}$ is right sub-term of
$\PrKM{}f~x~y\PrKO{}$. By using \emph{dest\_trm}, we can
define \textit{left} and \textit{right} as:
$$
\begin{array}{lcl}
\textit{left}(X,L) & \leftarrow & dest\_trm(X,L,\_). \\
\textit{right}(X,R) & \leftarrow & dest\_trm(X,\_,R). 
\end{array}
$$
Next, we introduce another atomic goal type
$$
const(X,C)
$$
which holds if and only if term $X$ is a the constant (name) $C$. 
For example, $const(\PrKM{}f\PrKO{},f)$ holds. Instead of treating 
\textit{top\_symbol(T,Y)} as an atomic goal type we can define it using these more primitive 
atomic goal types and recursion:
$$
\begin{array}{lcl}
top\_symbol(T,Y) & \leftarrow & const(T,Y). \\
top\_symbol(T,Y) & \leftarrow & \textit{left}(T,Z), top\_symbol(Z,Y). \\
\end{array}
$$
This goal type will traverse the left side of the term until the end is reached and check if this is the correct constant (or bind it to variable $Y$).
Note that it will not work if the top-level function is higher-order (i.e. a lambda abstraction). We can also define a function that checks both the left and right
side of an application. This amounts to checking if a symbol is present at any place of the term\footnote{For simplicity, this definition does not work in presence of binders (lambda abstractions),
but can easily be extended to support this with a new atomic goal type.}:
$$
\begin{array}{lcl}
has\_symbol(T,Y) & \leftarrow &  const(T,Y). \\
has\_symbol(T,Y) & \leftarrow & \textit{right}(T,Z), has\_symbol(Z,Y). \\
has\_symbol(T,Y) & \leftarrow &  \textit{left}(T,Z), has\_symbol(Z,Y).
\end{array}
$$}

\rev{
These examples show that, as with PSGraph, we can work with the goal types at different levels of abstraction. This is illustrated by treating \emph{top\_symbol}
as an atomic goal type or by defining it in the language in terms of lower-level more primitive atomic goal types.
}

%

\subsection{Graphical breakpoints}\label{sec:breakpoint}

There are two ways to apply a PSGraph to a goal in Tinker: (1) in the \emph{automatic mode} it is applied as a black box and all you see is the final subgoals on the output wires; (2)
in the \emph{interactive mode} the user can step through and guide the proof of the goal. When debugging a large proof, such as our current work with {\DRisQ}'s tactic  \cite{ABZ16},
one often wants to combine these modes: one would like to use an automatic/black-box execution until the problematic part of the proof strategy is reached, and at that point enter an interactive mode where the user can step through the proof.

\begin{figure}[h]
\vspace{-10pt}
$$
R_B = 
     \vcenter{\hbox{\includegraphics[width=0.75cm]{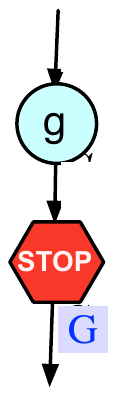}}} 
    \;\;\rewritesto\; 
     \vcenter{\hbox{\includegraphics[width=0.75cm]{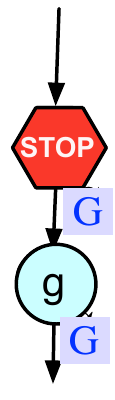}}}
$$
\vspace{-15pt}
\caption{Breakpoint rule}\label{fig:break}
\end{figure}

This is essentially how \emph{breakpoints} of modern IDEs work: the user inserts a breakpoint in the program text, and the debugger will execute the code until the breakpoint is reached. At that point the user can manually step through the code. Inspired by this idea for debugging programs, we extend PSGraph with a new special \emph{breakpoint node}, which can be seen in Fig.~\ref{fig:break}.

We also introduce a third mode called \emph{debug mode}. The intuition behind this mode is to achieve exactly the requirement above: the graph is executed as in automatic mode until it cannot execute any further, either because it has successfully terminated or because the goals are followed by a debug node. In order to keep the semantics of PSGraph, we only need to update the termination condition for the debug mode:
\begin{definition}[(Termination in debug mode)]
A graph has \textit{terminated} in \emph{debug mode}, if for all goals $g$ of the graph, $g$ is either on a graph output wire or it is wired to another goal, or is wired to a debug node.
\end{definition}


If the graph has successfully terminated in debug mode, it will enter interactive mode and the user can step through the graph manually. In this case, goals needs to be able to ``step over'' breakpoints, which is achieved by adding the rule 
 $R_B$ from Fig.~\ref{fig:break} to the ruleset $R_{eval}$ (Fig.~\ref{fig:eval}) when we are in \emph{interactive} and \emph{automatic} modes, whilst omitting it from $R_{eval}$ in \emph{debug} mode. 
 
This very small extension turns out to be a very powerful aid for debugging PSGraphs. We will see it in action in the case studies in \S \ref{sec:case_studies}.


\subsection{A logging mechanism}\label{sec:logging}

\begin{figure}
\begin{minipage}[b]{.35\textwidth}
\begin{center}
\includegraphics[width=0.7\textwidth]{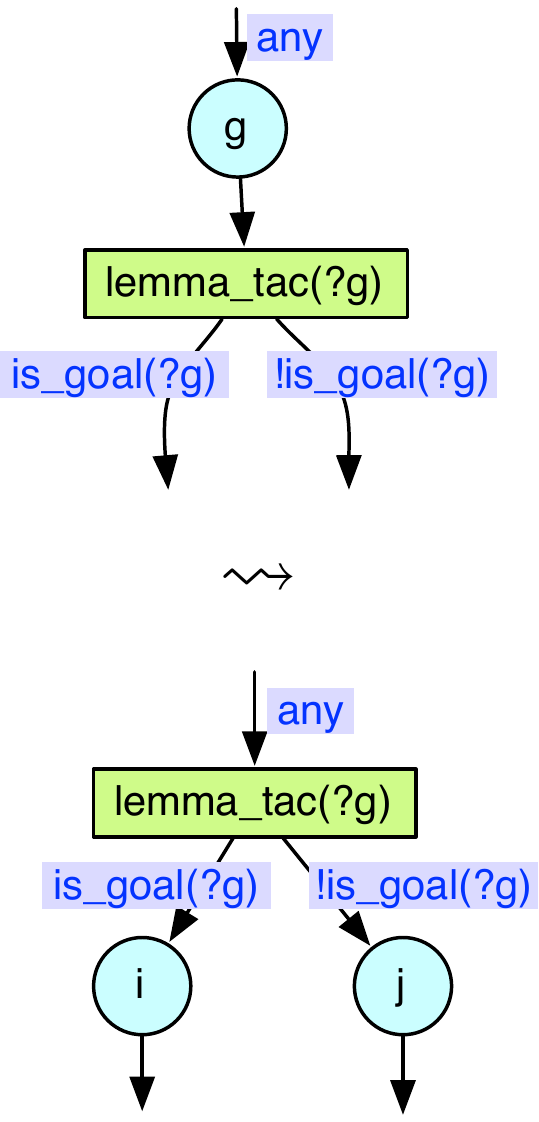} 
\end{center}\end{minipage}
   \hfill
\begin{minipage}[b]{.65\textwidth}\begin{verbatim}
> ENV_DATA : 
   g: E_Trm (...)
> GOAL : Open goals 
   [Goal i] ...
   [Goal j] ...
> GOALTYPE : evaluating is_goal(?g) with 
 pnode i env g => E_Trm(...)
> SUCCESS
> GOALTYPE : evaluating !is_goal(?g) with 
 pnode i env g => E_Trm(...)
> FAILURE
> GOALTYPE : evaluating is_goal(?g) with 
 pnode j env g => E_Trm(...)
> FAILURE 
> GOALTYPE : evaluating !is_goal(?g) with 
 pnode j env g => E_Trm(...)
> SUCCESS
> EVAL : Branch(goals on the output edges):  | i | j |
\end{verbatim}\end{minipage}

\caption{A logging example}
\label{fig:log exmaple}
\end{figure}

\rev{Another recent extension, which as we will show later has been very useful in our case studies, is a \emph{logging} mechanism.
Introducing such a mechanism is an engineering problem rather than a scientific
one, but as logging forms part of the user experience, it merits a brief discussion
here.}



To illustrate logging, consider the example in Fig.~\ref{fig:log exmaple} (left), where $lemma\_tac$ applies the cut rule with the term bound in $?g$.
The goal types $is\_goal(?g)$ and $is\_not\_goal(?g)$ check if the goal is or is not the same as $?g$ respectively. The logging mechanism will then print the logging messages as shown in Fig.~\ref{fig:log exmaple} (right). First, it prints the information about the environment of goal $g$, which says it has a variable $?g$ bound. The next two lines shows the open goals afterwards, which are $i$ and $j$. It then displays the results from evaluating the goal types: First we see that $i$ success for the wire labelled by $is\_goal(?g)$. As all possible combinations will be generated, it also checks if $i$ succeeds for the other wire, which fails. It then does the same for $j$, which will only succeeds for $is\_not\_goal(?g)$. The final line states that one branch was generated, with $i$ and $j$ on separate wires.

\rev{Full logging of a complex strategy with many branches can be very
verbose. Our logging mechanism allows the user to use the tags
such as} \verb|ENV_DATA|, \verb|GOAL| \rev{etc. seen in Fig.~\ref{fig:log exmaple}
to filter the types of message that are displayed.}



\section{Case studies}\label{sec:case_studies}

This section will address our hypothesis through three case studies. The first
example re-engineers a tautology-proving tactic into PSGraph.  We will express
the high-level ideas behind the tactic in an abstract way and then obtain an
efficient implementation by a sequence of refactorings adding goal types to
direct the proof search. Tinker's debugging capabilities are utilised to find
and correct mistakes in the encoding. The second example looks at a set of
\textit{ad hoc} domain-specific tactics developed to finesse the proof of a
lemma forming part of  the proof of security of a database system. We will see
how to use PSGraph to represent proof patterns involving
tacticals (tactic combinators). The final case study considers a
decision procedure for problems such as proving continuity of real-valued
functions. We will see how PSGraph can be used to express complex
recursive rewriting strategies.


\rev{A case-study approach for evaluation was chosen as the work is exploratory and improvement-driven.
The three case studies have different, yet relevant, challenges 
and thus provide us with necessary armoury for larger scale problems found in industrial settings. They also enables analyses of PSGraph from different
aspects, which is known as \emph{triangulation} in software engineering \cite{runeson2009guidelines}. }
\rev{
We have deliberately addressed unfamiliar problems, as opposed to types of problems that we know that PSGraph will excel for. For reasons discussed
in \S \ref{sec:related}, our analysis is qualitative in nature.}


\subsection{A tautology tactic for propositional logic}



\input{tauttac}

\rev{Looking at even a simple example of the higher-order programming style
bring several questions to mind:
is the high-level proof plan visible to a non-expert looking at
the implementation? How easy would it be to locate a mistake in the code if
it failed to prove a tautology? How would we go about refactoring the code?}
In the rest of this section, we will illustrate how to encode
\tac{simple\_taut\_tac} in PSGraph using the Tinker system.  We will show how
to support developing a correct and optimised PSGraph implementation through a
set of refactoring and analysis supported by the Tinker framework.  \rev{For
the most readable version of this tactic, we refer to the final version (Fig.
\ref{fig:taut full})\footnote{\rev{It may be easier to understand the example
by working backwards from this final version.}}.}

\subsubsection{Version 1: A generic PSGraph of the tautology tactic}

\begin{figure}
\begin{center}
\includegraphics[width=0.15\textwidth]{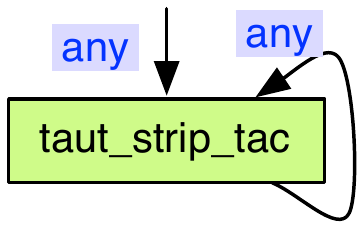} 
\end{center}
\vspace{-15pt}
\caption{\stac{simp\_taut\_tac} version 1
}
\label{fig:taut1}
\end{figure}

\tac{simple\_taut\_tac} repeats \tac{taut\_strip\_tac} until it is no longer applicable; if all subgoals are discharged at this point then the tactic succeeds, and it fails otherwise. Fig.~\ref{fig:taut1} implements this tactic at a very high level of atomicity where
\tac{taut\_strip\_tac} is treated as an atomic tactic.  

Graphically, repetition is simply represented as a feedback loop. By making this feedback loop the only output wire we achieve the same termination semantics as \tac{simple\_taut\_tac}. 
This can be justified as follows: if \tac{taut\_strip\_tac} fails on any subgoal then the overall tactic will fail: in \tac{simple\_taut\_tac} this means that the \code{REPEAT} combinator will terminate with a subgoals which result in failure. If the tactic produces subgoals then \tac{taut\_strip\_tac} is re-applied, as is the case for the \code{REPEAT} combinator. Finally, if there are no more subgoals, then the PSGraph will successfully terminate; this is also the success case for PSGraph.

\subsubsection{Version 2: From sequential to parallel tactic application}

\begin{figure}
\begin{center}
\includegraphics[width=1\textwidth]{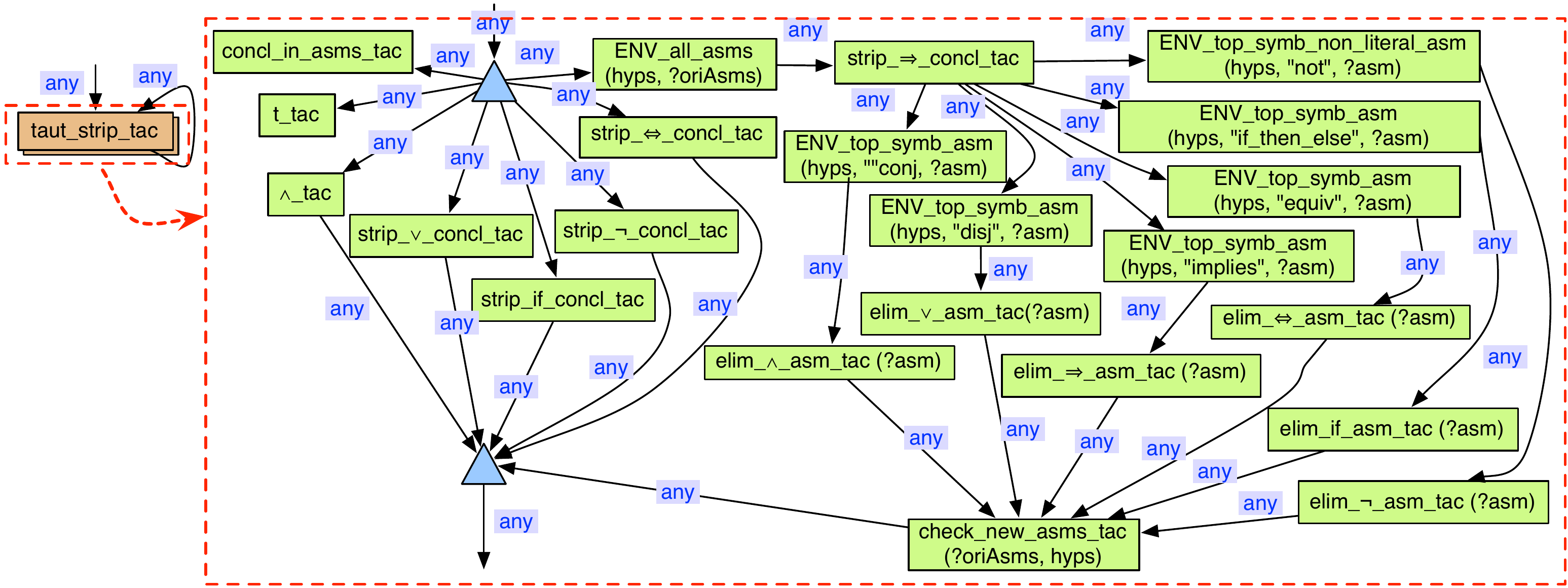} 
\end{center}
\vspace{-15pt}
\caption{Version 2: Flat and parallel tactic application.}
\label{fig:taut22}
\end{figure}

The example of Fig.~\ref{fig:taut1} does not show sufficient details to understand how \tac{simple\_taut\_tac} works. This require a further ``unfolding'' of \tac{taut\_strip\_tac} into a graph. Fig.~\ref{fig:taut22}  shows the same tactic as a graph tactic and its subgraph. 

This is achieved by ``unpacking'' all the tacticals, and represent each of the components as a tactic, with some minor modifications. To illustrate, the left part of the subgraph in Fig.~\ref{fig:taut22} corresponds to the  \tac{taut\_strip\_concl\_ts} tactic, with the conversions in the list of \tac{taut\_strip\_concl\_conv} represented by the 4 atomic tactics starting with `strip'. 

The right part of the subgraph corresponds to the work conducted on the hypothesis by \tac{taut\_strip\_thm\_thens}
when an implication introduction rule is applied. It first uses the \emph{ENV\_all\_asms} environment tactic to store all hypothesis
in a variable $?oriAsms$, before applying the introduction rule. For each propositional combinators, there is then a case adapted from \tac{taut\_strip\_thm\_thens} and the conversions in \tac{taut\_strip\_thm\_conv}. In most cases, they follow the pattern illustrated in Fig.~\ref{fig:env tac exmaple} (\S \ref{sec:envtactic}), where the hypothesis is first bound by one environment tactic and then the elimination rule is applied. At the end of this branch of the graph, \emph{check\_new\_asms\_tac} will get the lists of new hypothesis by comparing the current hypothesis (\emph{hyps}) with the hypothesis on entry ($?oriAsms$) to this part of the graph.

A conceptual difference between Fig.~\ref{fig:taut22} and the \tac{simple\_taut\_tac} ProofPower tactic is that in PSGraph we no longer need to enforce a sequential order; if two or more tactics are mutually independent, we can put them next to each other using identity tactics as necessary to split inputs and merge outputs.


Note that each wire is labelled by \gt{any}, meaning it will always succeed. This means that for a given subgoal generated by a given tactic, all possible output wires will be attempted in a separate branch of the search space. Thus, this graph can be seen as a generalisation of \tac{taut\_strip\_tac} in the sense that it will succeed if \tac{taut\_strip\_tac}  succeeds (albeit it is not as efficient).

\subsubsection{Version 3: Modularising the graph through hierarchies}

\begin{figure}
\begin{center}
\includegraphics[width=0.9\textwidth]{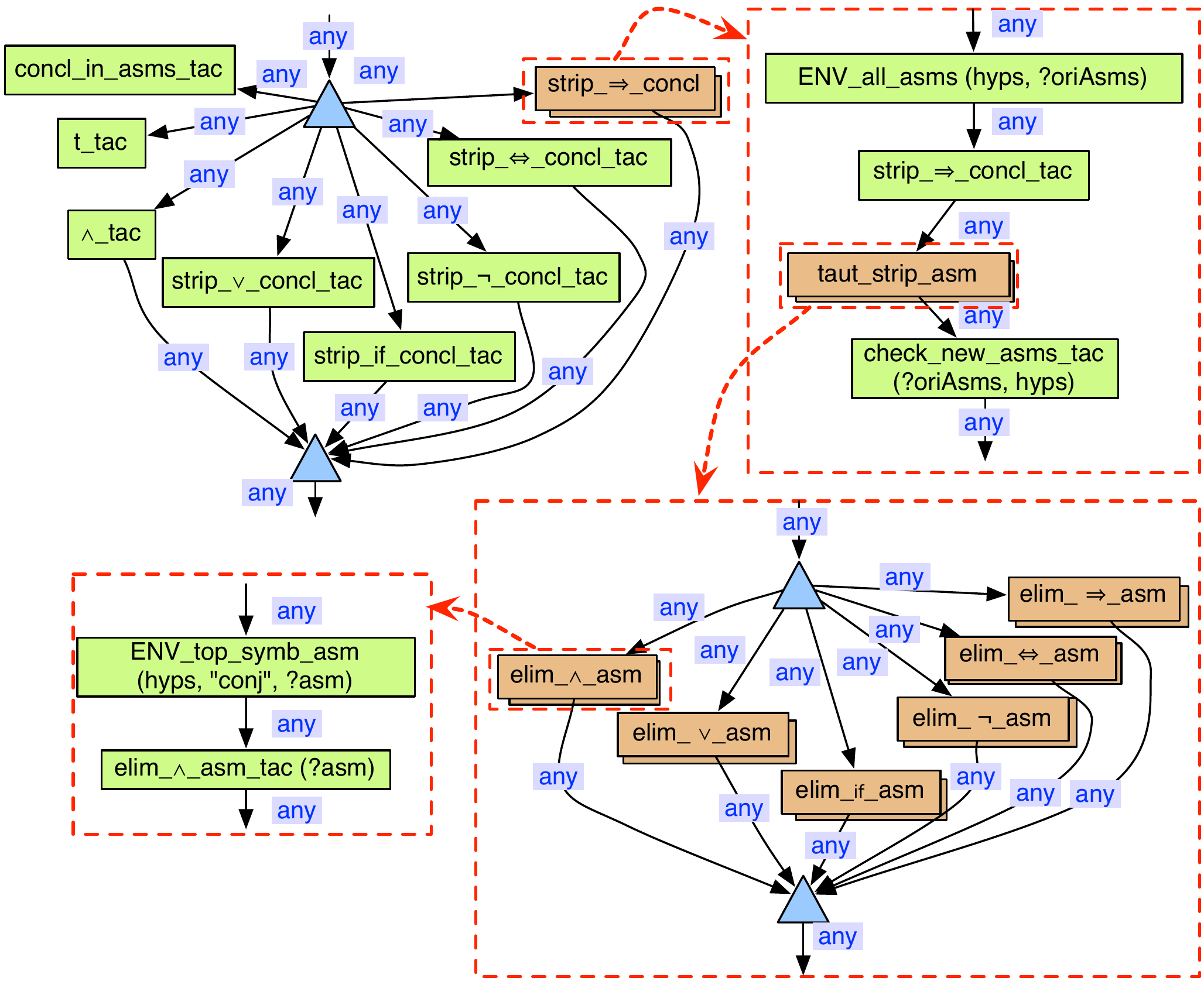} 
\end{center}
\vspace{-15pt}
\caption{\stac{simp\_taut\_tac} version 3: with hierarchies}
\label{fig:taut3}
\end{figure}

PSGraph aims to support development of proof strategies that are easy understand, maintain and refactor. 
To achieve this it should be intuitive to see what the proof strategy is meant to do. Whilst 
the flat graph of Fig.~\ref{fig:taut22} gives a detailed account of how the goals flow, it mixes high-level 
descriptive details of the proof strategy with low-level implementation details that are required to run it.
It also ``merges'' different operations which are best to split, e.g. operations on the conclusion and operations
on the hypothesis. 
This should be avoided when a more declarative and readable strategy is sought. 

For tactic languages modularity is handled by  sub-tactics, as is the case for \tac{taut\_strip\_tac}. Within PSGraph such \emph{modularity} is achieved through hierarchical graph tactics. Fig.~\ref{fig:taut3} refactors 
the graph of Fig.~\ref{fig:taut22} into a more modular graph. 

The top-level graph (top-left) contains the atomic operations on the goal, but has refactored the case that 
handles implications (and following operations on the hypothesis) into a graph tactic called \node{strip\_$\Rightarrow$\_concl}. This is shown in the top-right corner of Fig.~\ref{fig:taut3}. 
Within  \node{strip\_$\Rightarrow$\_concl}, the actual operations on the hypothesis are refactored into a graph tactic
\node{taut\_strip\_asm}, which is comparable to the \tac{taut\_strip\_thms\_thens} tactic, shown on the bottom left corner of
Fig.~\ref{fig:taut3}. 
Each case of this level corresponds to a propositional operator and is a nested graph tactic, where each of these follow the structure shown on the bottom right side. Here, an environment tactic first bind the operator and then the tactic is applied. 
The example illustrates the case for a conjunction, but the other cases are similar.

\subsubsection{Version 4: The use of goal types to explain and optimise the tactic}


The hierarchies help in exposing the high-level proof idea by hiding lower-level details and ``grouping'' together sub-strategies, such as separating operations on hypothesis from operations on the conclusion.
However, all wires are labelled by the \gt{any} goal type, which always succeeds. This use of \gt{any} has at least threes problems:
\begin{itemize}
\item \textit{Explanation:} the proof strategy does not explain \emph{why} a goal should choose a particular path. This is crucial in order to understand the proof strategy.
\item \textit{Evaluation:} the use of \gt{any} means is that all paths are attempted, which is in-efficient. 
\item \textit{Debugging:} a side-effect of the evaluation is that it debugging becomes hard as:
\begin{itemize} 
\item there more (failed) branches in the search space to analyse;
\item it is not clear what the intention of a particular path is which makes it hard (and time consuming) to find the ``correct" branch;
\item the error may manifest itself at different place further down the ``flow'' of the strategy.
\end{itemize}
\end{itemize}

\begin{figure}
\begin{center}
\includegraphics[width=0.9\textwidth]{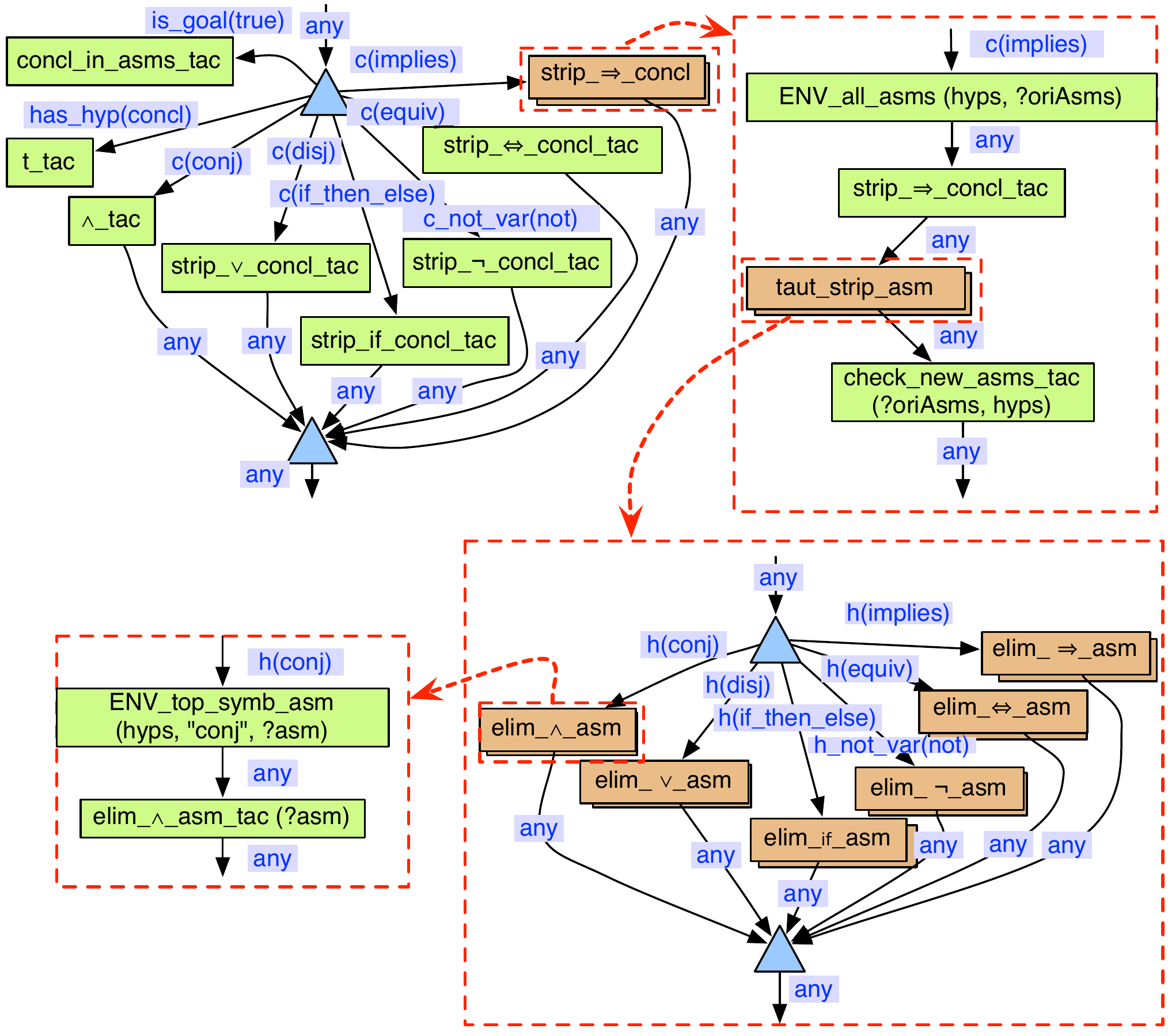} 
\end{center}
\vspace{-15pt}
\caption{\stac{simp\_taut\_tac} version 4: with hierarchies and goal types}
\label{fig:taut4}
\end{figure}

\noindent Developing goal types is one of the more challenging tasks of developing PSGraphs, and is also where development deviates most from standard tactic developments. In tactics we often end up trying one tactic first (e.g. if it is very quick or normally works) and if it fails we try something else. This is essentially what the tacticals in the list used by \tac{taut\_strip\_tac} does. 
Although this is possible in PSGraph, it is better to think about \emph{why} a particular tactic (or sub-strategy) should be applied. Moreover, when it fails it is hard to analyse where and why the failure happened. If the tactic/PSGraph also contains the ``reason'' for why a tactic is applied, in form of a goal type, then any failure is likely to show up at the right place and not several tactic applications later.
This will create much more \emph{maintainable} proof strategies; as we will illustrate below, it also becomes easier to analyse and patch a mistake in a proof strategy.

Fig.~\ref{fig:taut4} updates the graph of Fig.~\ref{fig:taut3} with goal types. Some of these goal types were introduced in \S \ref{sec:goaltype}, while we will introduce some new ones here. \rev{Firstly, recall the atomic goal type
$$dest\_trm(X, Y, Z)$$
from \S \ref{sec:goaltype}, which holds if term $X$ is ``destructed'' into its left $Y$ and right $Z$ sub-terms.} 
We use $c(X)$ and $h(X)$ as shorthand for the top symbol of the conclusion and a hypothesis, respectively:
$$
\begin{array}{lcl}
 c(X) & \leftarrow & top\_symbol(concl,X). \\
 h(X) & \leftarrow & member(hyps, Z), top\_symbol(Z,X).
 \end{array}
$$
The conversions applied to deal with negations by \emph{strip\_$\lnot$\_concl\_tac} (conclusion) and
\emph{elim\_$\lnot$\_asm} (hypothesis) require that the top level symbol is a negation, and the body is not just 
a variable (i.e. it is either compound or a constant). These properties are expressed by the goal types:
$$
\begin{array}{lcl}
 c\_not\_var() & \leftarrow & c(not), dest\_trm(concl,\_,Z), !trm\_var(Z). \\
 h\_not\_var() & \leftarrow & member(hyps, Y), top\_symbol(Y,not), dest\_trm(Y,\_,Z), \\ && !trm\_var(Z). \\
\end{array}
$$
With the goal types one can see in which cases a tactic should be applied, and evaluation will only try the branches
where a goal satisfies the goal type.


\subsubsection{Version 5: Discovery and patching of a bug}

\begin{figure}
\begin{center}
$\vcenter{\hbox{\includegraphics[width=0.3\textwidth]{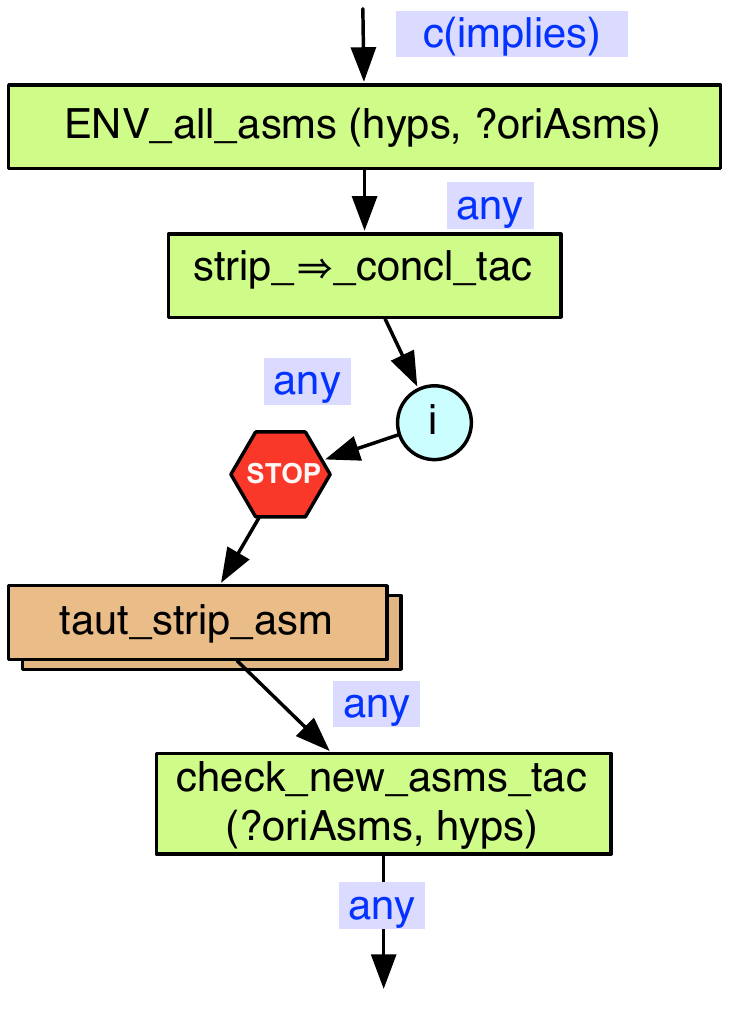}}}$
  \large{$\rightsquigarrow \quad$} 
$\vcenter{\hbox{\includegraphics[width=0.3\textwidth]{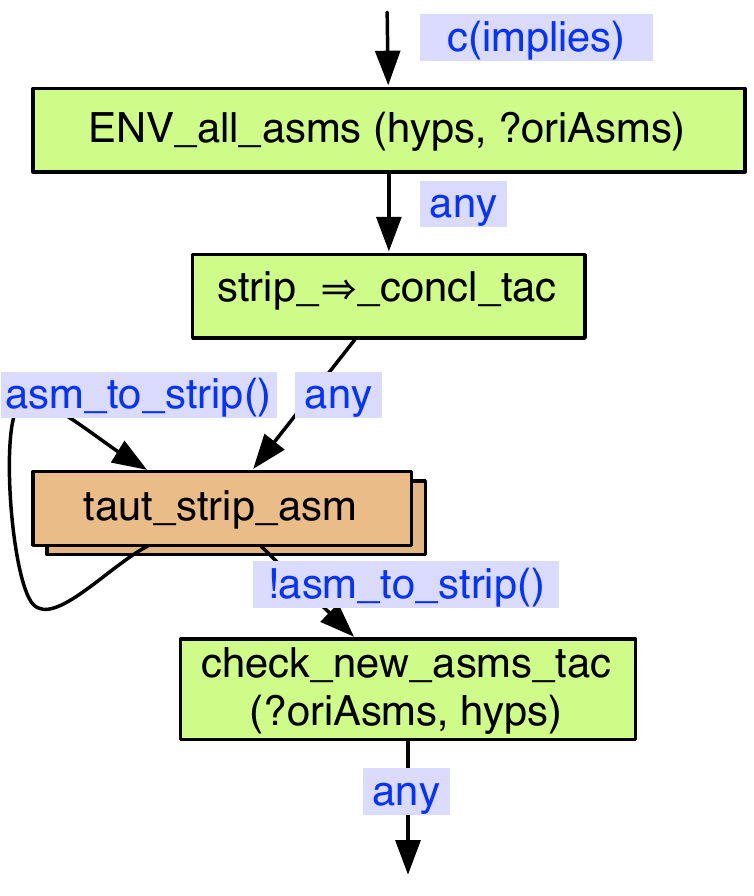}}}$
\end{center}
\vspace{-15pt}
\caption{Left:  \node{strip\_$\Rightarrow$\_concl} (version 4) illustrating a bug with a breakpoint. Right:
 \node{strip\_$\Rightarrow$\_concl} (version 5) with bug fixed by feedback loop.}
\label{fig:taut5}
\end{figure}

The proof strategy of Fig.~\ref{fig:taut4} will succeed for a large set of propositional tautologies, such as: 
\begin{GFT}{GOAL}
\+(* ?\PrPE{} *)\PrKM{}A \PrLB{} B \PrLE{} B \PrLB{} A\PrKO{}\\
\end{GFT}
\noindent However, it fails for the following (correct) goal:
\begin{GFT}{GOAL}
\+(* ?\PrPE{} *)\PrKM{}A \PrLB{} B \PrLB{} C \PrLE{} C \PrLB{} B \PrLB{} A\PrKO{}\\
\end{GFT}
As we have an idea where the problem is, we insert a breakpoint and automatically evaluate the strategy until the break point is
reached\footnote{If we had no idea where the problem may have been we could just have evaluated the strategy from the beginning -- the same approach as described in the rest of the section would still be applicable.}.
This is shown in Fig.~\ref{fig:taut5} (left). At this point the goal, labelled by $i$, has been simplified to:
\begin{GFT}{GOAL}
\+(*  1 *)\PrKM{}A \PrLB{} B \PrLB{} C\PrKO{}\\
\+(* ?\PrPE{} *)\PrKM{}C \PrLB{} B \PrLB{} A\PrKO{}\\
\end{GFT}
We can now step through the proof from that point in the nested \tac{taut\_strip\_asm} graph tactic. The goal
satisfies $h(conj)$, as the top level symbol of an hypothesis is a conjection, 
and correctly splits up the conjunction in the hypothesis \textit{(* 1 *)}:  
\begin{GFT}{GOAL}
\+(*  1 *)\PrKM{}A\PrKO{}\\
\+(*  2 *)\PrKM{}B \PrLB{} C\PrKO{}\\
\+(* ?\PrPE{} *)\PrKM{}C \PrLB{} B \PrLB{} A\PrKO{}\\
\end{GFT}
\noindent At this point it will exit the graph tactic, and (via \tac{check\_new\_asms\_tac}) return 
to the top of the top-level graph.  The problem is that one hypothesis \textit{(* 2 *)} contains a conjunction, and 
the aim of the overall proof plan is that all connectives in the hypothesis should have been eliminated.
One could still continue to run the proof, where it will eventually will have the goal:
\begin{GFT}{GOAL}
\+(*  1 *)\PrKM{}A\PrKO{}\\
\+(*  2 *)\PrKM{}B \PrLB{} C\PrKO{}\\
\+(* ?\PrPE{} *)\PrKM{}C\PrKO{}\\
\end{GFT}
\noindent which will not satisfy the goal type (of the top level graph), and thus fail.  The problem is that \tac{taut\_strip\_asm} has to be repeated until there are no more 
connectives. This is reflected in the updated proof strategy shown in Fig.~\ref{fig:taut5} (right). Here, we need a goal type to identify when there are more goals to be satisfied and label the loop with this:
$$
\begin{array}{lcl} 
 asm\_to\_strip() & \leftarrow & h(conj). \\
 asm\_to\_strip() & \leftarrow & h(disj). \\
 asm\_to\_strip() & \leftarrow & h(equiv). \\
 asm\_to\_strip() & \leftarrow & h(implies). \\
 asm\_to\_strip() & \leftarrow & h(if\_then\_else). \\
 asm\_to\_strip() & \leftarrow & h\_not\_literal(not).
\end{array}
$$
This is essentially a disjunction of all the possible symbols. The output wire, representing termination of the loop, is labelled by its negation: $!asm\_to\_strip()$. Note that if we had this goal type
instead of $any$ as output of \tac{taut\_strip\_asm}  in Fig.~\ref{fig:taut4} (left), then the error would manifested itself at the correct place, which illustrates the importance of goal types.
The above goal will now succeed.

\subsubsection{Version 6 \& final version: Discovery and patching of another bug}

\begin{figure}
\begin{center}
 $\vcenter{\hbox{\includegraphics[width=0.3\textwidth]{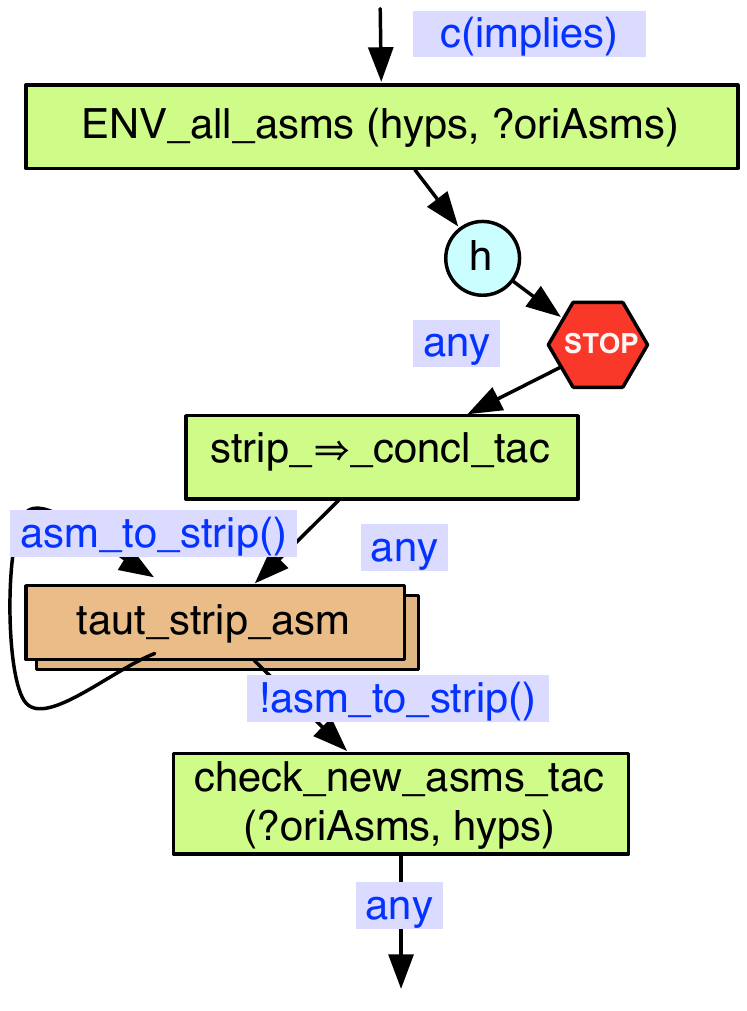}}}$
 \large{$\rightsquigarrow \quad$} 
 $\vcenter{\hbox{\includegraphics[width=0.32\textwidth]{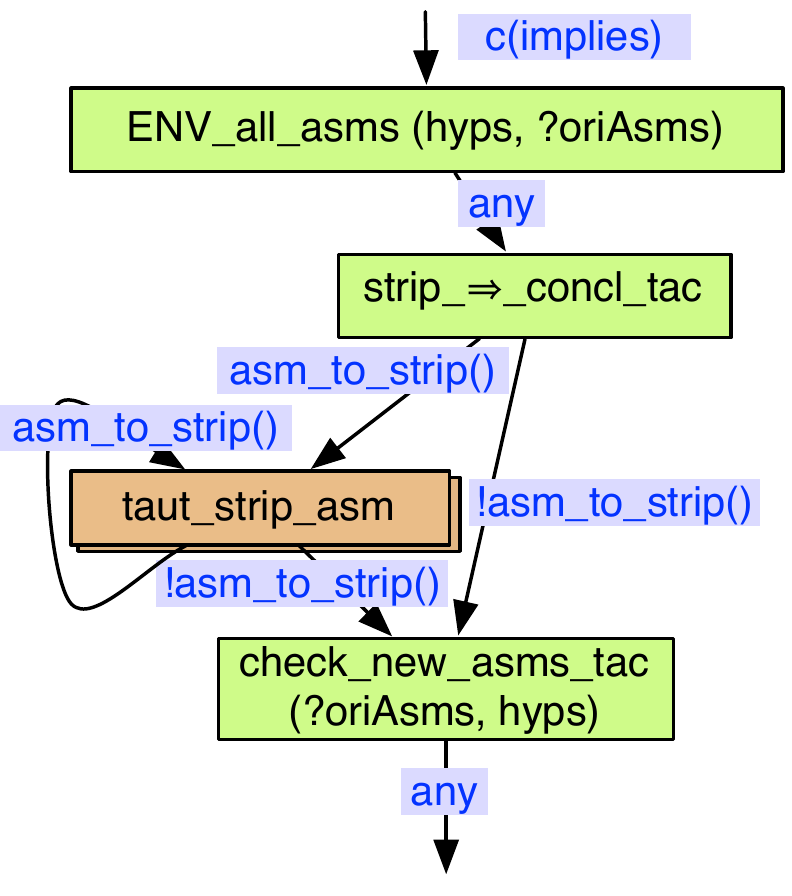}}}$
\end{center}
\vspace{-15pt}
\caption{Left:  \node{strip\_$\Rightarrow$\_concl} (version 5) illustrating bug with goal node and breakpoint. Right:
 \node{strip\_$\Rightarrow$\_concl} (final version 6) with bug fixed by new wire.}
\label{fig:taut6}
\end{figure}
Next, we try to prove the following goal:
\begin{GFT}{GOAL}
\+(* ?\PrPE{} *)\PrKM{}A \PrLE{} A \PrLB{} A\PrKO{}\\
\end{GFT}
Again, the tautology strategy fails. As with the previous case, one would suspect the issues is related to the hypothesis, thus we insert a breakpoint just before 
this part as shown in Fig.~\ref{fig:taut6} (left). At this point, the goal is the same as the original goal. In the next step, \tac{strip\_\Rightarrow\_concl\_tac} 
is applied generating:
\begin{GFT}{GOAL}
\+(*  1 *)\PrKM{}A\PrKO{}\\
\+(* ?\PrPE{} *)\PrKM{}A \PrLB{} A\PrKO{}\\
\end{GFT}
It will then enter the \tac{taut\_strip\_asm} graph tactic, but it will then fail when stepping over the identity tactic. 
At this point, we can use the \emph{logging mechanism} of Tinker, which gives the following message:
\begin{verbatim}
   FAILURE : Fail to match any Loop for the output goal node: 
            [Goal i : A |- A & A]
\end{verbatim}
In this case, the only assumption is 
\begin{GFT}{ASSUMPTIONS}
\+(*  1 *)\PrKM{}A\PrKO{}\\
\end{GFT}
\noindent which does not have any logical connectives and should therefore not be further simplified. The problem is that we have forgotten to bypass \tac{taut\_strip\_asm} when there are no
assumptions to simplify. To rectify the strategy, this missing case is added, using the $asm\_to\_strip()$ goal type to separate the two cases. This is shown in Fig.~\ref{fig:taut6} (right)

\begin{figure}
\begin{center}
\includegraphics[width=\textwidth]{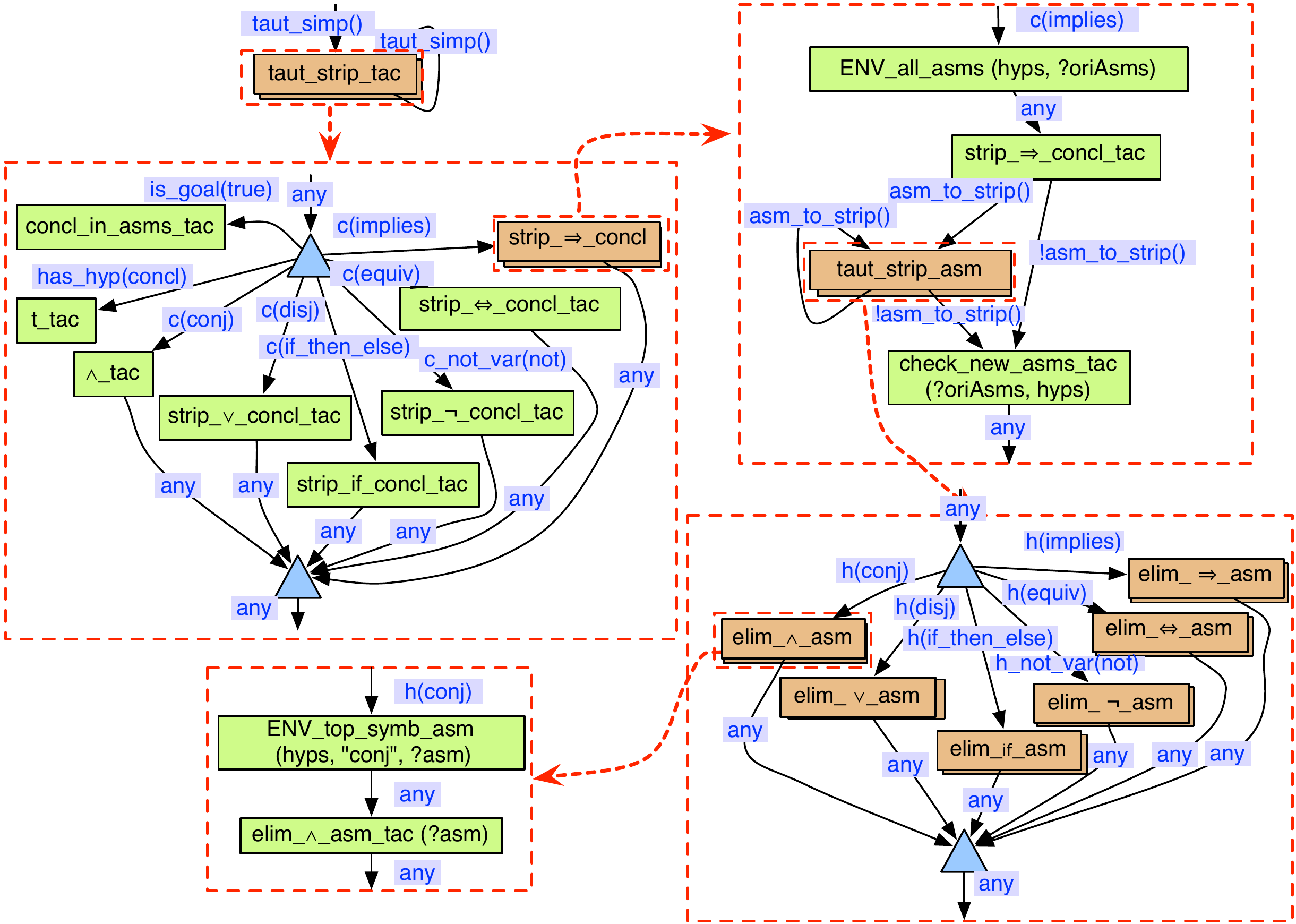} 
\end{center}
\vspace{-15pt}
\caption{\stac{simp\_taut\_tac} completed version}
\label{fig:taut full}
\end{figure}

This completes the development of \tac{simple\_taut\_tac} as a PSGraph. A complete version of the tautology tactic is shown in Fig.~\ref{fig:taut full}. 
In this final version we have also added a goal type to the input of of the overall strategy to show which type of goals it will work for:
$$
\begin{array}{lcl} 
 taut\_simp() & \leftarrow & is\_goal(true). \\
 taut\_simp() & \leftarrow & has\_hyp(concl). \\
 taut\_simp() & \leftarrow & has\_no\_hyp(concl), c(conj). \\
 taut\_simp() & \leftarrow & has\_no\_hyp(concl), c(disj). \\
 taut\_simp() & \leftarrow & has\_no\_hyp(concl), c(if\_then\_else). \\
 taut\_simp() & \leftarrow & has\_no\_hyp(concl), c(equiv). \\
 taut\_simp() & \leftarrow & has\_no\_hyp(concl), c(implies). \\
 taut\_simp() & \leftarrow & has\_no\_hyp(concl), c\_not\_literal(not). \\
\end{array}
$$

\subsubsection{Discussion}

A simple and instructive example has been used to illustrate many features of PSGraph and Tinker, which can be contrasted to 
the existing sentential encoding of \code{simple\_taut\_tac} in ML.  One of the key advantages is that a non-expert can read the
overall proof plan directly from the graph. Compare that for example Fig.~\ref{fig:taut full} (starting from top-left) with the
use of the \emph{REPEAT\_TTCL} tactical used in \tac{taut\_strip\_tac}. Without a deep understanding of underlying semantics of this (and other) tactical, it is very hard to see what the tactic does. For example, it matters if \emph{REPEAT\_TTCL} is applied 1 or more times or 0 or more times.  Our original version only applied it a single time; the second version 1 or more times; and the final version
got it right and applied it 0 or more times. Through the debugging features of Tinker and PSGraph,
this was fairly easy to find\footnote{Note that these bugs where not artificial by any means: they were genuine mistakes we did during development and the graphical representation help to locate.}. To find a similar mistake in the original code would not be as easy, and  would very likely require us to tear the tactic apart so that we can step through the execution, which in itself is not always an easy task.  For a subset of ML, the \emph{Tactician} tool for HOL light \cite{Adams2015} can automate such tearing apart of tacticals, but in most cases it has to be done manually. To summarise, 
while short ``one liners'', such as \tac{taut\_strip\_tac}, are elegant, they are not necessarily that easy to understand. Anecdotally, getting the ML code right is hard even for an expert -- then what about novel users and non-experts? Would you show the one-liner to your (non-technical) line manager or would you draw up a high-level diagram?

We have also found other advantages of the graphical representation. One example is that PSGraph allows us to delay decision on control flow, and make it efficient later (by adding goal types). This is not possible in ML, 
where we must give the order straightaway. Secondly, hierarchies enable hiding of low-level (implementation) details: we can use the
top-level graph(s) to show the high-level (often declarative) proof ides, while the graph tactic contains implementation details. Consider for example how \emph{elim\_$\wedge$\_asm} hides how we use an environment tactic to apply conjunction elimination in 
Fig.~\ref{fig:taut full}.  Thirdly, meta-level properties of the proof strategies 
can be read directly from the graph -- which is not as obvious in the ML code. One example of this is the symmetry between the proof steps in the conclusion (top-left) and the hypothesis (bottom-right) of Fig.~\ref{fig:taut full}.



As a general  \emph{guidance}, picking sensible names for goal types and tactics is crucial if the PSGraph should act as an explanation of a proof strategy. It is also useful to try to combine parts that ``belong'' together in graph tactics. To work with PSGraph, one need to change to a more  \emph{declarative} way of thinking compared with the more \emph{procedural} way of developing LCF tactics.
It is important to think about \emph{why} certain tactics should be applied, and encode this knowledge into the goal types.
 It is also important  to be aware that there is still work required at the ML level: one has to develop atomic tactics and atomic goal types, and also think about the arguments (if any) of them.
Through goal types, PSGraph ensures \emph{locality} when changing part of the proof strategy. You can safely assume that 
changes you make will only effect goals that satisfy the goal types leading to the sub-graph that is change: it will not have any
impact on parallel sub-graphs.

This case study has highlighted a current \emph{limitation} with respect to \emph{parametrised graph tactics}. All the tactics within \tac{taut\_strip\_asm} of Fig.~\ref{fig:taut full} have
the structure illustrated by \emph{elim\_$\wedge$\_asm}. The only difference is that logical connective and tactic used. It would have been desirable to be able to make this an argument for a generic graph tactic, to avoid having to re-implement each version. The next
example will illustrate how we can parametrise over tactic and goal type arguments, but not the actual tactics. 



\subsection{Domain specific tactics from the \emph{Front End Filter} (FEF) project}


\input{feftacs}



Representing these application-specific tactics in PSGraph prevents novel
challenges.  In the tautology example we were faced with mature production
level code, implementing a clear underlying proof plan. In this example, we are
faced with a set of \textit{ad hoc} tactics that ``did the job". An interesting
problem is to understand the underlying proof idea, with the ultimate aim of
providing a tactic that is robust to changes and can be re-used for similar
proofs. In the remainder of this section we will make steps by showing a
systematic and modular way of transferring tactics to PSGraph to improve
understandability and maintainability.

\subsubsection{Encoding common proof patterns}

\begin{figure} 
\begin{center}
\includegraphics[width=\textwidth]{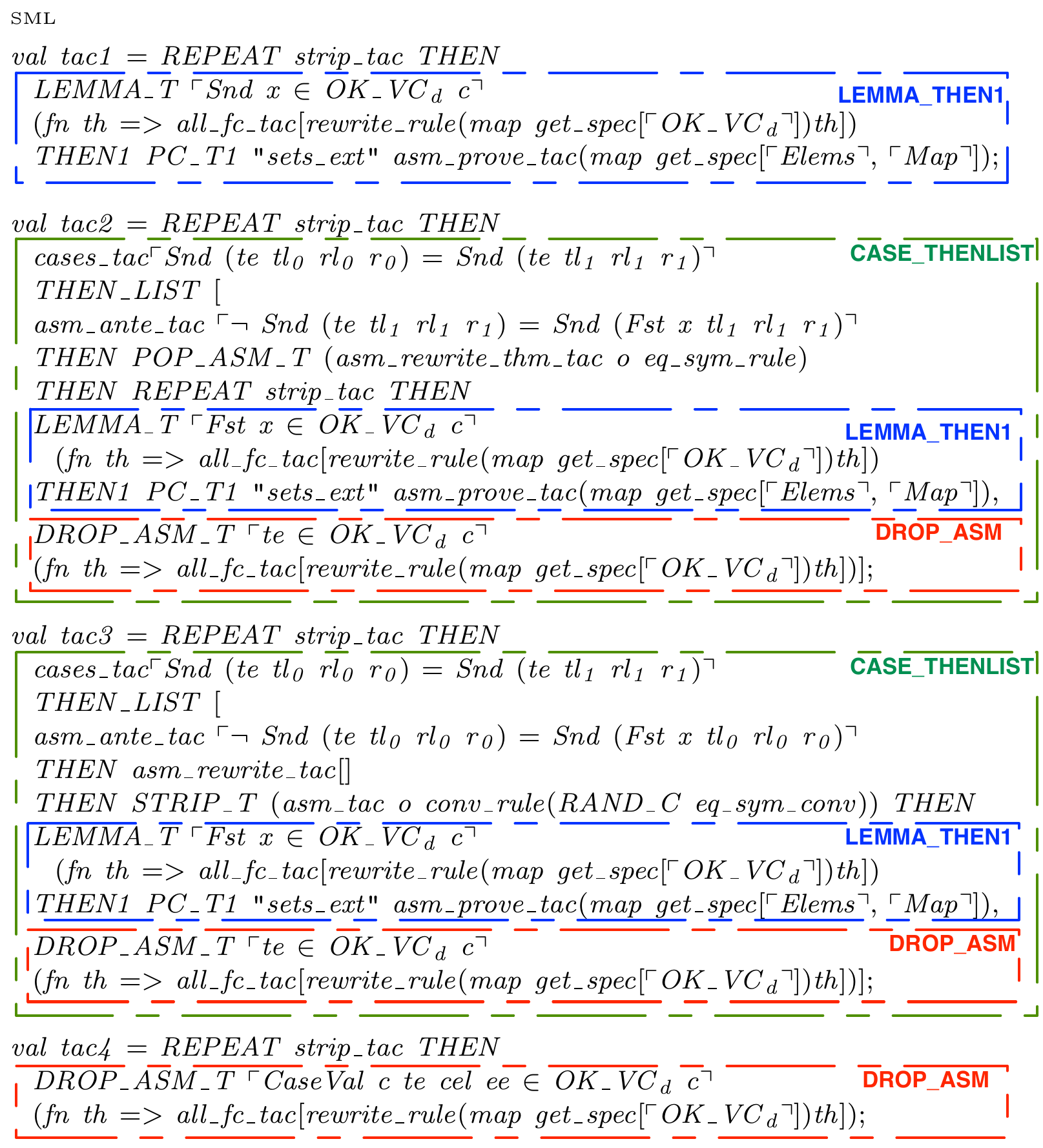} 
\end{center}
\vspace{-15pt}
\caption{FEF tactics with patterns highlighted}\label{fef:patterns}
\end{figure}

The first observation we make about the four \emph{FEF} tactics is that of code repetition, which we can extract into 
separate proof patterns. Our first step is to develop these patterns in PSGraph; this will enable us to develop a library and re-use 
the patterns using the \emph{library} functionality of Tinker (see \S \ref{sec:tinker})
We have found three patters, which we call \emph{LEMMA\_THEN1}, \emph{CASE\_THENLIST} and \emph{DROP\_ASM}. These are highlighted in Fig.~\ref{fef:patterns}, which shows the ProofPower code of the tactics. Note that code is only included to
show the repetition: we will not go into details of the actual tactics except for the encoding of the patterns.


\paragraph{The \emph{LEMMA\_THEN1} pattern}

\begin{figure}
\begin{center}
\includegraphics[width=0.65\textwidth]{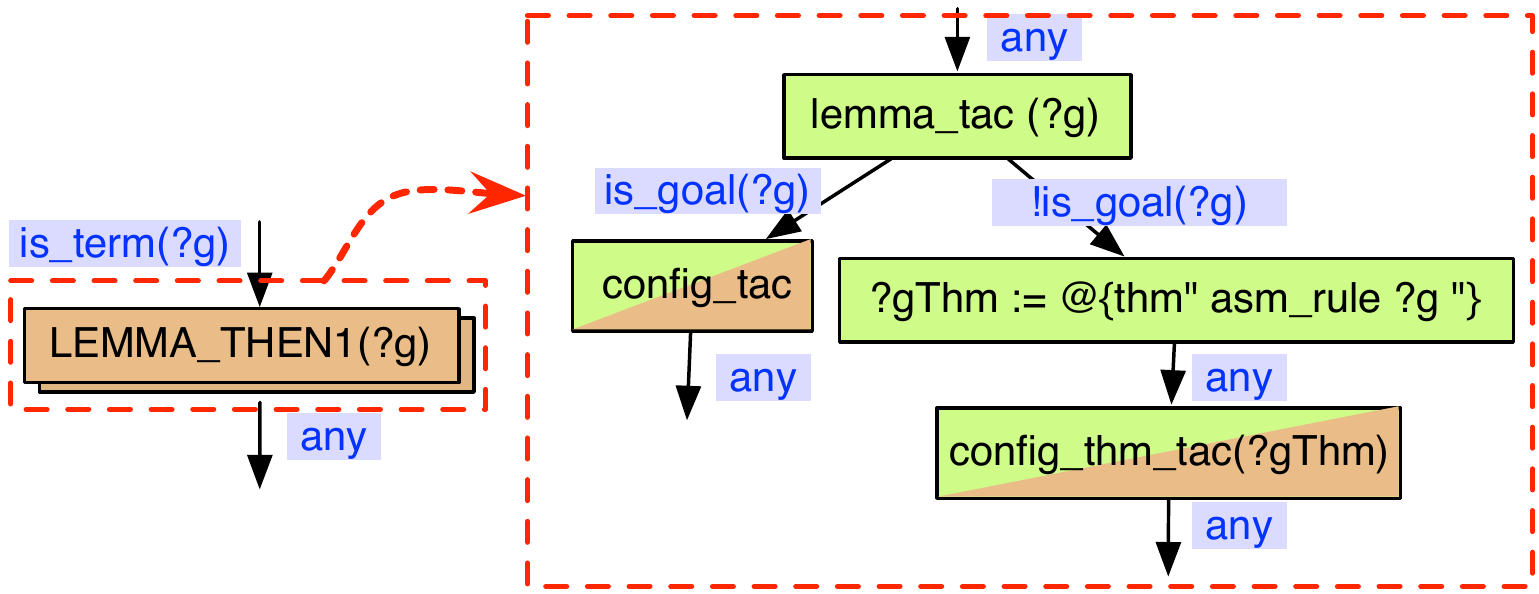} 
\end{center}
\vspace{-15pt}
\caption{PSGraph encoding of \emph{LEMMA\_THEN1}}
\label{fig:fef:lemmat}
\end{figure}

The \emph{LEMMA\_THEN1} can be implemented as follows:
\begin{GFT}{}
\+fun LEMMA\_THEN1 g thm\_tac tac = (LEMMA\_T g thm\_tac) THEN1 tac; \\
\end{GFT}
\noindent It will apply the cut rule by adding a new subgoal $g$. This is followed by an application of tactic
\tac{thm\_tac} to the original goal (with $g$ added to the list of hypothesis), and tactic \emph{tac} to the new subgoal. This pattern can for example be seen in \emph{tac1} of Fig.~\ref{fef:patterns}, where we can instantiate the pattern as follows:
\begin{GFT}{}
\+ LEMMA\_THEN1 \\
\+	\PrKM{}Snd x \PrIN{} OK\_VC\PrIJ{d} c\PrKO{} \\
\+ 	PC\_T1 "sets\_ext" asm\_prove\_tac(map get\_spec[\PrKM{}Elems\PrKO{}, \PrKM{}Map\PrKO{}])\\
\+ 	(fn th => all\_fc\_tac[rewrite\_rule(map get\_spec[\PrKM{}OK\_VC\PrIJ{d}\PrKO{}])th])\\
\end{GFT}
%

Fig.~\ref{fig:fef:lemmat} shows the PSGraph encoding of the \emph{LEMMA\_THEN1} pattern. The tactic is encapsulated in a graph tactic \emph{LEMMA\_THEN1(?g)} (left), where $?g$ is the subgoal in which the pattern is parametrised over. 
The $?g$ argument of the \emph{LEMMA\_THEN1} graph tactic is used to make $?g$ available for the nested graph, while 
the incoming goal type $is\_term(?g)$ ensures that $?g$ is bound to a term in the environment before the pattern is applied. 

The right hand side of the figure shows the body of the tactic, which ``unfolds'' the \emph{LEMMA\_T} and \emph{THEN1} tacticals. First, \emph{lemma\_tac(?g)} is applied, which applies the cut rule. The graph then depicts the parallel nature of how the new subgoal $?g$ and the existing subgoal are handled separately. The goal type is used to guide the goal to the correct tactic. However, the example highlights the limitation of parametrised graph tactics we have already discussed (and return to in \S \ref{sec:fef:discussion}).  
We need to introduce two dummy tactics, \emph{config\_tac} and \emph{config\_thm\_tac}, which has to be manually replaced (renamed) when this pattern is used (as shown below).

\paragraph{The \emph{CASE\_THENLIST} pattern}

\begin{figure}
\begin{center}
\includegraphics[width=0.65\textwidth]{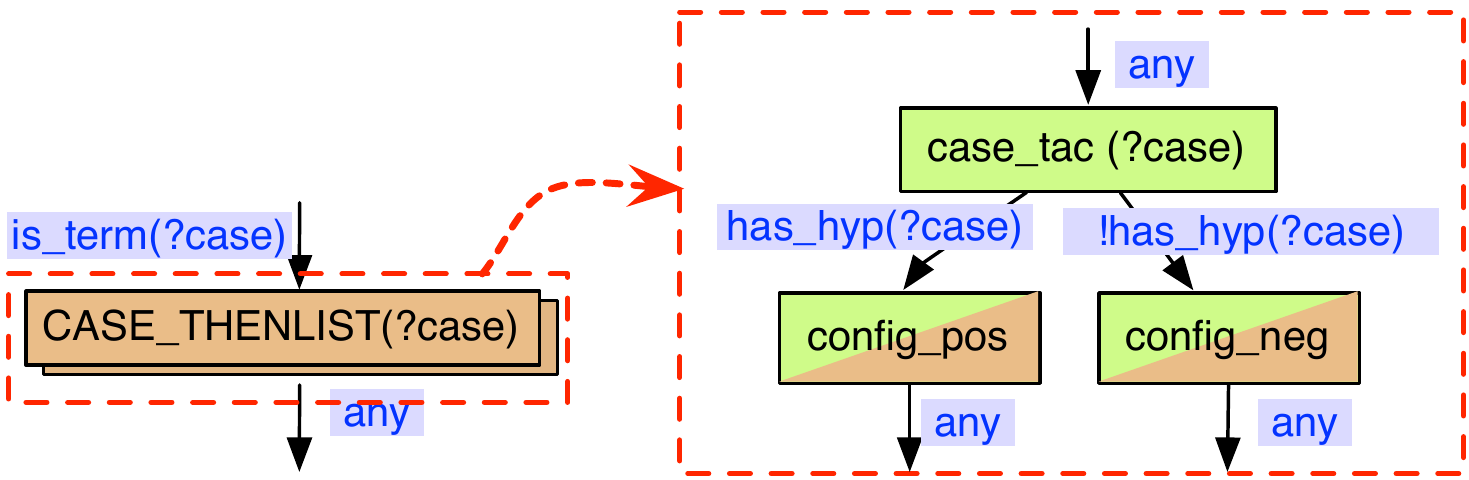} 
\end{center}
\vspace{-15pt}
\caption{PSGraph encoding of \emph{CASE\_THENLIST}}
\label{fig:fef:case}
\end{figure}

The second pattern applies a case-split on a given variable \emph{case}, followed by tactic \tac{tac1} when
\emph{case} holds, and tactic \tac{tac2} for its negation. This is called \emph{CASE\_THENLIST}:
\begin{GFT}{}
\+fun CASE\_THENLIST case tac1 tac2 = case\_tac case THEN\_LIST[tac1, tac2]; \\
\end{GFT}
The PSGraph encoding of this pattern can be seen in Fig.~\ref{fig:fef:case}. It is similar in structure of \emph{LEMMA\_THEN1}, and also illustrates how naturally PSGraph highlights that the two cases should be handled separately. It should be noted that, albeit sufficient in this case, \emph{case\_tac} may simplify the hypothesis thus $!has\_hyp(?case)$ will not work as expected in all cases.


\paragraph{The \emph{DROP\_ASM} pattern}

\begin{figure}
\begin{center}
\includegraphics[width=0.55\textwidth]{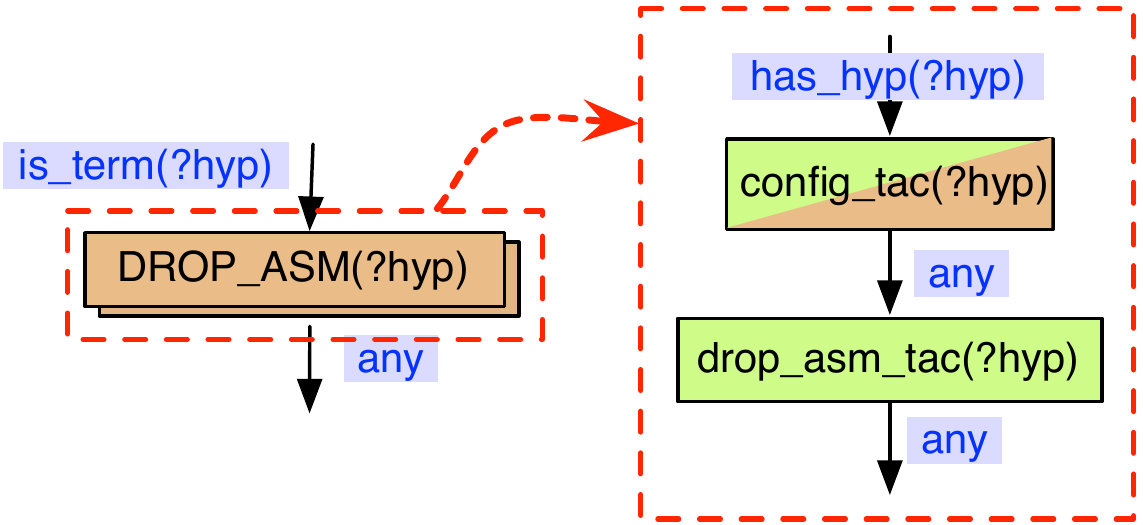} 
\end{center}
\vspace{-15pt}
\caption{PSGraph encoding of \emph{DROP\_ASM}}
\label{fig:fef:drop}
\end{figure}

In Fig.~\ref{fef:patterns}, the final pattern \emph{DROP\_ASM}, are instances of 
the \tac{DROP\_ASM\_T} tactical. For purposes of the underlying pattern representation in PSGraph, 
we unfold the meaning of \tac{DROP\_ASM\_T}, creating the ML function:
\begin{GFT}{}
\+fun DROP\_ASM hyp thm\_tac = thm\_tac hyp THEN drop\_asm\_tac hyp;\\
\end{GFT}
It will first apply a tactic \tac{thm\_tac} using the given hypothesis \emph{hyp}, and then remove the hypothesis by the \tac{drop\_asm\_tac} tactic. Fig.~\ref{fig:fef:drop} shows the PSGraph version.

\subsubsection{Encoding the FEF tactics in PSGraph}


\begin{figure}
\begin{center}
\includegraphics[width=0.35\textwidth]{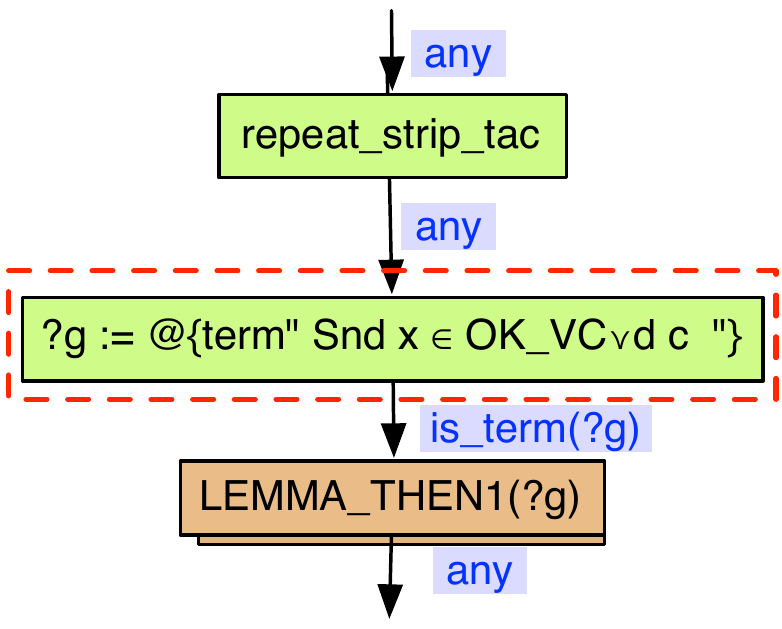} 
$\quad \quad \quad \quad$
\includegraphics[width=0.45\textwidth]{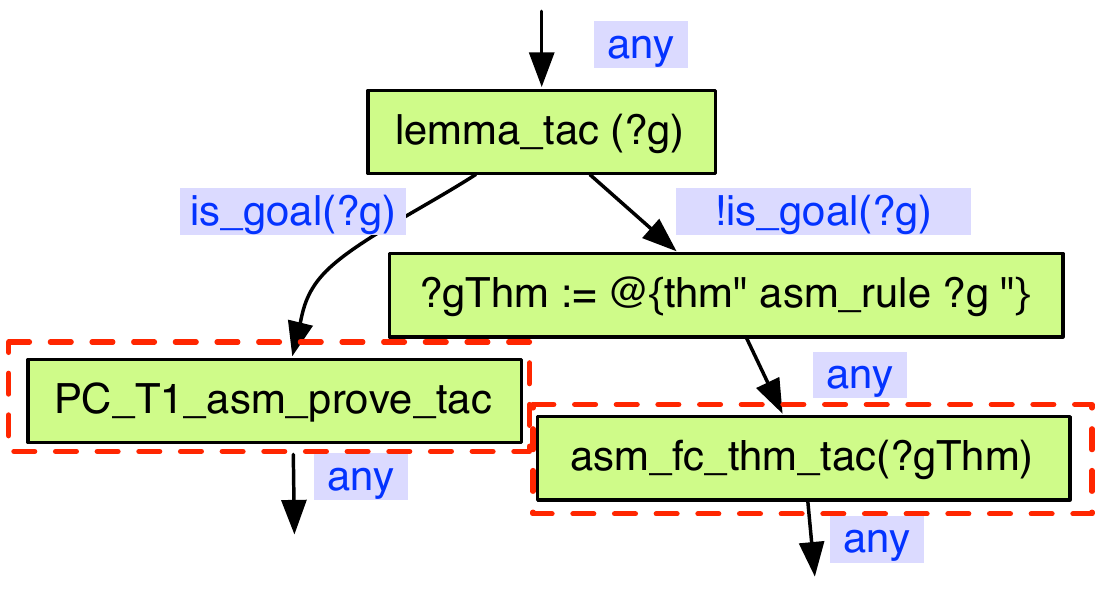} 
\end{center}
\vspace{-15pt}
\caption{PSGraph \emph{tac1}: top level (left) and the sub-graphs of node \emph{LEMMA\_THEN1(?g)}(right)}
\label{fig:fef:tac1}
\end{figure}

The patterns can easily be used with Tinker's library functionality in a drag-and-drop manner\footnote{See the dedicate web page
\cite{webpage} for a screencast of the library functionality.}.

We can then develop the tactics $tac1, \cdots, tac4$ in a modular way. The two smallest tactics are $tac1$ and $tac4$. They 
only use one single pattern. To illustrate, Fig.~\ref{fig:fef:tac1} shows $tac1$ encoded in PSGraph. In the figure, the stippled
boxes illustrates the instantiation of the parametrised components of the pattern. 
ProofPower's \emph{strip\_tac} is first applied, followed by an environment tactics that binds variable $?g$ to 
$\PrKM{}Snd\; x \PrIN{} OK\_VC\PrIJ{d} c\PrKO{}$\footnote{Note that within the Tinker tool, we have followed the syntax used by Isabelle's anti-quotation mechanism where a term $t$ is written $@\{term~"t"\}$.}.
This is followed by the \emph{LEMMA\_THEN1} pattern, where \emph{config\_tac} is replaced by 
\emph{PC\_T1\_asm\_prove\_tac}, which is defined as
\begin{GFT}{SML}
\+ PC\_T1 "sets\_ext" asm\_prove\_tac(map get\_spec[\PrKM{}Elems\PrKO{}, \PrKM{}Map\PrKO{}]);\\
\end{GFT}
\noindent and \emph{dummy\_thm\_tac} is replaced by \emph{asm\_fc\_thm\_tac(?gThm)}, which is defined as
\begin{GFT}{SML}
\+ all\_fc\_tac[rewrite\_rule(map get\_spec[\PrKM{}OK\_VC\PrIJ{d}\PrKO{}]) ?gThm]\\
\end{GFT}

\begin{figure}
\begin{center}
\includegraphics[width=\textwidth]{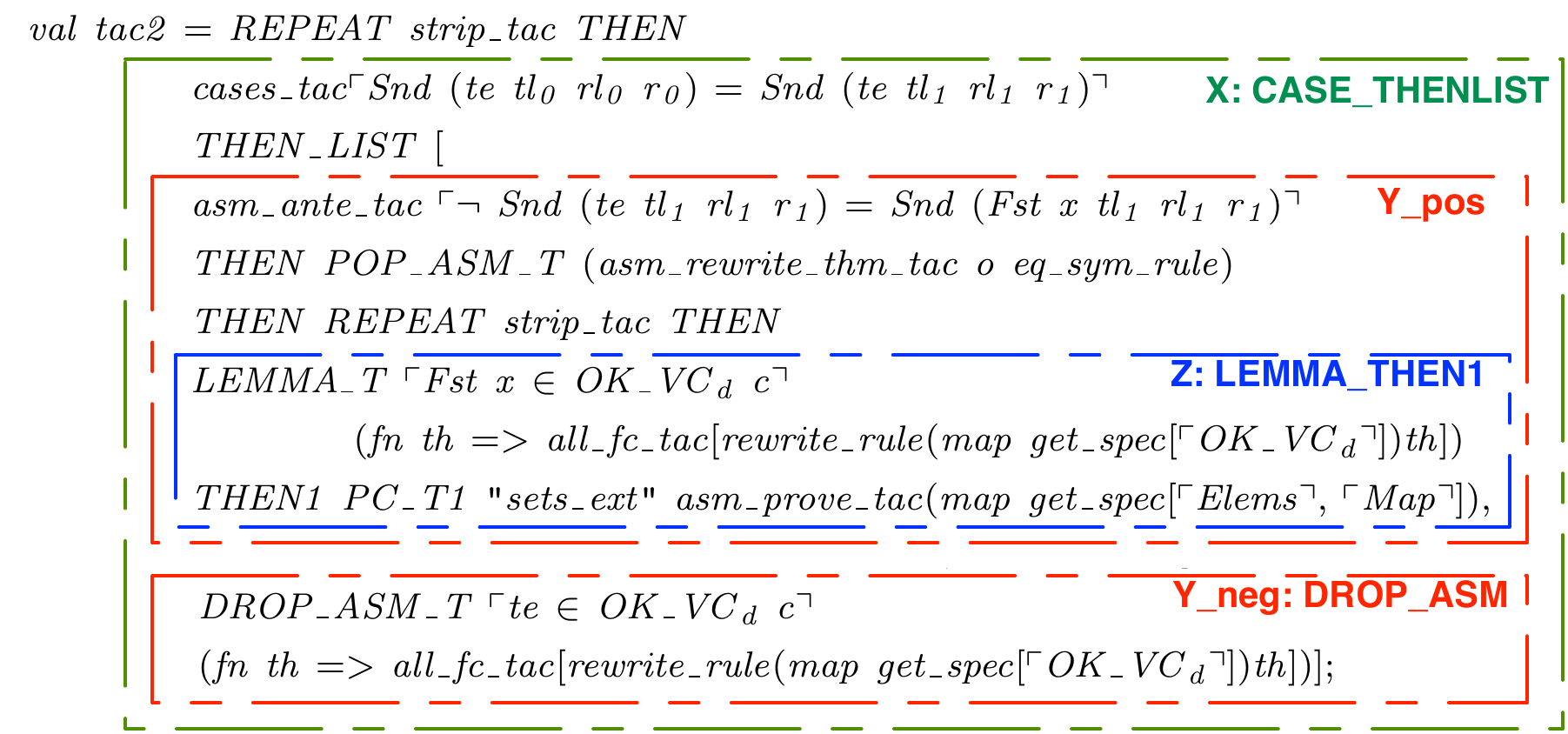} 
\end{center}
\vspace{-15pt}
\caption{ProofPower code of FEF \emph{tac2} tactic}\label{fig:fef:tac2}
\end{figure}

For more complicated tactics, such as \emph{tac2} and \emph{tac3}, we have broken the tactic up into components representing
the identified proof patterns (where appropriate). We will demonstrate this with 
\emph{tac2}. Fig.~\ref{fig:fef:tac2} gives the ProofPower code for this tactic, where the patterns have been highlighted as follows:
the green box shows the \emph{CASE\_THENLIST} pattern; the red boxes show the positive and negative cases for \emph{CASE\_THENLIST}. Within the positive case, the blue box shows the \emph{LEMMA\_THEN1} pattern, while the negative case
uses the \emph{DROP\_ASM} pattern.

\begin{figure}
\begin{center}
\includegraphics[width=0.25\textwidth]{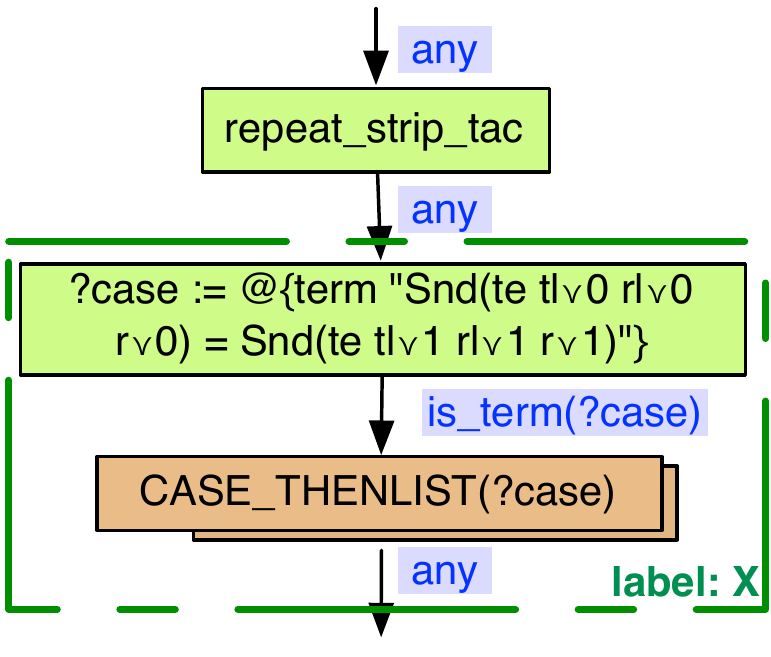} 
$\quad$
\includegraphics[width=0.4\textwidth]{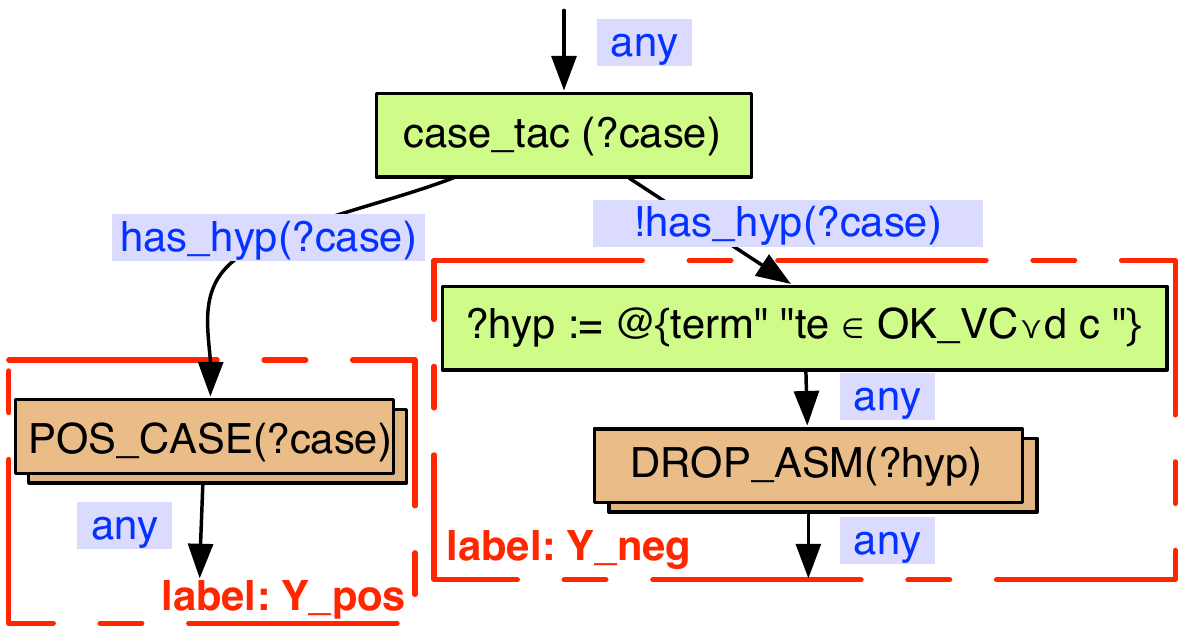} 
$\quad$
\includegraphics[width=0.25\textwidth]{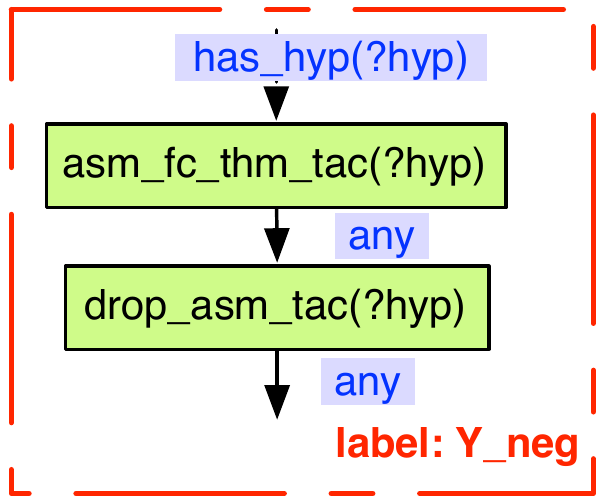} 
\end{center}
\vspace{-15pt}
\caption{The top-level, second-level and subgraphs of \emph{DROP\_ASM} of \emph{tac2} in PSGraph}
\label{fig:fef:tac2 1st and 2nd}
\end{figure}

Fig.~\ref{fig:fef:tac2 1st and 2nd} (left) shows the top-level view of the graph, which follows the same structure as \emph{tac1}, albeit 
with the \emph{CASE\_THENLIST} pattern applied. The body of this graph tactic is shown in the middle of the figure. Here, 
\emph{?case} is first instantiated to 
$
\PrKM{}Snd (te tl\PrIJ{0} rl\PrIJ{0} r\PrIJ{0}) = Snd (te tl\PrIJ{1} rl\PrIJ{1} r\PrIJ{1})\PrKO{}
$.
Note the use of stippled labelled boxes to show which part of the ProofPower code each sub-graph corresponds to.

In the instantiation of \emph{CASE\_THENLIST} (middle),  the tactics for the positive and negative cases, i.e. \tac{config\_pos} and \tac{config\_neg}, are replaced by two sub-components. The positive case is an instance of the \emph{DROP\_ASM} tactic, shown at the right hand side of Fig.~\ref{fig:fef:tac2 1st and 2nd}. Here,  \emph{?hyp} is initialised to
$
\PrKM{}te \PrIN{} OK\_VC\PrIJ{d} c\PrKO{}
$
\noindent and \emph{config\_tac(?hyp)} is replaced by \emph{asm\_fc\_thm\_tac(?hyp)}, which is defined to be
\begin{GFT}{SML}
\+  all\_fc\_tac[rewrite\_rule(map get\_spec[\PrKM{}OK\_VC\PrIJ{d}\PrKO{}]) ?hyp])\\  
\end{GFT}


\begin{figure}
\begin{center}
\includegraphics[width=0.3\textwidth]{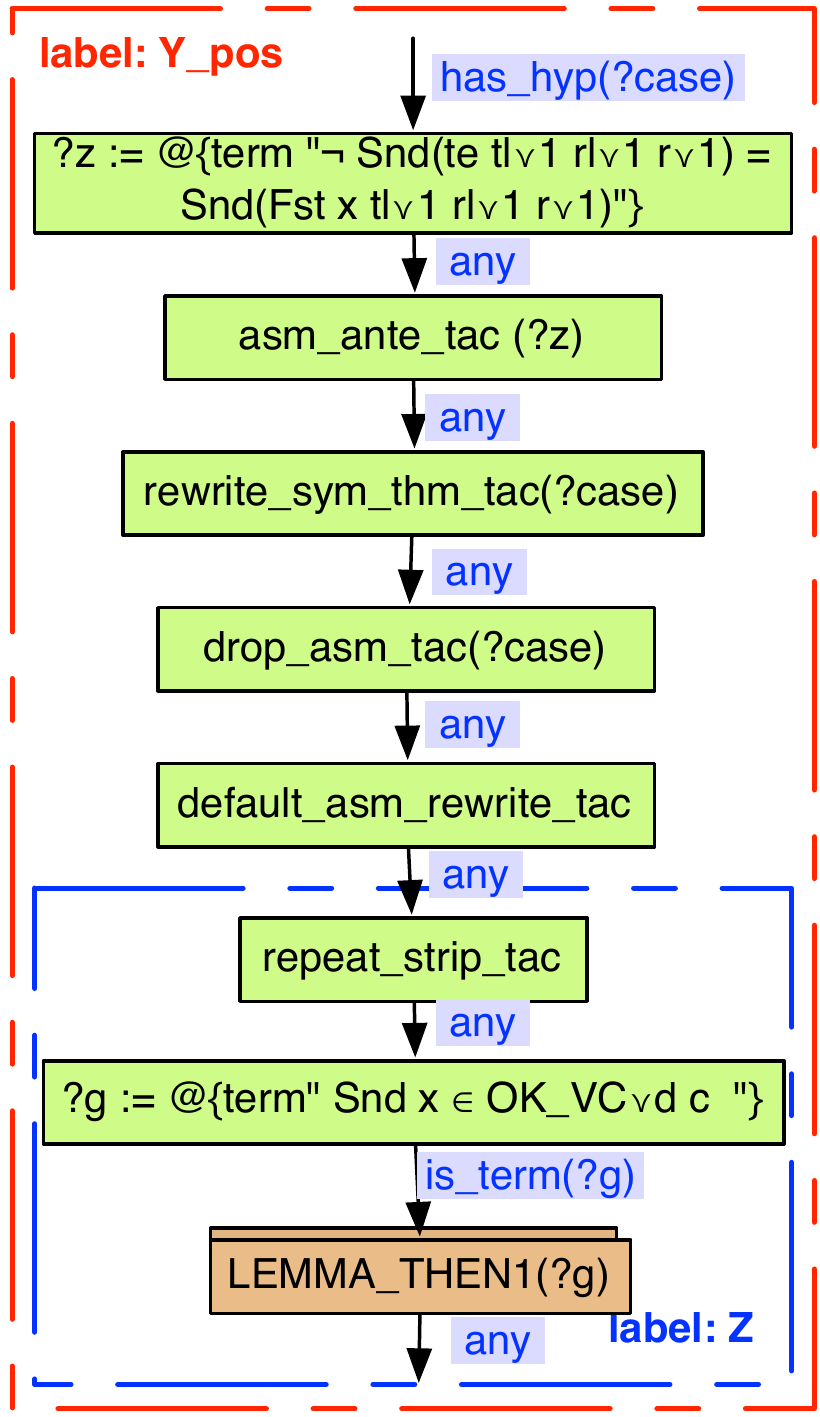} 
$\quad$
\includegraphics[width=0.65\textwidth]{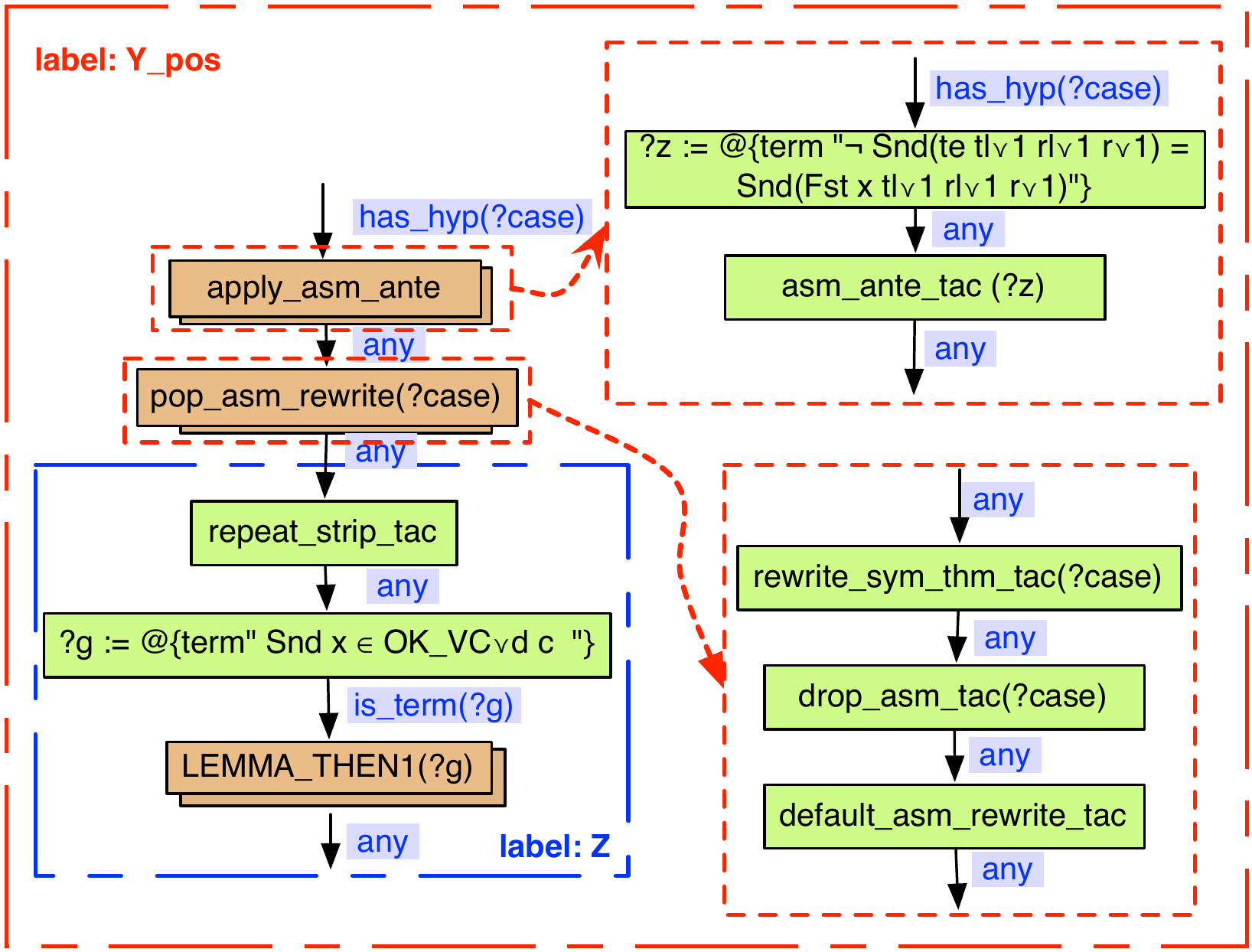} 
\end{center}
\vspace{-15pt}
\caption{\emph{POS\_CASE} of $tac2$: flat version (left) and a hierarchical version (right)}
\label{fig:fef:tac22pos}
\end{figure}

The positive case is wrapped in a graph tactic \emph{POS\_CASE} shown in Fig.~\ref{fig:fef:tac22pos}. The left hand side shows a flat version. Here, $?z$ is first bound to the term
$ \PrKM{}\PrLD{} Snd (te tl\PrIJ{1} rl\PrIJ{1} r\PrIJ{1}) = Snd (Fst x tl\PrIJ{1} rl\PrIJ{1} r\PrIJ{1})\PrKO{}$.
The \tac{asm\_ante\_tac(?z)} tactic is then applied.  This is followed by a sequential application of 
\tac{rewrite\_sym\_thm\_tac(?case)}, \tac{rewrite\_sym\_thm\_tac(?case)} and \tac{default\_asm\_rewrite\_tac}. The remaining part are the same as \emph{tac1}, i.e. it can be developed by using the \emph{LEMMA\_THEN1} pattern. The right hand of Fig.~\ref{fig:fef:tac22pos} is the same as the left hand side, with hierarchies introduced for increased readability.


\subsubsection{Discussion}\label{sec:fef:discussion}

The FEF case study is both \textit{ad hoc} and domain specific, which is very different from the tautology example in the previous section.
It has provided useful principles of how to apply PSGraph with a new perspective, where the goal is to extract/understand rather than reflect the proof structure. 
As a result, the proof structure follows more from instantiation of discovered proof patterns, rather than an overall proof plan.

The case study has provided us with a workbench to test robustness, and our next step will be to change the FEF specification and see how to fix the proofs in both ML and PSGraph and then compare the relative advantages. We could also re-create the real world issue 
that the proof is part of a big induction and the induction hypothesis has to be strengthened. This is a very common problem for inductive proofs.

The example has shown use of Tinker's library function to reuse patterns and how graph tactics can be parametrised by its input.
However, it has also further indicated the need to parametrise graph tactic by actual tactics rather than their arguments, as we saw in the tautology example. Here, we provided dummy graph nodes that had to be manually replaced. One example to overcome this, is to allow tactics to be bound in the goal node environment; another example is to configure/replace tactics in the graph tactics. For example,
\begin{GFT}{}
\+  LEMMA\_THEN(config\_tac := another\_tac,?g)\\
\end{GFT}
would replace tactic \tac{config\_tac} with a tactic \tac{another\_tac} for this particular use of the \textit{LEMMA\_THEN} graph tactic.

As the Tinker tool is foremost a research vehicle we have by design made it generic to work with multiple provers. Whilst we still think this is the right choice, a more specialised version for a particular prover could offer a closer level of integration, e.g. within the tool we have used $@\{term~``-"\}$, more familiar for Isabelle users, compared with ProofPower's  $\PrKM{} - \PrKO{}$ representation.


\subsection{Developing conversions}


\input{ctsconv}

\subsubsection{The problem of transforming conversions into tactics}

As with the other case studies, we would like to transfer \tac{rec\_conv} into PSGraph in order to achieve a more intuitive representation to support future maintenance.  However, PSGraph works with \emph{tactics} while \tac{rec\_conv}  is a \emph{conversion}, and
in order to use a conversion in PSGraph a conversion must first be transformed into a tactic. Functionality to achieve this
is provided by ProofPower via 
\begin{GFT}{SML}
\+ val conv\_tac : CONV -> TACTIC\\
\end{GFT}
Using this function we can for example turn our \emph{rec\_conv} conversion into a tactic:
\begin{GFT}{SML}
\+ val conv\_rec\_tac : TACTIC = conv\_tac rec\_conv\\
\end{GFT}
As we have seen, conversions are combined via a set of \emph{conversion combinators}, comparable to tacticals for tactics. To follow
the same approach as in the previous examples, we need to turn these into tacticals instead.  For some examples
we can define algebraic laws to support this. For example, sequential composition:  
$$
conv\_tac (conv1~\textit{THEN\_C}~conv2) = (conv\_tac~conv1)~\textit{THEN}~(conv\_tac~conv2)
$$
We can then use the same approach as before and turn \tac{conv\_tac conv1} and \tac{conv\_tac} \tac{conv2} into atomic tactics with a wire between them. The problem with this approach is in the presence of combinators such as \emph{RAND\_C}, which doesn't apply a conversion to a term, but to the operands of a function of the term. E.g.  given a 
term $\PrKM{}f x\PrKO{}$, \tac{RAND\_C conv} will apply conversion \tac{conv} to  $\PrKM{}x\PrKO{}$.
We will call the family of conversion combinators that changes the ``focus" of a term to a sub-term, such as \emph{RAND\_C} and \emph{RANDS\_C},  for \emph{term focus} combinators.  

Now, consider a conversion 
\begin{GFT}{SML}
\+ \emph{RAND\_C} (conv1 \textit{THEN\_C} conv2)\\
\end{GFT}
\noindent This will first change the focus of the term to the operand, and then apply \tac{conv1} followed by \tac{conv2} to 
the operand. We cannot naively break up the \textit{THEN\_C} combinator into the \textit{THEN} tactical, as 
$conv2$ should work on the sub-term as a result of the use of \emph{RAND\_C}.
For this case, we could again develop suitable algebraic laws and ``push'' \emph{RAND\_C} inwards, i.e. 
\begin{GFT}{SML}
\+ (\emph{RAND\_C} conv1) \textit{THEN\_C} (\emph{RAND\_C} conv2) )\\
\end{GFT}
\noindent We can then use the same approach as above to turn this into a tactic.

A deeper problem for \tac{rec\_conv} is the recursion: in several cases a conversion is applied, followed by a recursive call wrapped in a term focus combiner, before \emph{simp\_conv} is applied to the pre-recursive focus. To achieve this in PSGraph, where recursion becomes a feedback loop, each call needs to keep track of the ``current'' term focus, together with the focus of the pre-call in order to apply \emph{simp\_conv} correctly.

\subsubsection{Encoding recursion with ``term focusing'' in PSGraph}

\begin{figure}
\begin{center}
\includegraphics[width=0.7\textwidth]{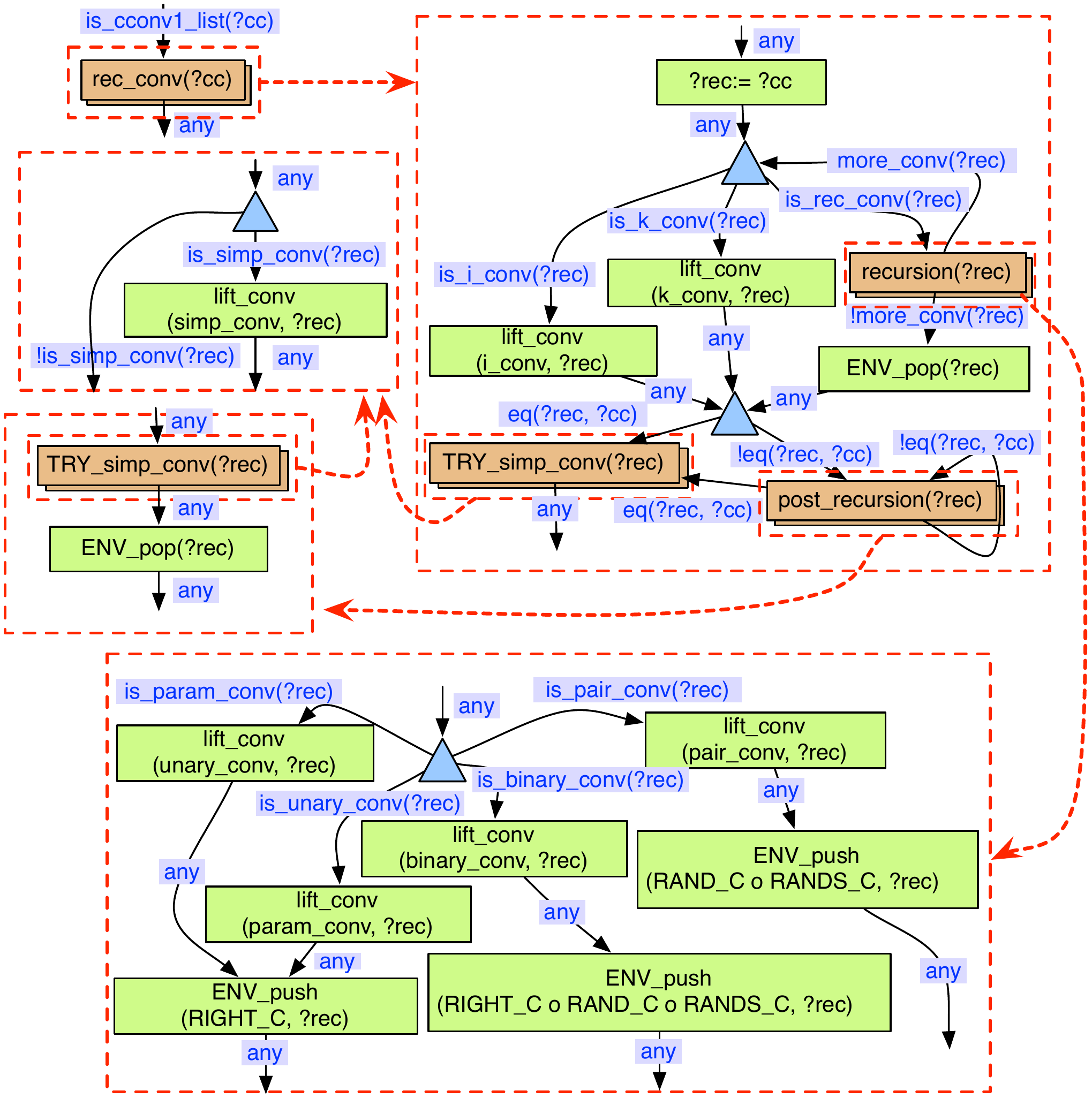} 
\end{center}
\vspace{-15pt}
\caption{Encoding recursion with ``term focusing'' in PSGraph}
\label{fig:conv:rec}
\end{figure}


Our solution when implementing the \tac{rec\_conv} conversion, is to augment the PSGraph with a variable $?rec$. This
will act as a stack (frame) that keeps track of each change of focus to a term when entering a recursive step. This variable has type\footnote{We have updated the allowed types in an environment for simplicity -- we could just
as well used a list of names, and developed a parser to turn this into a conversion.}
\begin{GFT}{}
\+  ?rec : (CONV -> CONV) list\\
\end{GFT}
The idea is as follows. For each application of a conversion, we first sequentially compose all elements of $?rec$ and apply this to focus the correct place in a term followed by the
application of a conversion. We specialise \emph{conv\_tac} with a tactic that takes these two arguments:
\begin{GFT}{ML}
\+ lift\_conv: (CONV -> CONV) list -> CONV -> tactic\\
\end{GFT}
\noindent Here, the first argument is a list of term focusing combinators and the second argument is desired conversion.
\emph{lift\_conv} has the following semantics:
$$
\emph{lift\_conv}~ [f_1;\ldots;f_n] \emph{conv} =  \emph{conv\_tac}~\big((f_n \circ \ldots \circ f_1)~\emph{conv}\big)
$$

If this conversion is followed by a recursive application, then the ``term focus combinators'' are pushed to the $?rec$ stack. This is achieved by an environment tactic 
\emph{ENV\_push(conv,var)}, which takes one of these conversion combinators and a variable as argument. 

Once the recursion has completed \emph{simp\_conv} should be applied to each operand, in the inverse order of the recursion (i.e. the last sub-term that a conversion was applied to should be simplified first). This is achieved by using $?rec$, and apply 
\emph{simp\_conv}, followed by popping an element of it until $?rec$ is empty.  The environment tactic \emph{ENV\_pop(var)} applies such a single `pop' operation.

The encoding of this strategy is shown in Fig.~\ref{fig:conv:rec}. It takes an input goal with an environment containing a variable $?cc$,
which has the ``term focus" on entry (and exit). It initialises $?rec$
to this value. The $i\_conv$ and $k\_conv$ conversions are applied without any recursion, as described above. For the other cases, the \emph{recursion} graph tactic is added. For each case
conversion $c$, i.e. each element of the list of \emph{rec\_conv} we develop an atomic goal type, called \emph{is\_c\_conv}. For the graph tactic, \emph{is\_rec\_conv} is just the disjunction
of the four recursive cases:
$$
\begin{array}{lcl}
is\_rec\_conv(X) & \leftarrow &  is\_unary\_conv(X).\\
is\_rec\_conv(X) & \leftarrow &  is\_param\_conv(X).\\
is\_rec\_conv(X) & \leftarrow &  is\_binary\_conv(X).\\
is\_rec\_conv(X) & \leftarrow &  is\_pair\_conv(X).
\end{array}
$$
For each of the cases, the conversion is applied and then the focus of the term is changed. For example \emph{pair\_conv} is followed by two applications of \emph{RAND\_C} and then a recursive call
again to \emph{rec\_conv}. Thus, for this case, $\textit{RAND\_C} \circ \textit{RAND\_C}$ is added to $?rec$. 

The recursion stops when there are no further applicable conversions, i.e. when none of the goal types succeed. This is the negation of the case where any of the conversion goal types are satisfied, i.e.:
$$
\begin{array}{lcl}
more\_conv(X) & \leftarrow &  is\_i\_conv(X).\\
more\_conv(X) & \leftarrow &  is\_k\_conv(X).\\
more\_conv(X) & \leftarrow &  is\_rec\_conv(X).
\end{array}
$$
On exit, the last change of focus is removed (as it failed for this application), and we enter the \emph{post\_recursion} graph tactic, unless the case where $?rec$ is still the same as $?cc$. This indicates that none of the recursive calls succeeded.
In this graph tactic, the \emph{simp\_conv} will be applied for the same sub-terms as in the recursive case (due to the feedback loop around the tactic). When $?rec$ is the same as $?cc$, the loop is terminated, which ends the proof.

\subsubsection{Discussion}

When contrasting the ML code of \emph{rec\_conv} in the start of the section with the corresponding PSGraph
in Fig.~\ref{fig:conv:rec}, one can see that choices made are much more declarative; the parallel view of the choice of conversion, which
one had in mind in the first place, is clear to see compared with the sequential ordering of the ML code.
One of the most common bugs in such development is to get the scheduling wrong, and, as we saw with the tautology example, PSGraph is good at supporting this type of debugging. 

The use of conversions is very common: they are used heavily in our work with {\DRisQ}'s powerful tactic \cite{ABZ16}, where we are developing conversions to port the tactic from verification of Ada programs to C. The work presented here is a starting point which has exposed
a lot interesting questions when dealing with low-levels issue involving terms and rewriting. In particular, how to evaluate sub-parts of terms within a given context across multiple boxes
in an efficient way. In the presented solution we had to re-build data structures from outside and inwards for each box before actually applying the rewriting, which is very expensive.

One way to overcome this \rev{in the future} is to augment the goal node with the \emph{current focus} of a goal (conclusion and/or hypothesis). Huet's 
Zippers \cite{huet1997zipper}, which combines a tree with a sub-tree of focus, including operations to this focus move up/down/sideways
could for example be applied.  This has been implemented for a term representation similar to ProofPower's in 
 \cite{paper:Dixon:03}. Another option is Grundy's \emph{window inference}\cite{grundy1991window}, which will also build up contextual information when focus is shifted to a sub-term. 

\rev{Another question is whether tactics and conversions should be treated
uniformly or separately.  To treat conversions separately would be a more
radical extension to PSGraph, perhaps  involving a new \emph{conversion graph
notation} devoted to equational reasoning, in which the boxes would denote
conversions rather than tactics.}



\section{Related work}\label{sec:related}

Motivated mainly by large scale proof developments, such as the Coq proof of the four colour theorems \cite{gonthier2005computer}, the HOL light proof of the Kepler conjecture \cite{hales2015formal}
and the Isabelle verification of the seL4 micro-kernel \cite{klein2014comprehensive}, ideas from software engineering have started to make their way into proof development. The term
\emph{proof engineering} has been used for this new discipline \cite{klein2014proof}. This includes topics such as \emph{productivity} of proofs \cite{staples2014productivity} 
as well as proof \emph{metrics} \cite{Aspinall2016}.

Whilst we put the work presented here in this context of proof engineering, our motivation and focus is different from the above work.
Instead of being motivated by large proof developments and their proofs,
our motivation has been development and maintenance of a large proof strategy (i.e. \emph{tactic}) in an industrial setting, using a graphical representation \cite{ABZ16}.
There has been a considerable amount of work on visualising \emph{proof trees}, including: L$\Omega$UI \cite{SiekmannHBCFHKKMMPS99}
for $\Omega$mega; XIsabelle \cite{Ozols97} for Isabelle; ProveEasy \cite{Burstall2000} and Jape \cite{bornat1997jape} for teaching; and some more recent work for Mizar \cite{Libal14,pkak2010algorithms}. However, none of these
visualise the high-level strategy. This raises the question of what the difference between a proof and proof strategy is (in a mechanical theorem proving setting)? If one think of tactics as proof strategies, then a proof strategy is really just a procedural description of how to conduct the search for a proof. Bundy \cite{bundy1998science} on the other hand, has argued for the additional role of \emph{explanation}.  We hope that we have shown how PSGraph can help explaining the strategy of a proof in addition to 
be used to guide the search. Such explanation is important for maintenance when team changes, and our work with {\DRisQ} \cite{ABZ16} has had very promising initial results when porting their proof tactics from Ada to C verification.

\rev{Through case studies, we have conducted a comprehensive comparison of PSGraph with ProofPower's tactic language. This language is very similar to the tactic languages found in LCF-based provers such as HOL4, HOL light, and Isabelle. LTac for Coq \cite{Delahaye02} and Eisbach for Isabelle \cite{Matichuk2014} provide support for tactic development within the proof scripts. They support matching features similar to the environment tactic and goal types of PSGraph. However, the composition of tactics is closer to the more traditional tactic languages and they do not contain debugging features as presented for PSGraph here.
In \emph{proof planning}~\cite{Bundy91} tactics are given pre-conditions and post-conditions, which can be achieved by the goal types of PSGraph. IsaPlanner \cite{paper:Dixon:03} is a proof planner for Isabelle, and the PSGraph project started out as a new composition (tactic) language for IsaPlanner (which was similar to ProofPower's tactic language).
There are also several \emph{typed tactic} languages, e.g. VeriML \cite{Stampoulis10}, Beluga \cite{pientka2008type}, Delphin \cite{poswolsky2009system} and 
Mtac \cite{ziliani2013mtac}. Here the types contain information about the relationship between tactics and the proofs produced. For VeriML, Beluga and Delphin there is a clear separation between the tactic and the base logic. It is not clear how they can be incorporated in an established theorem prover, as we have detailed for PSGraph and ProofPower here. Mtac is an extension to Coq and, as with LTac, does not address the issues of maintenance and usability we are focusing on, albeit with improved static (composition) properties through types. 
Autexier and Dietrich \cite{Autexier10} have developed \emph{a declarative tactic language on top of a declarative proof language} where a strategy is represented 
as a \emph{schema} which needs to be instantiated. Their work is more declarative than PSGraph, whilst PSGraph handles tactic compositions in a more declarative way. However, it is not clear how issues such as debugging is helped by such schemas. There have also been several attempts to create \emph{declarative tactic languages on top of procedural tactic languages} \cite{Harrison96,Giero:07}. Asperti et al \cite{Asperti09} argues that these approaches suffer from two drawbacks: goal selection for multiple subgoals, and information flow between tactics -- both of these are addressed by goal types in PSGraph. 
Finally, HiTac is a tactic language with additional support for hiding complexities using hierarchies \cite{paper:Aspinall:2008}, which our notion of hierarchies is based upon. Hierarchies in HiTacs are restricted to single inputs, which probably makes the semantics simpler and more elegant, whereas our approach, which utilises goal types, has followed from a more practical and empirical approach. HiTac has predominantly been used for more theoretical work, and as with VeriML, Beluga and Delphin, it has not been integrated into an established theorem prover.
}

\rev{
As the evaluation was partly exploratory in nature, a case study approach was chosen as case studies are considered suitable for qualitative analysis \cite{runeson2009guidelines}.
The fact that the third author (Arthan) was new to PSGraph also provide new and more objective reflections in this work.
For quantitative results, an alternative approach would have been to set up a controlled experiments, e.g. where participants are given suitable training with both ProofPower's tactic languages and PSGraph, and are then asked to explain and identify/repair faulty proof strategies. 
We can then measure the time and compute and compare their \emph{mean time to repair} (MTTR) metric \cite[p.462]{fenton2014software}. There are also models for maintainability that are based upon \emph{activities}, which may be more suitable as they have been applied to both text-based code and (graphical)
Simulink models (in industrial settings) \cite{deissenboeck2007activity}. This can form the basis for future experiments.
We could also try to measure and compare structural factors of the two representations, however  metrics for both usability \cite[p.460]{fenton2014software} and maintainability (e.g. the \emph{maintainability index} (MI) \cite{oman1991definition}) are tailored for linear languages and would not be applicable to our graphical language. To illustrate, MI combines measures such as lines of code, cyclomatic complexity and Halstead effort\footnote{This is computed using the number of operators, operands and program length.} \cite{halstead1977elements}.
}

\rev{
We have studied and contrasted \emph{programs} written in different languages. However, what we are really interested is the usabilty and maintainability provided by
the \emph{languages} themselves. In \S \ref{sec:intro} we gave one example from cognitive science that shows that humans find it more natural to comprehend flow networks diagrammatically compared with linear 
(term-based) representations \cite{larkin1987diagram}. Block-based languages, such
as Logo \cite{feurzeig1969programming}, Scratch \cite{resnick2009scratch}, Alice \cite{cooper2000alice} and AppInventor \cite{wolber2011app}, are similar to 
PSGraph in that programs are inter-connecting blocks.
Block-based languages are popular within CS education and there have been some studies comparing them with traditional languages. 
E.g. \cite{weintrop2015block} found that the drag-and-drop mechanism and the ease of browsing block-based libraries were advantageous;
whilst perceived drawbacks relate to expressibility and scalability. Besides this, we are not familiar with any work comparing 
actual languages\footnote{There has been some more interest recently, as illustrated by a SIG meeting on 
\emph{usability of programming languages} at CHI 2016.}. Simulink's popularity for embedded systems shows industrial use of graphical languages
for certain types of systems.
}

We are not familiar with any work on case studies in refactoring proof tactics as comprehensive as the work presented in the present paper. 
Whiteside (with others) has developed a refactoring framework for hierarchical proofs (HiProofs) \cite{paper:Whiteside:11,phd:Whiteside:13}, however this work focused mainly on proofs rather than proof strategies, albeit including
some work on folding and unfolding tactics. The most relevant tool to ours, which we are familiar with, is the aforementioned \emph{Tactician} tool to refactor proofs in HOL light \cite{Adams2015}. In HOL light (as is also the case for ProofPower)
a proof can be a sequence of (interactive) `apply' step, or they can be combined into a single step (by means of tactic combinators) which is then applied. Tactician is a tool to fold sequences into a single tactic and unfold a tactic into a sequence of steps. This can then be used for debugging by enable users to step through a large tactic, similar to how this can be achieved with Tinker. However, it only work for a small subset of ML and it is not clear how this approach can be generalised to arbitrary tactics. Moreover, it unfolds only one particular branch of the proof which does not necessarily reflect the underlying proof strategy.

Another tool recently developed to support debugging is the new tracing mechanism for the \emph{simp} tactic in Isabelle \cite{Hupel2014}. This is implemented as plug-in for the 
Isabelle/jEdit Prover IDE. It supports hierarchical viewing of simplification traces, and, as with Tinker, it enables breakpoints to be inserted where the user can step through and interact with the 
tactic. The breakpoints can either be an application of certain theorems or if the subgoal matches certain patterns. Note that it is not used to debug the (sub-)tactics used to implement the simplifier: it will only show how the simplifier applies rewrite steps.
Our logging mechanism is considerably simpler, and closer to the
more rudimentary ones supported in other ITP systems (including the previous tracing mechanism for the \emph{simp} tactic in Isabelle). However, in practice we have found that our logging mechanisms is sufficient as it only relates to a step at a time, while the \emph{simp} tactic could involve hundreds of steps.

\section{Conclusion \& future work}\label{sec:concl}

We have extended our graphical approach for tactic development with features to support development and debugging of proof strategies. Through examples, we have shown relative advantages when it comes to \emph{understanding} and \emph{debugging}
proof strategies compared with ML code, which is part of the hypothesis addressed here. We would argue that \emph{maintenance} 
is a consequence of this, and aim to further \rev{study this} this by completing the tasks set out in the FEF case study.

The experiments have provided some valuable lessons. 
The inclusion of PSGraph into a theorem prover should be seen to be \emph{evolutionary} rather than revolutionary; it should be seen as a new tool in the ecosystem for LCF-style theorem proving systems. 
An important feature is that you can still develop standard tactics using tacticals and treat them as atomic in PSGraph. This is a positive, and, with some better integration with particular theorem proving and SML system, you will still have the old version with an additional tool. \rev{It should also be noted that Tinker has been connected to a non-SML based prover:
Rodin is developed using Java \cite{Liang2016}.}
 
To use PSGraph effectively, one needs to change from a purely procedural view of tactics to more declarative thinking. 
There are trade-offs to be made  between a graphical, declarative and parallel approach on the one hand and the (hopefully) concise higher-order functional programming approach on the other.
Whilst losing some of the power of functional programming, the graphical approach is intrinsically more accessible. 
In a world where teams change, a solution perceived as elegant by the original developer may be perceived as incomprehensible and unmaintainable by his or her successors, particularly as the raw code of a tactic does not document \emph{why} design choices were made. If this is true for experts on tactics -- then what would `Joe Engineer' prefer if this technology is to become mainstream?
One compromise between procedural and declarative styles may be to think about how to represent tacticals graphically. PSGraph is essentially a way of composing tactics. A way forward may be to first just replace tacticals with similar graphical tactics:
this was illustrated this in the FEF case study, while \cite{grov13} shows some common tacticals graphically.

Technically, in order to use PSGraph with a new theorem prover all that has to be done is to implement a ML signature which tells PSGraph how to work with the proof state of that prover. Detailed information of how this is done in ProofPower can be found in \cite{webpage}. 

The case studies have shown some limitations with PSGraph that need to be addressed. This includes, more \emph{parametric graph tactics} to enable better re-use and modularity; closer integration with the theorem proving system to make it more natural for users; and low level term manipulations and rewriting, as illustrated by our encoding of conversions. Building on the FEF case study we would also like to test the robustness and maintainability, and contrast this with the linear tactic language such as ML.
Another experiment we would like to conduct is to start from scratch and implement known (or even unknown) decision procedures in PSGraph (e.g. quantifier elimination), and start building up a library  that can help in general.
It will also be interesting to see how PSGraph can be used as explanation of existing large libraries (e.g. the AFP for Isabelle or the Mizar library) -- or even as a good representation when writing papers on new decision procedures.

As PSGraph is implemented as a generic tool, and not specialised to a particular theorem prover, it may provide a way of translating proof strategies between theorem proving systems. While OpenTheory \cite{hurd2009} relies on the low-level similarities of the LCF kernels to translate proofs, PSGraph could potentially help exporting high-level proof strategies. This would require some  minimal set of atomic tactics and goal types and just use them (with a defined semantics).
Short term we could at least carry across structures of proof strategies, where the user needs to fill in the details in terms of atomic tactics and goal types. PSGraph could also be an experimental way of developing proof procedures, acting as the ``source code" with tactic languages as the ``target code" as they are likely to be more efficient than our graph interpreter.  It is however not clear how to do such compilation; possibly bottom-up in an interactive manner.


We have shown how our graphical representation can help to explain and support maintenance of selected examples from a real-world system.
The next challenge is to continue our work with D-RisQ \cite{ABZ16} on their 10K LoC tactic. Here, we believe that PSGraph can provide real 
economic advantages in terms of reduced development costs, reduced maintenance costs and improved communication with the user community.

\begin{footnotesize}

\end{footnotesize}

\end{document}

%% file: proofpower.tex
\def\VCDate{2016/12/05}\def\VCVersion{28fb919}

{\vertbarfalse
\Hide{
\begin{GFT}{SML}
\+open\_theory"hol";\\
\+new\_theory"pp\_examples";\\
\+new\_parent"\PrKP{}";\\
\+set\_pc"predicates";\\
\end{GFT}
}
ProofPower \cite{Arthan-Jones05} is a suite of tools supporting specification and proof.
At its heart is an implementation in Standard ML of Mike Gordon's HOL logic (the same logic
as is implemented in HOL4 \cite{hol4} and HOL Light \cite{hol-light} and the core of the logic implemented in Isabelle/HOL \cite{isabelle-isar}).
It is implemented in the well-known LCF style \cite{Gordon79,Gordon00}.
In this paper, we assume some familiarity with basic LCF concepts, but we will briefly review how these concepts are realised in ProofPower.
\rev{The purpose of this section is to provide a miniature primer on ProofPower
and how it is implemented. This is intended to give a feel for the user experience using and programming
an LCF style system in the traditional way for comparison with the Tinker approach.
This section also provides  technical background
for those interested in the details of how Tinker is connected to the theorem
prover. Readers familiar with some member of the LCF family
are invited to skip or skim this section on a first reading.}

Recall that in LCF terminology, the programming language used to implement the system is
referred to as the metalanguage (or ML) while the language of the logic implemented by the
system is referred to as the object language.  
ProofPower is implemented as a large library of functions that are invoked from its
metalanguage ProofPower-ML, which is the interactive functional programming language Standard ML with extensions to support convenient entry of object language constructs.
Object language terms are represented by an abstract data type {\it TERM}
with a constructor for each syntactic category in the object language.
Values of type {\it TERM} can be entered using object language concrete syntax
enclosed in  Quine corners, `\PrKM{}' and `\PrKO{}'. So, for example, the following ML command line:
\begin{GFT}{SML}
\+val tm1 = \PrKM{}(\PrMM{}x\PrLH{}\PrMM{}f\PrLH{} f x) 1\PrKO{};\\
\end{GFT}
\noindent
causes the string of symbols between the Quine corners to be parsed and type-checked
resulting in a value of type {\it TERM} that is bound to the ML variable {\it tm1}.
\rev{(For customer-oriented reasons, the design of the concrete syntax for HOL in ProofPower
was heavily influenced by the Z notation \cite{Spivey92}, hence the rather heavyweight
bullets in the $\lambda$-abstraction.)}

In the HOL logic, a theorem is a sequent $\phi_1, \ldots, \phi_n \vdash \phi$,
asserting that, if the hypotheses $\phi_1, \ldots, \phi_n$ hold, then so does the conclusion $\phi$
(here $\phi$ and the $\phi_i$ are propositions, i.e., terms of type {\it BOOL}).
Theorems are implemented as an abstract data type {\it THM} with a constructor for
each primitive inference rule schema of the logic, parametrised by the antecedents of
the rule schema and any other information needed to instantiate the schema.
For example, one primitive inference rule schema is an axiom schema
asserting that any $\beta$-redex is equal to its $\beta$-reduct.
It is implemented by a constructor
{\it simple\_\PrMC{}\_conv}
\ with a single parameter that identifies the $\beta$-reduct.
If (after executing the command above), we execute:
\begin{GFT}{SML}
\+val thm1 = simple\_\PrMC{}\_conv tm1;\\
\end{GFT}
\noindent
the system responds with:
\begin{GFT}{ProofPower Output}
\+val thm1 = \PrPE{} (\PrMM{} x f\PrLH{} f x) 1 = (\PrMM{} f\PrLH{} f 1): THM\\
\end{GFT}
\noindent
indicating that a value of type {\it THM}
with the appropriate instance of the $\beta$-reduction axiom as its conclusion
has been bound to the ML variable {\it thm}.
(Note that the pretty-printer has used a short-hand form for the nested $\lambda$-abstraction,
as shall we in future examples.)

The constructor {\it asm\_rule} implements the axiom schema containing all theorems
of the form $\phi \vdash \phi$ and the constructor {\it eq\_trans\_rule} implements
the rule schema for transitivity of equality. Putting these together, if we execute:
\begin{GFT}{SML}
\+val thm2 = eq\_trans\_rule (asm\_rule \PrKM{}H = (\PrMM{}x f\PrLH{}f x) 1\PrKO{}) thm1;\\
\end{GFT}
\noindent
the system responds with
\begin{GFT}{ProofPower Output}
\+val thm2 = H = (\PrMM{} x f\PrLH{} f x) 1 \PrPE{} H = (\PrMM{} f\PrLH{} f 1): THM\\
\end{GFT}
Typically, ProofPower users do not use the primitive inference rules directly, but instead
use derived proof procedures that operate at a higher-level.
One widely used abstraction supporting equational reasoning is the notion of conversion
\cite{Paulson83}. A conversion is a function of type
\begin{INLINEFT}%
\+TERM -> THM\\
\end{INLINEFT}%
, which, by convention, when passed a term $t$, returns a theorem with conclusion
of the form $t = t'$. The primitive inference rule
{\it simple\_\PrMC{}\_conv}
\ discussed above is an example of a conversion, which proves all theorems of the
form
\begin{INLINEFT}%
\+\PrPE{} (\PrMM{}x\PrLH{}t)u = t[u/x]\\
\end{INLINEFT}%
.
Conversions are often used to package various kinds of normal form.
For example, the conversion {\it anf\_conv} implements a normal form
for natural number arithmetic expressions. If we execute: 
\begin{GFT}{SML}
\+val thm3 = anf\_conv\PrKM{}7*(11 + x)*(13 + y)\PrKO{};\\
\end{GFT}
\noindent
the system responds with:
\begin{GFT}{ProofPower Output}
\+val thm3 = \PrPE{} 7 * (11 + x) * (13 + y) =\\
\+	1001 + 91 * x + 77 * y + 7 * x * y: THM\\
\end{GFT}
There is a heavily used family of conversions that work by rewriting with equational theorems
such as the definitions of library functions:
\begin{GFT}{SML}
\+val thm4 = rewrite\_conv[map\_def, pair\_ops\_def]\\
\+			\PrKM{}Map Fst [(1, 2); (3, 4); (5, 6)]\PrKO{};\\
\end{GFT}
\noindent
Here {\it map\_def} and {\it pair\_ops\_def} refer to the theorems representing the
definitions of the {\it Map} combinator and the constructor $(,)$ and destructors
{\it Fst} and {\it Snd} for pairs. (The ProofPower syntax follows the old HOL tradition
of using semi-colons to separate the elements of lists.)
This results in:
\begin{GFT}{ProofPower Output}
\+val thm4 = \PrPE{} Map Fst [(1, 2); (3, 4); (5, 6)] = [1; 3; 5]: THM\\
\end{GFT}
Many of the standard proof procedures provided in ProofPower are parametrised
by what is called a {\it proof context}: a named collection of standard
transformations to apply to a problem. The proof context allows the standard
proof procedures to be tailored to particular problem domains and proof techniques.
\rev{Higher-order functions are provided to allow proof procedures of various types to be executed
in a specified proof context. For example, the function
\begin{INLINEFT}%
\+PC\_C1\\
\end{INLINEFT}%
\ executes a function returning a conversion in a specified proof context.
Thus the following command performs a single step of rewriting
in the proof context {\it sets\_ext} designed for reasoning about
sets using extensionality:}
\begin{GFT}{SML}
\+val thm5 = PC\_C1 "sets\_ext" once\_rewrite\_conv[]\\
\+			\PrKM{}\{x | x < 20\} = \{x | x \PrLM{} 19\}\PrKO{};\\
\end{GFT}
\begin{GFT}{}
\+val thm5 = \PrPE{} \{x|x < 20\} = \{x|x \PrLM{} 19\} \PrKE{}\\
\+		(\PrLF{} x\PrLH{} x \PrIN{} \{x|x < 20\} \PrKE{} x \PrIN{} \{x|x \PrLM{} 19\}): THM\\
\end{GFT}
\noindent
The following command rewrites in the proof context {\it sets\_ext} until no more rewriting is possible:
\begin{GFT}{SML}
\+val thm5 = PC\_C1 "sets\_ext" rewrite\_conv[] \PrKM{}\{x | x < 20\} = \{x | x \PrLM{} 19\}\PrKO{};\\
\end{GFT}
\noindent
which reduces the problem to pure arithmetic:
\begin{GFT}{ProofPower Output}
\+val thm5 = \PrPE{} \{x|x < 20\} = \{x|x \PrLM{} 19\} \PrKE{} (\PrLF{} x\PrLH{} x < 20 \PrKE{} x \PrLM{} 19): THM\\
\end{GFT}
Conversions can be composed using the infix combinator {\it THEN\_C}.
If we compose rewriting in the proof context {\it sets\_ext} with rewriting
in a proof context designed to deal with linear natural number arithmetic
the problem reduces to truth and we can derive a proof of our original equation: 
\begin{GFT}{SML}
\+val thm7 = (PC\_C1 "sets\_ext" rewrite\_conv[] THEN\_C\\
\+		PC\_C1 "lin\_arith" rewrite\_conv[])\\
\+			\PrKM{}\{x | x < 20\} = \{x | x \PrLM{} 19\}\PrKO{};\\
\+val thm8 = \PrKE{}\_t\_elim thm7;\\
\end{GFT}
\noindent
yielding:
\begin{GFT}{ProofPower Output}
\+val thm7 = \PrPE{} \{x|x < 20\} = \{x|x \PrLM{} 19\} \PrKE{} T: THM\\
\+val thm8 = \PrPE{} \{x|x < 20\} = \{x|x \PrLM{} 19\}: THM\\
\end{GFT}
While forward proof using conversions and other inference rules gives a powerful
approach to programming proof procedures, a more natural and productive approach to
finding proofs interactively is a goal-directed search, starting with the assertion you
wish to prove as the initial goal and transforming each goal into subgoals that entail
that the goal and are (hopefully) easier to prove. The transformations are effected by
what Milner christened {\em tactics}: ML functions that map a goal to a pair comprising
{\em(i)} the list of subgoals and {\em(ii)} a proof, i.e., a function that will prove
the goal given theorems validating the subgoals.
As in other HOL systems, goals in ProofPower comprise a list of
assumptions and a conclusion, so this simple but powerful idea is
captured in the following type declarations:
\begin{GFT}{}
\+	type GOAL	= TERM list * TERM;\\
\+	type PROOF	= THM list -> THM;\\
\+	type TACTIC	= GOAL -> GOAL list * PROOF;\\
\end{GFT}
\noindent
\rev{For example, consider the goal:
\begin{INLINEFT}%
\+([], \PrKM{}1 < 2 \PrLB{} 2 < 3\PrKO{})\\
\end{INLINEFT}%
\noindent (with an empty list of assumptions).
A tactic (namely
\begin{INLINEFT}%
\+\PrLB{}\_tac\\
\end{INLINEFT}%
) might reduce this goal to:}
\begin{GFT}{}
\+	([([], \PrKM{}1 < 2\PrKO{}) , ([], \PrKM{}2 < 3\PrKO{})],\\
\+		fn [th1, th2] => \PrLB{}\_intro th1 th2) : GOAL list * PROOF\\
\end{GFT}
\noindent
\rev{I.e., it gives us two subgoals with conclusions $1 < 2$ and $2 < 3$ respectively,
together with a function, which, given a list comprising two theorems that validate these subgoals,
will use $\land$-introduction to return a theorem validating our original goal.}

This approach to interactive proof was supported from
the earliest days of LCF (see \cite{Gordon00} for the history)
and came into its own when Paulson implemented the first interactive package
for managing the subgoal state \rev{during the user's search for a combination of
tactics that will prove their goal.}

In the ProofPower subgoal package, the goal with assumptions $t_1, \ldots, t_n$ and conclusion
$t$ is internally represented 
as a single term that is logically equivalent to the universal closure
of $t_1 \land \ldots \land t_n \Rightarrow t$.
The logical state of the proof search is captured
in a theorem whose conclusion represents the original goal and
whose assumptions represent the outstanding subgoals.
When a tactic is applied to a goal, the corresponding assumption is
replaced by terms representing the list of subgoals returned by the
tactic. Assumptions are labelled by dot-separated lists of natural numbers,
representing a position in a tree whose root corresponds to
the original goal and whose nodes correspond to tactic applications which
result in more than one subgoal.  In any state there is a current subgoal
that tactics are applied to. A function {\it set\_labelled\_goal}
is provided to allow the user to navigate around the outstanding subgoals.

A session with the subgoal package may be initiated with the {\it set\_goal} command: 
\begin{GFT}{SML}
\+set\_goal([], \PrKM{}\PrLF{}x\PrLH{} (1, x) \PrIN{} \{(a, b) | a = 1 \PrLC{} b \PrLO{} 2\} \PrLB{} (x \PrLO{} 3 \PrLC{} x \PrLM{} 4)\PrKO{});\\
\end{GFT}
\noindent
The system responds by printing out the state of the proof search.
There is 1 goal and its label is the empty string:
\begin{GFT}{ProofPower Output}
\+Now 1 goal on the main goal stack\\
\+\\
\+(* *** Goal "" *** *)\\
\+\\
\+(* ?\PrPE{} *)\PrKM{}\PrLF{} x\PrLH{} (1, x) \PrIN{} \{(a, b)|a = 1 \PrLC{} b \PrLO{} 2\} \PrLB{} (x \PrLO{} 3 \PrLC{} x \PrLM{} 4)\PrKO{}\\
\end{GFT}
\noindent \rev{Here the symbol $?\vdash$ is used in the display of a goal as indicative of
a sequent that has not yet been proved. It is included as an ML comment to facilitate
copying and pasting the output as executable code in an ML script.}

At this point, we have several choices about the tactic to apply.
If we want to take a fine-grained approach, we could apply tactics that exactly
match the outer two layers of the logical structure:
\begin{GFT}{SML}
\+a(\PrLF{}\_tac THEN \PrLB{}\_tac);\\
\end{GFT}
\noindent
\rev{Here $THEN$ is a {\em tactical}, i.e., an operator that constructs new tactics from old,
in this case by a form of sequential composition.}
This results in:
\begin{GFT}{ProofPower Output}
\+Tactic produced 2 subgoals:\\
\+\\
\+(* *** Goal "2" *** *)\\
\+\\
\+(* ?\PrPE{} *)\PrKM{}x \PrLO{} 3 \PrLC{} x \PrLM{} 4\PrKO{}\\
\+\\
\+\\
\+(* *** Goal "1" *** *)\\
\+\\
\+(* ?\PrPE{} *)\PrKM{}(1, x) \PrIN{} \{(a, b)|a = 1 \PrLC{} b \PrLO{} 2\}\PrKO{}\\
\end{GFT}
Alternatively, we could repeatedly apply the general purpose tactic {\em strip\_tac}
which applies a standard simplification to logical connectives if possible
and proof context-dependent transformations to atomic formulas.
The following command does this after undoing what we have just done:
\begin{GFT}{SML}
\+undo 1; a(REPEAT strip\_tac);\\
\end{GFT}
\begin{GFT}{ProofPower Output}
\+Tactic produced 2 subgoals:\\
\+\\
\+(* *** Goal "2" *** *)\\
\+\\
\+(*  1 *)\PrKM{}\PrLD{} x \PrLO{} 3\PrKO{}\\
\+\\
\+(* ?\PrPE{} *)\PrKM{}x \PrLM{} 4\PrKO{}\\
\+\\
\+\\
\+(* *** Goal "1" *** *)\\
\+\\
\+(* ?\PrPE{} *)\PrKM{}(1, x) \PrIN{} \{(a, b)|a = 1 \PrLC{} b \PrLO{} 2\}\PrKO{}\\
\end{GFT}
\noindent
Note how in goal 2, the disjunction has been dealt with by asking us to
prove its right-hand side on the assumption that its left-hand side is false.
\rev{(The assumptions are displayed above the conclusion of the goals with numbers
in ML comments to identify them. In this case there is just one assumption.)}
Let us assume that this is not quite what we wanted; so we undo it and try again but only
stripping off two layers of connective:
\begin{GFT}{SML}
\+undo 1; a (strip\_tac THEN strip\_tac);\\
\end{GFT}
\noindent
This gives us:
\begin{GFT}{ProofPower Output}
\+Tactic produced 2 subgoals:\\
\+\\
\+(* *** Goal "2" *** *)\\
\+\\
\+(* ?\PrPE{} *)\PrKM{}x \PrLO{} 3 \PrLC{} x \PrLM{} 4\PrKO{}\\
\+\\
\+\\
\+(* *** Goal "1" *** *)\\
\+\\
\+(* ?\PrPE{} *)\PrKM{}(1, x) \PrIN{} \{(a, b)|a = 1 \PrLC{} b \PrLO{} 2\}\PrKO{}\\
\end{GFT}
Looking at goal 1, we see it should become trivial once the set notation has been
eliminated \rev{using standard properties of set comprehensions and pairs.
So iterating {\it strip\_tac} should do} just what we want in the {\it sets\_ext} proof context.
\rev{To do this we use the {\em tactical}
\begin{INLINEFT}%
\+PC\_T1\\
\end{INLINEFT}%
, which does for tactics what
\begin{INLINEFT}%
\+PC\_C1\\
\end{INLINEFT}%
, discussed above, does for conversions:}
\begin{GFT}{SML}
\+a(PC\_T1 "sets\_ext" REPEAT strip\_tac);\\
\end{GFT}
\noindent
which results in:
\begin{GFT}{ProofPower Output}
\+Tactic produced 0 subgoals:\\
\+Current goal achieved, next goal is:\\
\+\\
\+(* *** Goal "2" *** *)\\
\+\\
\+(* ?\PrPE{} *)\PrKM{}x \PrLO{} 3 \PrLC{} x \PrLM{} 4\PrKO{}\\
\end{GFT}
We recognise that this problem is entirely in the domain of linear natural number arithmetic.
The proof context {\it lin\_arith} for this domain includes a decision procedure that we
can access via a generic tactic {\it prove\_tac}:
\begin{GFT}{SML}
\+a(PC\_T1 "lin\_arith" prove\_tac[]);\\
\end{GFT}
\noindent
This completes the proof search:
\begin{GFT}{ProofPower Output}
\+Tactic produced 0 subgoals:\\
\+Current and main goal achieved\\
\end{GFT}
\noindent
We can now extract our theorem from the subgoal package:
\begin{GFT}{SML}
\+val thm9 = pop\_thm();\\
\end{GFT}
\begin{GFT}{ProofPower Output}
\+Now 0 goals on the main goal stack\\
\+val thm9 = \PrPE{} \PrLF{} x\PrLH{} (1, x) \PrIN{} \{(a, b)|a = 1 \PrLC{} b \PrLO{} 2\} \PrLB{} (x \PrLO{} 3 \PrLC{} x \PrLM{} 4): THM\\
\end{GFT}
} 

%% file: tauttac.tex
\Hide{
\begin{GFT}{SML}
\+set\_pc"predicates";\\
\+val local\_if\_thm = taut\_rule \PrKM{}\PrLF{} a t1 t2\PrLH{} (if a then t1 else t2) \PrKE{} (a \PrLE{} t1) \PrLB{} (\PrLD{} a \PrLE{} t2)\PrKO{};\\
\+val a\_\PrLC{}\_\PrLD{}b\_thm = taut\_rule \PrKM{}\PrLF{}a b\PrLH{}(a \PrLC{} \PrLD{}b) \PrKE{} (b \PrLE{} a)\PrKO{};\\
\+val a\_\PrLC{}\_b\_thm = taut\_rule \PrKM{}\PrLF{} a b\PrLH{} a \PrLC{} b \PrKE{} \PrLD{} a \PrLE{} b\PrKO{};\\
\+val \PrLD{}a\_\PrLC{}\_b\_thm = taut\_rule \PrKM{}\PrLF{} a b\PrLH{} \PrLD{} a \PrLC{} b \PrKE{} a \PrLE{} b\PrKO{};\\
\end{GFT}
} 
{\vertbarfalse
\makeatletter
\def\prelim@label#1{}
\makeatother

\rev{For the purposes of this section,
a tautology is defined to be a substitution instance of any formula $\chi$ formed from
boolean variables and the boolean constants $T$ and $F$ using the connectives
\begin{INLINEFT}%
\+\PrLD{}, \PrLB{}, \PrLC{}, \PrLE{}, \PrKE{}, if \_ then \_ else \_\\
\end{INLINEFT}%
, such that $\chi$ evaluates to $T$ under any substitution of the constants $T$ or $F$ for
its propositional variables.
We will describe the design and implementation of a tactic
that takes a goal which we assume (for simplicity) has no assumptions:
\begin{INLINEFT}%
\+?\PrPE{} \PrMG{}\\
\end{INLINEFT}%
\ and will prove any such goal where $\phi$ is a tautology.}

\rev{The decision procedure underlying the tautology tactic transforms its goal
to a set $S$ of subgoals:}
\begin{center}
\begin{minipage}{0.75\hsize}
\begin{GFT}{}
\+\PrIH{}\PrIJ{1} ?\PrPE{} t\PrIJ{1}, ..., \PrIH{}\PrIJ{n} ?\PrPE{} t\PrIJ{n}\\
\end{GFT}
\end{minipage}
\end{center}
\noindent where the
\begin{INLINEFT}%
\+\PrIH{}\PrIJ{i}\\
\end{INLINEFT}%
\ and the
\begin{INLINEFT}%
\+t\PrIJ{i}\\
\end{INLINEFT}%
\ comprise only propositional literals,
\rev{i.e., atoms or negated atoms(but not $\lnot T$ or $\lnot F$). The transformation
ensures that $S$ is
is logically equivalent equivalent to the original goal when viewed as a conjunction of implications.
The original goal is then a tautology iff each of the subgoals in $S$} has one of the
following forms (which we refer to below as {\em structural tautological forms}):
\begin{center}
\begin{minipage}{0.75\hsize}
\begin{GFT}{}
\+\PrIH{}, t, \PrLD{}t ?\PrPE{} u\\
\+\PrIH{}, t ?\PrPE{} t\\
\+\PrIH{}, F ?\PrPE{} u\\
\+\PrIH{} ?\PrPE{} T.\\
\end{GFT}
\end{minipage}
\end{center}
\noindent
\rev{The implementation will realise these transformations as tactics and will apply
a tactic that will recognise and discharge structural tautological subgoals
as they are created. Realising the decision procedure using tactics in this way
converts it from an algorithm that merely recognises tautologies into
an algorithm that finds a proof.}

\rev{The tactic that implements the decision procedure} uses two rewrite systems.
The rewrite systems are defined by theorems giving universally quantified
bi-implications which are instantiated as appropriate and used as left-to-right
rewrite rules.
The first rewrite system is applied to the conclusions of subgoals:

\begin{center}
\begin{minipage}{0.75\hsize}
\begin{GFT}{}
\+\PrPE{} \PrLF{} a\PrLH{} \PrLD{} \PrLD{} a \PrKE{} a\\
\+\PrPE{} \PrLF{} a b\PrLH{} \PrLD{} (a \PrLB{} b) \PrKE{} \PrLD{} a \PrLC{} \PrLD{} b\\
\+\PrPE{} \PrLF{} a b\PrLH{} \PrLD{} (a \PrLC{} b) \PrKE{} \PrLD{} a \PrLB{} \PrLD{} b\\
\+\PrPE{} \PrLF{} a b\PrLH{} \PrLD{} (a \PrLE{} b) \PrKE{} a \PrLB{} \PrLD{} b\\
\+\PrPE{} \PrLF{} a b\PrLH{} \PrLD{} (a \PrKE{} b) \PrKE{} a \PrLB{} \PrLD{} b \PrLC{} b \PrLB{} \PrLD{} a\\
\+\PrPE{} \PrLD{} T \PrKE{} F\\
\+\PrPE{} \PrLD{} F \PrKE{} T\\
\+\PrPE{} \PrLF{} a b c\PrLH{} \PrLD{} (if a then b else c) \PrKE{} (if a then \PrLD{} b else \PrLD{} c)\\
\+\PrPE{} \PrLF{} a b\PrLH{} (a \PrKE{} b) \PrKE{} (a \PrLE{} b) \PrLB{} (b \PrLE{} a)\\
\+\PrPE{} \PrLF{} a t1 t2\PrLH{} (if a then t1 else t2) \PrKE{} (a \PrLE{} t1) \PrLB{} (\PrLD{} a \PrLE{} t2)\\
\+\PrPE{} \PrLF{} a b\PrLH{} a \PrLC{} \PrLD{} b \PrKE{} b \PrLE{} a\\
\+\PrPE{} \PrLF{} a b\PrLH{} \PrLD{} a \PrLC{} b \PrKE{} a \PrLE{} b\\
\+\PrPE{} \PrLF{} a b\PrLH{} a \PrLC{} b \PrKE{} \PrLD{} a \PrLE{} b\\
\end{GFT}
\end{minipage}
\end{center}

The following ProofPower idiom implements the above rewrite system:

\begin{GFT}{SML}
\+val taut\_strip\_concl\_conv : CONV = (\\
\+	eqn\_cxt\_conv(\\
\+	map thm\_eqn\_cxt\\
\+	[\PrLD{}\_\PrLD{}\_thm, \PrLD{}\_\PrLB{}\_thm, \PrLD{}\_\PrLC{}\_thm, \PrLD{}\_\PrLE{}\_thm,\\
\+	 \PrLD{}\_\PrKE{}\_thm, \PrLD{}\_t\_thm, \PrLD{}\_f\_thm, \PrLD{}\_if\_thm,\\
\+	\PrKE{}\_thm, local\_if\_thm,\\
\+	a\_\PrLC{}\_\PrLD{}b\_thm, \PrLD{}a\_\PrLC{}\_b\_thm, a\_\PrLC{}\_b\_thm]));\\
\end{GFT}
\noindent
\rev{Here 
\begin{INLINEFT}%
\+\PrLD{}\_\PrLD{}\_thm\\
\end{INLINEFT}%
,
\begin{INLINEFT}%
\+\PrLD{}\_\PrLB{}\_thm\\
\end{INLINEFT}%
\ etc. name the theorems of the rewrite system in the order given above.}
Repeated application of these rewrite rules will transform the conclusion of a subgoal
into either a propositional literal or a conjunction or an implication.
If the conclusion is a propositional literal, the tactic will test whether the subgoal
has one of the structural tautological forms, discharging the subgoal if it passes the
test and reporting an error if it fails.
If the conclusion is a conjunction, the subgoal splits into two subgoals, one for each conjunct.
\rev{If the conclusion is an implication say $\phi \Rightarrow \psi$, the subgoal will reduce to a set of subgoals
obtained from the original subgoal by adding certain literals to its assumptions. These literals are
 obtained by ``stripping'' the logical connectives out of the antecedent $\phi$ while making case splits as appropriate.}

\rev{The following list of functions captures the processing described above but defers
the stripping of new assumptions to a parameter that is a function of type}
\begin{GFT}{}
\+	THM\_TACTIC = THM -> TACTIC,\\
\end{GFT}
\noindent
\rev{(Functions of this type are referred to as {\em theorem continuations} and play an important role in the traditional approach to programming LCF style systems  \cite{Paulson87}.)
For a conclusion of the form $\phi \Rightarrow \psi$, the tactical
\begin{INLINEFT}%
\+\PrLE{}\_T\\
\end{INLINEFT}%
\ will carry out the stripping process by passing the theorem
\begin{INLINEFT}%
\+\PrMG{} \PrPE{} \PrMG{}\\
\end{INLINEFT}%
\ representing the new assumption to the parameter function.
For the cases other than implications, this parameter is ignored.}

\begin{GFT}{SML}
\+val taut\_strip\_concl\_ts : (THM\_TACTIC -> TACTIC) list = [\\
\+	fn \_ => \PrLB{}\_tac,\\
\+	\PrLE{}\_T,\\
\+	fn \_ => t\_tac,\\
\+	fn \_ => conv\_tac taut\_strip\_concl\_conv,\\
\+	fn \_ => concl\_in\_asms\_tac];\\
\end{GFT}

Stripping the antecedent $\phi$ of an implication $\phi \Rightarrow \psi$ into the assumptions is dual to the processing of a conclusion.
If $\phi$ is not a conjunction or a disjunction, it is rewritten using a second rewrite system.
This second system is like the first but with the rules for disjunctions replaced by
the following rule for implications:

\begin{center}
\begin{minipage}{0.75\hsize}
\begin{GFT}{}
\+\PrPE{} \PrLF{} a b\PrLH{} a \PrLE{} b \PrKE{} \PrLD{} a \PrLC{} b\\
\end{GFT}
\end{minipage}
\end{center}

\begin{GFT}{SML}
\+val taut\_strip\_thm\_conv : CONV = (\\
\+	eqn\_cxt\_conv(\\
\+	map thm\_eqn\_cxt\\
\+	[\PrLD{}\_\PrLD{}\_thm, \PrLD{}\_\PrLB{}\_thm, \PrLD{}\_\PrLC{}\_thm, \PrLD{}\_\PrLE{}\_thm,\\
\+	 \PrLD{}\_\PrKE{}\_thm, \PrLD{}\_t\_thm, \PrLD{}\_f\_thm, \PrLD{}\_if\_thm,\\
\+	\PrLE{}\_thm, \PrKE{}\_thm, local\_if\_thm]));\\
\end{GFT}
The new subgoals derived by stripping $\phi$ into the assumptions are then produced
by iterating around the following list of functions: if $\phi$ is a conjunction,
$\phi_1 \land \phi_2$, we strip $\phi_1$ and $\phi_2$ into the assumptions separately;
if $\phi$ is disjunction, $\phi_1 \lor \phi_2$, we get two subgoals, one with
$\phi_1$ stripped into its assumptions and one with $\phi_2$ stripped into its assumptions;
otherwise we attempt to apply the rewrite rules:
\begin{GFT}{SML}
\+val taut\_strip\_thm\_thens : THM\_TACTICAL list = [\\
\+	\PrLB{}\_THEN,\\
\+	\PrLC{}\_THEN,\\
\+	CONV\_THEN taut\_strip\_thm\_conv];\\
\end{GFT}
\noindent
\rev{Here
\begin{INLINEFT}%
\+\PrLB{}\_THEN\\
\end{INLINEFT}%
\ and
\begin{INLINEFT}%
\+\PrLC{}\_THEN\\
\end{INLINEFT}%
\ are operators on theorem continuations
\ that perform one logical transformation and pass the theorems representing the result on to
their operands.
Operators like this provide a powerful continuation-passing style
for programming tactics. This style was introduced and popularised by Paulson \cite{Paulson87}
and widely adopted by developers of tactics in LCF-style systems.}

The following tactic implements a single step in the above process as determined by
the principal connective of the conclusion of the goal.
\rev{The expression beginning
\begin{INLINEFT}%
\+REPEAT\_TTCL\\
\end{INLINEFT}%
\ is the
theorem continuation parameter for
\begin{INLINEFT}%
\+\PrLE{}\_T\\
\end{INLINEFT}%
\ mentioned above.
It uses
\begin{INLINEFT}%
\+taut\_strip\_thm\_thens\\
\end{INLINEFT}%
\ to strip a new assumption into atoms and then uses
the tactic
\begin{INLINEFT}%
\+check\_asm\_tac\\
\end{INLINEFT}%
\ to add these assumptions to the resulting subgoals while checking
for and discharging subgoals having one of the structural tautological forms.}

\begin{GFT}{SML}
\+val taut\_strip\_tac : TACTIC = (\\
\+	FIRST\\
\+	(map(fn t => t(REPEAT\_TTCL (FIRST\_TTCL taut\_strip\_thm\_thens)\\
\+			check\_asm\_tac))\\
\+		taut\_strip\_concl\_ts));\\
\end{GFT}
\noindent
\rev{Here
\begin{INLINEFT}%
\+REPEAT\_TTCL\\
\end{INLINEFT}%
\ and
\begin{INLINEFT}%
\+FIRST\_TTCL\\
\end{INLINEFT}%
\ are combinators on theorem continuations that provide repetition until failure
and selection of the first non-failing theorem continuation from a list.}

\rev{For an example of
\begin{INLINEFT}%
\+taut\_strip\_tac\\
\end{INLINEFT}%
\ in action, let us see how it will work given Peirce's law:
\begin{INLINEFT}%
\+((a \PrLE{} b) \PrLE{} a) \PrLE{} a.\\
\end{INLINEFT}%
$\;$ It will actually prove this immediately by stripping  the antecedent
\begin{INLINEFT}%
\+(a \PrLE{} b) \PrLE{} a\\
\end{INLINEFT}%
\ into the assumptions of a goal with conclusion
\begin{INLINEFT}%
\+a\\
\end{INLINEFT}%
. The new assumption will first be rewritten in the form
\begin{INLINEFT}%
\+\PrLD{}(a \PrLE{} b) \PrLC{} a\\
\end{INLINEFT}%
\ resulting in a case split into two goals, one with assumption
\begin{INLINEFT}%
\+\PrLD{}(a \PrLE{} b)\\
\end{INLINEFT}%
\ and one with assumption $a$. The assumption
\begin{INLINEFT}%
\+\PrLD{}(a \PrLE{} b)\\
\end{INLINEFT}%
\ will be rewritten as
\begin{INLINEFT}%
\+a \PrLB{} \PrLD{}b\\
\end{INLINEFT}%
\ which will be stripped into two new assumption literals,
\begin{INLINEFT}%
\+a\\
\end{INLINEFT}%
\ and
\begin{INLINEFT}%
\+\PrLD{}b\\
\end{INLINEFT}%
. In both cases, the resulting subgoal is a structural tautology that will be discharged by
\begin{INLINEFT}%
\+check\_asm\_tac\\
\end{INLINEFT}%
.}

The tautology tactic then simply repeats the single step tactic until there are no subgoals left
or until no further progress can be made, in which case it raises an exception.
\begin{GFT}{SML}
\+val simple\_taut\_tac : TACTIC = (fn gl =>\\
\+	case REPEAT taut\_strip\_tac gl of\\
\+		done as ([], \_) => done\\
\+	|	\_ => fail "simple\_taut\_tac" 28121 []);\\
\end{GFT}
\noindent
\rev{(The number $28121$ here is an error code identifying a message reporting that the conclusion of the goal is not a tautology.) }

\rev{Failure-driven higher-order functional programming using combinators to control
iteration and sequencing has proved very successful in programming LCF-style systems.
Here it enables us to code a rather complex recursion scheme in a compact
way that does reveal the structure of the algorithm to those familiar with the approach.
Although we have spent several pages here describing the tautology tactic, the actual source
code we have presented is only 32 lines of which all but 9 do little more than set up tables.
However, we agree with Paulson, who concedes that while higher-order functions
provide good control and efficiency, they can be hard to understand \cite{Paulson87}.}

} 

%% file: feftacs.tex
{\vertbarfalse

The FEF Project \cite{fef} was an early application of ProofPower to
verify the security properties of a multi-level secure database system called SWORD.
In this section we will look at some very domain-specific tactics taken from the FEF
proof scripts as an experiment
in porting an existing application proof to PSGraph with a view to making
future maintenance easier by presenting the proof at a higher level.

It would be inappropriate here to give a very detailed description of the four tactics
we investigate. Such a description would be long and not very instructive and would
not reflect the process by which the tactics came into existence, which was by interactive
trial and error. The best way to get a feeling for this kind of process is by replaying
the proofs interactively. See \cite{fef} for instructions for downloading the proof
scripts. The ProofPower documents that are most relevant to the present paper are {\tt fef032.doc}
for specifications and {\tt fef033.doc} for the proofs.
In the present paper, we will set the relevant proof in
context, which should give enough background to understand our transcription of
the tactics into PSGraph and should help anyone who is interested in more detail
to locate and work with the FEF documents. 

The query language for the SWORD database was Secure SQL (SSQL) an extension of standard
SQL whose semantics support a notion of security classification.
Data in the database would be a assigned a classification drawn from some lattice. e.g.,
SECRET $>$ COMPANY-RESTRICTED $>$ COMPANY-IN-CONFIDENCE $>$ UNCLASSIFIED.
If $c_1$ and $c_2$ are elements of the lattice we say $c_1$ {\em dominates} $c_2$
if $c_1 \ge c_2$.
Database users are assigned a security clearance which is also drawn from the lattice of classifications:
a user cleared at classification $c$ is only allowed to see data whose classification
is dominated by $c$.
SSQL was implemented via a preprocessor
(referred to as the Front End Filter or FEF) that translates SSQL queries to queries
in ordinary SQL on a database whose schema augmented the SSQL schema with security
classification for each item of data. The translated query uses these classifications
to prohibit information flows that would violate the security policy,
e.g., by revealing SECRET data to a user who is only cleared to see information at
COMPANY-IN-CONFIDENCE or UNCLASSIFIED.

The high-level security property for FEF requires that for every query $q$
and every user $u$, if two states $s_1$ and $s_2$ differ
only in respect of data that $u$ is not cleared to see, then, when executed
by $u$, $q$ will deliver the same result in state $s_1$ as it doesn in state $s_2$.
To achieve this, the semantics of SSQL label the result of any calculation with a
security classification. If the label on the end result of a query is $c$, then the
implementation will erase the information content of the result if the user is not
cleared to see data of classification $c$. The formal specification of the SSQL
semantics includes for each language construct both the derivation of the result
and the derivation of the classification label from the values and classifications
of the operands of the construct and of the database items it accesses.
There is then a proof obligation to show that the semantics satisfies the
high-level security property. Like conventional SQL, the syntax of SSQL queries
involves a mutual recursion between table-expressions and value-expressions.
These are given formal semantics as what we refer to as table computations and
value computations. To prove that the semantics satisfies the high-level security property
involves an induction over the syntax to prove a property
\begin{INLINEFT}%
\+OK\_TC\PrIJ{d}\\
\end{INLINEFT}%
\ on table computations that can be used
fairly  directly to prove the high-level security property.
\begin{INLINEFT}%
\+OK\_TC\PrIJ{d}\\
\end{INLINEFT}%
\ is parametrised by a security classification $c$ and is defined as follows:
\begin{GFT}{}
\+\PrLF{}c tc\PrLH{}	tc \PrIN{} OK\_TC\PrIJ{d} c \PrKE{}\\
\+	\PrLF{}tl\PrIJ{0} tl\PrIJ{1}\PrLH{}\\
\+		Map (HideDerTable c) tl\PrIJ{0} = Map (HideDerTable c) tl\PrIJ{1}\\
\+	\PrLB{}	\PrLD{}HideDerTable c (Snd(tc tl\PrIJ{0})) = HideDerTable c (Snd(tc tl\PrIJ{1}))\\
\+	\PrLE{}	\PrLD{}c dominates Fst(tc tl\PrIJ{0})\\
\end{GFT}
Here we see that the table computation $tc$ is a function with
one argument: a list of tables.
The result of the table computation is a pair whose first component is the classification label and whose
second is the computed result.
The function
\begin{INLINEFT}%
\+HideDerTable\\
\end{INLINEFT}%
\ is parametrised by the classification $c$ of a user and replaces all
items in its operand that the user is not cleared to see by dummy values.
The antecedent of the implication in the theorem therefore asserts that a value calculated by $tc$
has revealed information about the operands that a user at classification $c$ is
not cleared to see. A table computation $tc$ belongs to the set
\begin{INLINEFT}%
\+OK\_TC\PrIJ{d} c\\
\end{INLINEFT}%
iff whenever the antecedent holds $tc$ labels the return value with a classification
that will prevent a user with classification $c$ from seeing it.

To get the induction
to go through, the property
\begin{INLINEFT}%
\+OK\_TC\PrIJ{d}\\
\end{INLINEFT}%
\ needs to be strengthened by adding an additional property
\begin{INLINEFT}%
\+OK\_TC\PrIJ{c}\\
\end{INLINEFT}%
(which ensures that the classification labels do not provide a covert channel) and
we have to define analogous properties
\begin{INLINEFT}%
\+OK\_VC\PrIJ{d}\\
\end{INLINEFT}%
\ and
\begin{INLINEFT}%
\+OK\_TC\PrIJ{c}\\
\end{INLINEFT}%
on the value computations (the definition of
\begin{INLINEFT}%
\+OK\_VC\PrIJ{d}\\
\end{INLINEFT}%
\ is given below; see \cite[{\tt fef032.doc}]{fef} for the other definitions).

As we shall see in the example, these properties are actually repesented as sets and
are parametrised by a security classification (and the induction proves that the
SSQL table and value computations belong to the appropriate sets at every classification).
In this paper, we are going to look at some tactics defined to complete the proof of
a lemma about CASE-expressions that is needed in the induction. The goal is as follows:
\begin{GFT}{}
\+?\PrPE{} \PrLF{}c te cel ee\PrLH{}\\
\+	te \PrIN{} OK\_VC\PrIJ{d} c \PrLB{}\\
\+	Elems (Map Fst cel) \PrIA{} OK\_VC\PrIJ{d} c \PrLB{}\\
\+	Elems (Map Snd cel) \PrIA{} OK\_VC\PrIJ{d} c \PrLB{}\\
\+	ee \PrIN{} OK\_VC\PrIJ{d} c \PrLE{}\\
\+	CaseVal c te cel ee \PrIN{} OK\_VC\PrIJ{d} c\\
\end{GFT}
Here
\begin{INLINEFT}%
\+CaseVal\\
\end{INLINEFT}%
\ is the constant that captures the semantics of the SSQL CASE-expression.
As in SQL, this has the syntax:
\begin{verbatim}
CASE <te> WHEN <c1> THEN <e1> ... WHEN <cN> THEN <eN> ELSE <ee>
\end{verbatim}
In the goal, $c$ is the security classification of a user executing the query
and $te$, $cel$ and $ee$ give the semantic values of the operands
of the CASE-expression, with the WHEN/THEN pairs combined into a list of pairs $cel$.
\begin{INLINEFT}%
\+Elems\\
\end{INLINEFT}%
\ is the function that maps a list to its set of elements and
\begin{INLINEFT}%
\+Fst\\
\end{INLINEFT}%
\ and
\begin{INLINEFT}%
\+Snd\\
\end{INLINEFT}%
\ are the projections.
Hence the four conjuncts in the antecedent in the goal assert in turn that
the semantics of the test expression \verb|<te>|,
the condition expressions \verb|<c1>|\ldots\verb|<cN>|,
the result expressions \verb|<e1>|\ldots\verb|<eN>| and
the else expression \verb|<ee>| satisfy the
\begin{INLINEFT}%
\+OK\_VC\PrIJ{d}\\
\end{INLINEFT}%
\ part of the inductive hypothesis. So our lemma asserts that the CASE-expression
preserves this part of the inductive hypothesis.

The property
\begin{INLINEFT}%
\+OK\_VC\PrIJ{d}\\
\end{INLINEFT}%
has the following defining theorem:
\begin{GFT}{}
\+\PrLF{}c vc\PrLH{}	vc \PrIN{} OK\_VC\PrIJ{d} c \PrKE{}\\
\+	\PrLF{}tl\PrIJ{0} tl\PrIJ{1} rl\PrIJ{0} rl\PrIJ{1} r\PrIJ{0} r\PrIJ{1}\PrLH{}\\
\+		Map (HideDerTable c) tl\PrIJ{0} = Map (HideDerTable c) tl\PrIJ{1}\\
\+	\PrLB{}	Map (HideDerTableRow c) rl\PrIJ{0} = Map (HideDerTableRow c) rl\PrIJ{1}\\
\+	\PrLB{}	HideDerTableRow c r\PrIJ{0} = HideDerTableRow c r\PrIJ{1}\\
\+	\PrLB{}	\PrLD{}Snd(vc tl\PrIJ{0} rl\PrIJ{0} r\PrIJ{0}) = Snd(vc tl\PrIJ{1} rl\PrIJ{1} r\PrIJ{1})\\
\+	\PrLE{}	\PrLD{}c dominates Fst(vc tl\PrIJ{0} rl\PrIJ{0} r\PrIJ{0})\\
\end{GFT}
Here we see that the value computation $vc$ is a function with
three arguments: a list of tables (needed for nested SELECTs),
a list of table rows (needed for GROUPBY) and a table row (the row from which individual values are extracted by name in simple expressions).
The result of the value computation is a pair whose first component is the classification label and whose
second is the value of the expression.
The hide functions set items in their operand that the user is not cleared to see to dummy values.
So very like
\begin{INLINEFT}%
\+OK\_TC\PrIJ{d}\\
\end{INLINEFT}%
\ discussed above,
membership of
\begin{INLINEFT}%
\+OK\_VC\PrIJ{d} c\\
\end{INLINEFT}%
\ asserts that if a value calculated by $vc$
has revealed information about the operands that a user at classification $c$ is
not cleared to see then $vc$ must label the value with a classification
that will prevent that user seeing it.

A design goal of SSQL was to classify data as liberally as  the
security requirements and the desire for an efficient and maintainable
implementation permitted.  For the CASE-expression, the classification
is that of the test-expression if the user is not cleared to see
the test expression; if the user is not allowed to read one of
the condition expressions that comes before the selected condition, then
the classification is the classification of that condition expression
(since the fact that that condition was not selected reveals information
about the expression); otherwise the classification is the classification is
that of the selected result expression.
The main proof plan for the lemma is an induction on the list of WHEN/THEN
pairs. The inductive step is the following goal:
\begin{GFT}{}
\+(*  5 *)\PrKM{}CaseVal c te cel ee \PrIN{} OK\_VC\PrIJ{d} c\PrKO{}\\
\+(*  4 *)\PrKM{}te \PrIN{} OK\_VC\PrIJ{d} c\PrKO{}\\
\+(*  3 *)\PrKM{}Elems (Map Fst (Cons x cel)) \PrIA{} OK\_VC\PrIJ{d} c\PrKO{}\\
\+(*  2 *)\PrKM{}Elems (Map Snd (Cons x cel)) \PrIA{} OK\_VC\PrIJ{d} c\PrKO{}\\
\+(*  1 *)\PrKM{}ee \PrIN{} OK\_VC\PrIJ{d} c\PrKO{}\\
\+\\
\+(* ?\PrPE{} *)\PrKM{}CaseVal c te (Cons x cel) ee \PrIN{} OK\_VC\PrIJ{d} c\PrKO{}\\
\end{GFT}
To make the presentation more
readable, the calculation of the classification of a CASE-expression
is defined separately from the calculation of the value of the expression.
To make the inductive hypothesis (assumption 5) usable,
these definitions have to be combined (in the lemma
\begin{INLINEFT}%
\+CaseVal\_lemma\\
\end{INLINEFT}%
) into a single primitive recursion over the WHEN/THEN list.
Unfortunately, this combined definition involves 5 conditionals
and when we expand the definition of
\begin{INLINEFT}%
\+OK\_VC\PrIJ{d}\\
\end{INLINEFT}%
\ and do some normalisation, the goal ends up containing 15 conditionals:
5 in each expression corresponding to the 3 applications of {\it vc} in the
definition
\begin{INLINEFT}%
\+OK\_VC\PrIJ{d}\\
\end{INLINEFT}%
. The last few lines of the goal are as follows:
\begin{GFT}{}
\+...\\
\+\PrLE{} \PrLD{} c dominates (\\
\+	if Snd (te tl\PrIJ{0} rl\PrIJ{0} r\PrIJ{0}) = Snd (Fst x tl\PrIJ{0} rl\PrIJ{0} r\PrIJ{0})\\
\+	then\\
\+		if c dominates Fst (te tl\PrIJ{0} rl\PrIJ{0} r\PrIJ{0})\\
\+			\PrLB{} c dominates Fst (Fst x tl\PrIJ{0} rl\PrIJ{0} r\PrIJ{0})\\
\+		then Fst (Snd x tl\PrIJ{0} rl\PrIJ{0} r\PrIJ{0})\\
\+		else if \PrLD{} c dominates Fst (te tl\PrIJ{0} rl\PrIJ{0} r\PrIJ{0})\\
\+		then Fst (te tl\PrIJ{0} rl\PrIJ{0} r\PrIJ{0})\\
\+		else Fst (Fst x tl\PrIJ{0} rl\PrIJ{0} r\PrIJ{0})\\
\+	else if c dominates Fst (te tl\PrIJ{0} rl\PrIJ{0} r\PrIJ{0})\\
\+		\PrLB{} c dominates Fst (Fst x tl\PrIJ{0} rl\PrIJ{0} r\PrIJ{0})\\
\+	then Fst (CaseVal c te cel ee tl\PrIJ{0} rl\PrIJ{0} r\PrIJ{0})\\
\+	else if \PrLD{} c dominates Fst (te tl\PrIJ{0} rl\PrIJ{0} r\PrIJ{0})\\
\+	then Fst (te tl\PrIJ{0} rl\PrIJ{0} r\PrIJ{0})\\
\+	else Fst (Fst x tl\PrIJ{0} rl\PrIJ{0} r\PrIJ{0}))\\
\end{GFT}
A case split on the first 3 tests and some rewriting
eliminates 6 of the 8 cases leaving 2 outstanding subgoals `4.1' and `4.2'.
In each of these subgoals, a case split on what are now the first
3 tests and some rewriting simplifies away all the conditionals
leaving 16 outstanding subgoals `4.1.1' \ldots `4.1.8' and `4.2.1' \ldots `4.2.8'.
An interactive attack on `4.1.1' finds a combination of tactics that proves both it and
`4.1.2', so we package this up into an {\it ad hoc} tactic {\it tac1} and try it on
all of `4.1.1' \ldots `4.1.8' and find that it proves the first 4 of them leaving
`4.1.5' to prove. Continuing in this way we end up with 4 very application-specific
tactics each of which proves 4 of the 16 subgoals and with these we can complete the proof.
The ML code of the 4 tactics is shown in Figure~\ref{fef:patterns}.

%% file: ctsconv.tex
\def\VCDate{2016/12/03}\def\VCVersion{a088e4e}
{
\def\Func#1{{\mathsf{#1}}}
\def\I{\Func{I}}
\def\K{\Func{K}}
\def\S{\Func{S}}
\def\Cos{\Func{cos}}
\def\Sin{\Func{sin}}
\def\Uncurry{\Func{Uncurry}}
\def\Frees{\Func{frees}}
\def\Vars{\Func{vars}}
\def\Constant{\Func{Constant}}
\def\Unary{\Func{unary}}
\def\Binary{\Func{binary}}
\def\Parametrized{\Func{parametrized}}
\newcommand{\spot}{{\bullet}}
\newcommand{\lam}[1]{\lambda #1 \spot\:}
\newcommand{\eps}[1]{\varepsilon #1 \spot\:}
\newcommand{\all}[1]{\forall #1 \spot\:}
\newcommand{\ex}[1]{\exists #1 \spot\:}
\newcommand{\exu}[1]{\exists! #1 \spot\:}

In this section we consider part of the implementation of a proof procedure described in \cite{Arthan16} that automates proofs that a function formed from a given set of
atomic morphisms by composition and pairing is a morphism in a concrete
category. The proof procedure is itself parametrized by theorems that characterize the concrete
category of interest and identify the atomic morphisms. So, for example, when instantiated for
the category of topological spaces and continuous functions, with the arithmetic operators and the basic transcendental
functions as the atomic morphisms, the proof procedure will automatically prove
that the function
$\lam{(x, y)} (\Sin(x) + \Cos(y) + 1) ^ 2$ is continuous.

The first step in the decision procedure is to rewrite the $\lambda$-abstraction
into a combinator form. In our example, the first step will prove the following theorem
\rev{(which we state using the notation of \cite{Arthan16} rather than ProofPower concrete syntax)}:
\begin{align*}
\vdash & (\lam{(x, y)} (\Sin(x) + \Cos(y) + 1) ^ 2) = \tag*{(*)}\\
       & (\lam{x} x ^ 2) \circ \Uncurry (+) \circ
		\langle\Sin \circ \pi_1,
			\Uncurry (+) \circ \langle\Cos \circ \pi_2, \K\,1\rangle\rangle
\end{align*}
The proof procedure then proves that the right-hand side of (*) is a morphism
in our category (i.e., a continuous function) by a process of backchaining
with theorems stating that the combinators preserve morhphismhood and that the
atomic functions are morphisms. See \cite{Arthan16} for details. In this paper,
we are only concerned with the implementation of the first step.

The implementation proves (*) by first applying a conversion
\begin{INLINEFT}%
\+\PrMM{}\_unpair\_conv\\
\end{INLINEFT}%
\ to eliminate the paired abstraction\footnote{
The implementation described here is slightly different from the algorithm
described in \cite{Arthan16} in that it handles paired abstractions
by pre-processing rather than inside the rewrite system. 
}, yielding
$$
\vdash (\lam{(x, y)} (\Sin(x) + \Cos(y) + 1) ^ 2) = \\
	(\PrMM{} p\PrLH{} (\Sin (\pi_1(p)) + \Cos (\pi_2(p)) + 1) ^ 2)
$$
\noindent
(*) is then obtained by repeated application of the following system of rewrite rules to the right-hand side of the above equation.
\[
\begin{array}{rcl@{\quad\quad}l}
(\lam{x} x) &\leadsto& \I
	&  \\
(\lam{x} t) &\leadsto& \K\,t
	& \mbox{$x\not\in \Frees(t)$}  \\
(\lam{x} (t_1, t_2)) &\leadsto& \langle(\lam{x} t_1), (\lam{x} t_2)\rangle
	&  \\
(\lam{x} f\,t) &\leadsto& f \circ (\lam{x} t)
	& \mbox{$f \in \Unary$} \\
(\lam{x} g\,t_1\,t_2) &\leadsto& \Uncurry\,g \circ \langle(\lam{x} t_1), (\lam{x} t_2)\rangle
	& \mbox{$g \in \Binary$} \\
(\lam{x} h\,t\,j) &\leadsto& (\lambda x\bullet h\,x\,j) \circ (\lam{x} t)
	& \mbox{$h \in \Parametrized$} \\
\end{array}
\]
Here $\I$ and $\K$ are the identity and constant combinators, $\circ$
is functional composition, $\langle f, g\rangle$ is $\lam{(x, y)}(f\,x, g\,y)$ and  $\Uncurry\,g$ is $\lam{(x, y)}g\,x\,y$. $\Unary$, $\Binary$ and $\Parametrized$ denote sets of atomic morphisms that are parameters to the proof procedure: $\Unary$ and $\Binary$ contain functions of one and two arguments respectively and $\Parametrized$ contains functions of two arguments where the second argument is expected to be a constant. (The set $\Parametrized$ is used to deal with families of functions like exponentiation with natural number coefficients.)

The ProofPower code that implements this rewrite system is the following.
\begin{GFT}{ProofPower Code}
\+val rec rec\_conv : CONV = (fn t => (FIRST\_C [\\
\+	i\_conv,\\
\+	k\_conv,\\
\+	pair\_conv THEN\_C RAND\_C(RANDS\_C(TRY\_C rec\_conv)),\\
\+	unary\_conv THEN\_C RIGHT\_C (TRY\_C rec\_conv),\\
\+	binary\_conv THEN\_C\\
\+		RIGHT\_C (RAND\_C(RANDS\_C (TRY\_C rec\_conv))),\\
\+	parametrized\_conv THEN\_C RIGHT\_C (TRY\_C rec\_conv)]\\
\+		AND\_OR\_C simp\_conv) t);\\
\end{GFT}
Here we have a conversion for each rule in the rewrite system, these are combined using the function {\it FIRST\_C}
which applies its argument conversions to a term in turn until it finds one that does not fail.
Note the above definition local to a function {\it morphism\_conv}
which takes as parameters the sets of unary, binary and parametrized operators.
The conversions {\it i\_conv}, {\it k\_conv} etc. each perform one rewriting step.
{\it k\_conv}, for example, attempts to rewrite a term at the
outermost level using the theorem
$$
\vdash \all{c}(\lam{x}c) = \K\,c
$$
\noindent
while {\it binary\_conv} attempts to rewrite at the outermost level
using theorems obtained from the following theorem:
$$
\vdash \all{s\,t} (\lam{x} f (s x) (t x) = \Uncurry\,f \circ \langle s, t \rangle
$$
by instantiating $f$ to each of the binary atomic morphisms.
If either of the first two conversions succeeds there is nothing more
to do. In the other four cases, we must recursively apply the rewrite system to the $\lambda$-abstractions introduced by the conversion.
This is done using functions
{\it RAND\_C} and {\it RANDS\_C}
of type
\begin{INLINEFT}%
\+CONV -> CONV\\
\end{INLINEFT}%
, which apply their argument to the operand, respectively operands, of a function application.
I.e., if {\it conv} proves $\vdash t = s, \vdash t_1 = s_1, \ldots, \vdash t_m = s_m$,
then {\it RAND\_C conv} proves all theorems of the form $\vdash f\,t = f\,s$ and {\it RANDS\_C conv} proves all theorems of the form
$\vdash g\,t_1\ldots t_m = g\,s_1 \ldots s_m$, where $g$ is not itself
an application.
The reader will observe that {\it rec\_conv} also does something that is
not specified in the rewrite system: {\it simp\_conv} is a minor optimisation:
it was found in practice that the rewrite system is prone to
produce subterms of the form $f \circ \I$. While such subterms cause
no problems with later processing, it is tidy to rewrite away
the unnecessary composition and this is done by {\it simp\_conv},
which is combined with the implementation of the rewrite system using
the conversion combinator {\it AND\_OR\_C}, which does what its name suggests.

} 